\documentclass[aps,preprint]{revtex4}%
\usepackage{amssymb}
\usepackage{amsfonts}
\usepackage{amsmath}
\usepackage{graphicx}%
\providecommand{\U}[1]{\protect\rule{.1in}{.1in}}

\begin{document}
\title{Two-player quantum games: When player strategies are via directional choices}
\author{Azhar Iqbal and Derek Abbott}
\affiliation{School of Electrical \& Electronic Engineering, University of Adelaide, South
Australia 5005, Australia.}

\begin{abstract}
We propose a scheme for a quantum game based on performing an EPR type
experiment and in which each player's spatial directional choices are
considered as their strategies. A classical mixed-strategy game is recovered
by restricting the players' choices to specific spatial trajectories. We show
that for players' directional choices for which the Bell-CHSH inequality is
violated, the players' payoffs in the quantum game have no mapping within the
classical mixed-strategy game. The scheme provides a more direct link between
classical and quantum games.

\end{abstract}
\maketitle

\section{Introduction}

Broadly speaking, a quantum game \cite{Meyer,EWL,EW,Vaidman} can be considered
as a game \cite{Binmore,Rasmusen,Osborne} in which a player's payoff relations
involve a set of quantum probabilities \cite{Peres} that are obtained from
each player's strategic actions or strategies. For instance, in the quantum
version of a $2\times2$ game proposed in the Eisert Wilkens Lewenstein
(EWL)\ scheme \cite{EWL,EW}, each player's strategies are local unitary
transformations performed on a maximally entangled state. The state evolves
unitarily and the set of quantum probabilities is obtained by projecting the
final quantum state of the game to a basis in $2\otimes2$ Hilbert space, in
terms of which the payoff relations for each player are then expressed.
Quantum games are surveyed in Refs. \cite{Kolokoltsov,FSKhan} and recent works
in this area are in Refs.
\cite{Zhang,Brunner,Pappa,Ikeda,Ikeda1,Ikeda2,Passos,Santos,Frackiewicz}. An
extensive list of articles in this area are in Ref. \cite{GoogleScholar}.

A strategy profile is a Nash equilibrium (NE) \cite{Binmore,Rasmusen,Osborne}%
---with one strategy associated with each player---such that there remains no
motivation for any player for unilaterally deviating from that profile. In the
EWL scheme, a NE is a set of local unitary transformations that satisfies the
Nash conditions.

A quantization scheme can be viewed as a mechanism that establishes a
convincing link between each player's strategies---quantum or classical---and
a set of quantum probabilities, obtained from the players' strategies, and in
terms of which each player's payoffs are then expressed. As players have
access to much larger strategy sets in EWL scheme---relative to the strategy
sets available to them in the classical game---Enk and Pike \cite{EnkPike}
argued that a quantum game in that scheme can be considered as an extended
classical game. They argued that the quantized version of a game, in
EWL\ scheme, solves a new classical game---with players' strategy sets
extended---without solving the dilemma within the original game. This led to
suggestions for using EPR type experiment
\cite{Peres,Bell1,Bell2,Bell3,Aspect,CHSH} in constructing quantum games
\cite{Iqbalepr4,Iqbalepr7} and in which each player's strategy set remain
classical while resulting in a set of quantum probabilities---thus
circumventing Enk and Pike's argument.

It appears to us that historically there have been two distinct approaches in
the literature in the area of quantum games. The first approach considers
specially-designed classical games, for instance, the game proposed by Vaidman
in Ref. \cite{Vaidman} that involves a winning condition, in which a quantum
advantage can be demonstrated directly. The second approach, however, develops
quantization procedures for a whole class of classical games, as reported in
Refs. \cite{EWL,EW}. The second approach is distinct from the first in that a
game is not designed in order to demonstrate an advantage in its quantum
mechanical implementation---usually tied to crafting a winning condition---but
the objective, instead, is to determine how an implementation that allows
access to the resources of quantum superposition and entanglement, results in
a different outcome of the game. The present paper is along the lines of the
second approach.

Non-cooperative games using a tripartite EPR experiment with GHZ states are
discussed in Refs. \cite{Iqbalepr3,Iqbalepr5}, and in references therein. A
tripartite EPR setting using GHZ states is considered in Ref.
\cite{IqbalAbbottGHZ} that presents a quantum version of a three player
non-cooperative game. Each player's strategic choices are three directions
$\mathbf{\hat{a}},$ $\mathbf{\hat{b}},$ and $\mathbf{\hat{c}}$ along which the
dichotomic observables $\mathbf{n}\cdot\mathbf{\sigma}$\ are measured, where
$\mathbf{n}=\mathbf{\hat{a}},\mathbf{\hat{b}},\mathbf{\hat{c}}$ and
$\mathbf{\sigma}$ is a vector whose components are the standard Pauli matrices
$\sigma_{x},$ $\sigma_{y},$ and $\sigma_{z}$.

In this paper, we present a scheme for playing a two-player quantum game in
which each player's (classical) strategy sets---consist of orientating his/her
unit vector along any direction in three dimensions---and dichotomic
measurement outcomes of $\pm1$ along those directions. This scheme therefore
uses each player's classical strategies to obtain a set of quantum
probabilities in terms of which each player's payoff relations are then
expressed. As the players' strategies are directional choices, Nash equilibria
of the game emerge as directional pairs. For the players' directional choices
for which the Bell-CHSH inequality is violated, the payoffs in the quantum
game cannot be mapped to a classical mixed-strategy game. As the players in
our scheme have access to classical strategy sets, it provides a more direct
link between classical and quantum games.

The mixed-strategy version of a classical game is to be faithfully imbedded
within the corresponding quantum game. When each player's strategies are
spatial directions, we find that requiring a classical mixed strategy game to
be imbedded in the corresponding quantum game results in placing constraints
on each player's available directional choices. That is, we place restrictions
on allowed trajectories on the surface of a unit sphere of the heads of the
unit vectors representing each player's strategies.

\section{Quantized Prisoners' Dilemma game}

Consider the symmetric bimatrix game%

\begin{equation}%
\begin{array}
[c]{c}%
\text{Alice}%
\end{array}%
\begin{array}
[c]{c}%
S_{1}\\
S_{2}%
\end{array}
\overset{\overset{%
\begin{array}
[c]{c}%
\text{Bob}%
\end{array}
}{%
\begin{array}
[c]{ccc}%
S_{1}^{\prime} &  & S_{2}^{\prime}%
\end{array}
}}{%
\begin{tabular}
[c]{ll}%
$(\alpha,\alpha)$ & $(\beta,\gamma)$\\
$(\gamma,\beta)$ & $(\delta,\delta)$%
\end{tabular}
}\label{GameMatrix}%
\end{equation}
in which $S_{1}$ and $S_{2}$ are Alice's moves and $S_{1}^{\prime}$ and
$S_{2}^{\prime}$ are Bob's pure strategies and the entries in the brackets are
the players' payoffs. For instance, when Alice plays $S_{1}$ whereas Bob plays
$S_{2}^{\prime}$, Alice's payoff is $\beta$ and Bob's payoff is $\gamma$. Let
the players have access to mixed strategies and $p$ is Alice's probability of
playing $S_{1}$, and thus $(1-p)$ is the probability of she playing $S_{2}$.
Likewise, $q$ is Bob's probability of playing $S_{1}^{\prime},$ and thus
$(1-q)$ is the probability of he playing $S_{2}^{\prime}.$ For the game matrix
(\ref{GameMatrix}) each players' payoffs in the mixed-strategy game are then
obtained as%

\begin{align}
\Pi_{\text{\textrm{A}}}(p,q) &  =\alpha pq+\beta p(1-q)+\gamma(1-p)q+\delta
(1-p)(1-q),\nonumber\\
\Pi_{\text{\textrm{B}}}(p,q) &  =\alpha pq+\gamma p(1-q)+\beta(1-p)q+\delta
(1-p)(1-q),\label{Mixed_strategy_payoffs}%
\end{align}
where subscripts \textrm{A} and \textrm{B} are for Alice and Bob, respectively.%

\begin{figure}[ptb]%
\centering
\includegraphics[
height=3.2958in,
width=4.3976in
]%
{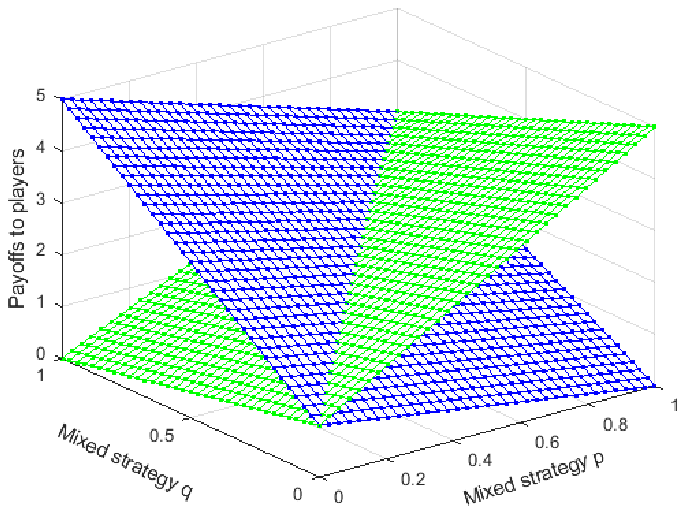}%
\caption{Plots of the mixed-strategy payoff relations of Eqs.
(\ref{Mixed_strategy_payoffs}) with $\alpha=3,$ $\beta=0,$ $\gamma=5,$ and
$\delta=1$ for the Prisoners' Dilemma game. Here $p$ and $q$ are independent
variables in the horizontal plane and the blue plane represents Alice's payoff
whereas the green plane represents Bob's payoff.}%
\end{figure}

For the strategy pair $(p^{\ast},q^{\ast})$ to be a NE---corresponding to the
two players---neither player is left with any motivation to unilaterally
deviate from it, and this is defined by Nash inequalities%

\begin{equation}
\Pi_{\mathrm{A}}(p^{\ast},q^{\ast})-\Pi_{\mathrm{A}}(p,q^{\ast})\geq0,\text{
}\Pi_{\mathrm{B}}(p^{\ast},q^{\ast})-\Pi_{\mathrm{B}}(p^{\ast},q)\geq
0.\label{NEs}%
\end{equation}
For the game of Prisoners' Dilemma considered in Ref. \cite{EWL} we have%

\begin{equation}
\alpha=3,\text{ }\beta=0,\text{ }\gamma=5,\text{ }\delta
=1,\label{PD_coefficients}%
\end{equation}
and the inequalities (\ref{NEs}) result in obtaining $p^{\ast}=0=q^{\ast}$ and
$(S_{2},S_{2}^{\prime})$ emerges as the unique NE of the game at which
$\Pi_{\mathrm{A},\mathrm{B}}(0,0)=1$.

\subsection{EWL scheme}

In the quantized version of the game (\ref{GameMatrix}) developed in Ref.
\cite{EWL}---henceforth referred to as the EWL scheme---each player's
strategies consist of local unitary transformations performed on a maximally
entangled state. The state evolves and after passing through an unentangling
gate, it is measured in a suitable basis. The game (\ref{GameMatrix}) is
played with two qubits whose quantum state is described in a $2\otimes2$
dimensional Hilbert space.

For this game, a measurement basis for the quantum state of two qubits is
chosen as $\left\vert S_{1}S_{1}^{\prime}\right\rangle ,$ $\left\vert
S_{1}S_{2}^{\prime}\right\rangle ,$ $\left\vert S_{2}S_{1}^{\prime
}\right\rangle ,$ $\left\vert S_{2}S_{2}^{\prime}\right\rangle $. An entangled
initial quantum state $\left\vert \psi_{i}\right\rangle $ is obtained by using
a two-qubit entangling gate $\hat{J}$ i.e. $\left\vert \psi_{i}\right\rangle
=\hat{J}\left\vert S_{1}S_{1}^{\prime}\right\rangle $ where $\hat{J}%
=\exp\left\{  i\gamma S_{2}\otimes S_{2}^{\prime}/2\right\}  $ and $\gamma$
$\in\lbrack0,\pi/2]$ is a measure of the game's entanglement. A separable or a
product game has $\gamma=0$ whereas a maximally entangled game has $\gamma
=\pi/2$. The players perform their local unitary transformations $\hat
{U}_{\mathrm{A}}$ and $\hat{U}_{\mathrm{B}}$ on an initial maximally entangled
state $\left\vert \psi_{i}\right\rangle $. The transformations $\hat
{U}_{\mathrm{A}}$ and $\hat{U}_{\mathrm{B}}$ were from the set%

\begin{equation}
U(\theta,\phi)=\left(
\begin{tabular}
[c]{ll}%
$e^{i\phi}\cos(\theta/2)$ & $\sin(\theta/2)$\\
$\text{-}\sin(\theta/2)$ & $e^{-i\phi}\cos(\theta/2)$%
\end{tabular}
\ \right)  ,\label{Eisert's unitary operators}%
\end{equation}
where $\theta\in\lbrack0,\pi],$ \ \ $\phi\in\lbrack0,\pi/2].$ Note that EWL
defined the unitary operator $\hat{J}=\exp\left\{  i\gamma S_{2}\otimes
S_{2}/2\right\}  $ with $\gamma\in\lbrack0,\pi/2]$ representing a measure of
the game's entanglement. Each player's actions change $\left\vert \psi
_{i}\right\rangle $ to $(\hat{U}_{\mathrm{A}}\otimes\hat{U}_{\mathrm{B}}%
)\hat{J}\left\vert S_{1}S_{1}^{\prime}\right\rangle $ and the state then
passes through an untangling gate $\hat{J}^{\dagger}$ and the state changes to
the final state i.e. $\left\vert \psi_{f}\right\rangle =\hat{J}^{\dagger}%
(\hat{U}_{\mathrm{A}}\otimes\hat{U}_{\mathrm{B}})\hat{J}\left\vert S_{1}%
S_{1}^{\prime}\right\rangle $. The state $\left\vert \psi_{f}\right\rangle $
is now measured in the basis $\left\vert S_{1}S_{1}^{\prime}\right\rangle ,$
$\left\vert S_{1}S_{2}^{\prime}\right\rangle ,$ $\left\vert S_{2}S_{1}%
^{\prime}\right\rangle ,$ $\left\vert S_{2}S_{2}^{\prime}\right\rangle $. With
the quantum probability rule, the players' payoffs are then obtained as%

\begin{align}
\Pi_{\mathrm{A}}(\hat{U}_{\mathrm{A}},\hat{U}_{\mathrm{B}}) &  =\alpha
\left\vert \left\langle S_{1}S_{1}^{\prime}\mid\psi_{f}\right\rangle
\right\vert ^{2}+\beta\left\vert \left\langle S_{1}S_{2}^{\prime}\mid\psi
_{f}\right\rangle \right\vert ^{2}+\gamma\left\vert \left\langle S_{2}%
S_{1}^{\prime}\mid\psi_{f}\right\rangle \right\vert ^{2}+\delta\left\vert
\left\langle S_{2}S_{2}^{\prime}\mid\psi_{f}\right\rangle \right\vert
^{2},\nonumber\\
\Pi_{\mathrm{B}}(\hat{U}_{\mathrm{A}},\hat{U}_{\mathrm{B}}) &  =\alpha
\left\vert \left\langle S_{1}S_{1}^{\prime}\mid\psi_{f}\right\rangle
\right\vert ^{2}+\gamma\left\vert \left\langle S_{1}S_{2}^{\prime}\mid\psi
_{f}\right\rangle \right\vert ^{2}+\beta\left\vert \left\langle S_{2}%
S_{1}^{\prime}\mid\psi_{f}\right\rangle \right\vert ^{2}+\delta\left\vert
\left\langle S_{2}S_{2}^{\prime}\mid\psi_{f}\right\rangle \right\vert
^{2}.\label{EWLpayoffs}%
\end{align}
As discussed above, Eqs. (\ref{EWLpayoffs}) show the link that this
quantization scheme establishes between each player's strategies---consisting
of unitary transformations---and the set of four quantum probabilities i.e.
$\left\vert \left\langle S_{1}S_{1}^{\prime}\mid\psi_{f}\right\rangle
\right\vert ^{2},$ $\left\vert \left\langle S_{1}S_{2}^{\prime}\mid\psi
_{f}\right\rangle \right\vert ^{2},$ $\left\vert \left\langle S_{2}%
S_{1}^{\prime}\mid\psi_{f}\right\rangle \right\vert ^{2},$ and $\left\vert
\left\langle S_{2}S_{2}^{\prime}\mid\psi_{f}\right\rangle \right\vert ^{2}$.
The NE for the quantum game consists of a pair $(\hat{U}_{\mathrm{A}}^{\ast
},\hat{U}_{\mathrm{B}}^{\ast})$---corresponding to the two players---of local
unitary transformations that satisfy the inequalities%

\begin{equation}
\Pi_{\mathrm{A}}(\hat{U}_{\mathrm{A}}^{\ast},\hat{U}_{\mathrm{B}}^{\ast}%
)-\Pi_{\mathrm{A}}(\hat{U}_{\mathrm{A}},\hat{U}_{\mathrm{B}}^{\ast}%
)\geq0,\text{ \ \ }\Pi_{\mathrm{B}}(\hat{U}_{\mathrm{A}}^{\ast},\hat
{U}_{\mathrm{B}}^{\ast})-\Pi_{\mathrm{B}}(\hat{U}_{\mathrm{A}}^{\ast},\hat
{U}_{\mathrm{B}})\geq0.\label{NashInequalities}%
\end{equation}
That is, it is a pair $(\hat{U}_{\mathrm{A}}^{\ast},\hat{U}_{\mathrm{B}}%
^{\ast})$ from which any unilateral deviation no longer improves player
payoff. For (\ref{PD_coefficients}) a unique quantum NE $(\hat{Q},\hat{Q})$
was realized where $\hat{Q}=\left(
\begin{array}
[c]{cc}%
i & 0\\
0 & -i
\end{array}
\right)  =\hat{U}(0,\pi/2).$ Benjamin and Hayden \cite{Benjamin1} noted that
when their two-parameter set is extended to include all local unitary
operations, i.e. all of $SU(2)$ \cite{Peres}, the strategy $\hat{Q}$ does not
remain an equilibrium and in the full space of deterministic quantum
strategies there exists no equilibrium for the quantum Prisoners' Dilemma.
This was also discussed further in Ref. \cite{FlitneyHollenberg}.

\section{Quantum probabilities from players' directional choices}

In EWL scheme, the players' unitary transformations $\hat{U}_{\mathrm{A}}$ and
$\hat{U}_{\mathrm{B}}$ along with the subsequent quantum measurements result
in the quantum probability set:%

\begin{equation}
\left\vert \left\langle S_{1}S_{1}^{\prime}\mid\psi_{f}\right\rangle
\right\vert ^{2},\text{ \ \ }\left\vert \left\langle S_{1}S_{2}^{\prime}%
\mid\psi_{f}\right\rangle \right\vert ^{2},\text{ \ \ }\left\vert \left\langle
S_{2}S_{1}^{\prime}\mid\psi_{f}\right\rangle \right\vert ^{2},\text{ \ \ and
}\left\vert \left\langle S_{2}S_{2}^{\prime}\mid\psi_{f}\right\rangle
\right\vert ^{2}.\label{Qprobabilities}%
\end{equation}
The players' payoff relations (\ref{EWLpayoffs}) are then expressed as
expectation values of entries in the game matrix (\ref{GameMatrix}) over the
quantum probability set (\ref{Qprobabilities}).

For a three-player symmetric game, a more direct approach in obtaining a set
of quantum probabilities is proposed in Ref. \cite{IqbalAbbottGHZ}. More
specifically, this approach cosiders tripartite EPR experiment performed on a
GHZ state as a three-player non-cooperative quantum game. Each player's
strategies are the three directions $\mathbf{\hat{a}},$ $\mathbf{\hat{b}},$
and $\mathbf{\hat{c}}$ along which the dichotomic observables $\mathbf{\hat
{n}}\cdot\mathbf{\sigma}$\ are measured, with the eigenvalues $+1$ or $-1$
where $\mathbf{\hat{n}}=\mathbf{\hat{a}},\mathbf{\hat{b}},$ or $\mathbf{\hat
{c}}$ and $\mathbf{\sigma}$ is a vector whose components are the standard
Pauli matrices $\sigma_{x},$ $\sigma_{y},$ and $\sigma_{z}$. A three-player
quantum game is developed whose underlying setup is the tripartite EPR experiment.

In the present paper---instead of each player's strategies consisting of local
unitary transformations $\hat{U}_{\mathrm{A}}$ and $\hat{U}_{\mathrm{B}}$---we
consider player \textrm{A} and \textrm{B} strategies as their directional
choices $\mathbf{\hat{a}}$ and $\mathbf{\hat{b}}$. In an EPR setting, the
measurement outcomes along $\mathbf{\hat{a}}$ and $\mathbf{\hat{b}}$ are
denoted by $m=\pm1$ and $n=\pm1$, respectively. That is, the considered
setting requires that a pair of unit vectors $(\mathbf{\hat{a}},\mathbf{\hat
{b}})$ results in a set of quantum probabilities:%

\begin{equation}
\Pr_{{\small Q}}(S_{1},S_{1}^{\prime}),\text{ }\Pr_{{\small Q}}(S_{1}%
,S_{2}^{\prime}),\text{ }\Pr_{{\small Q}}(S_{2},S_{1}^{\prime}),\text{ }%
\Pr_{{\small Q}}(S_{2},S_{2}^{\prime}),\label{QprobabilitiesG}%
\end{equation}
where $%
{\displaystyle\sum}
\Pr_{{\small Q}}(S_{1},S_{1}^{\prime})+\Pr_{{\small Q}}(S_{1},S_{2}^{\prime
})+\Pr_{{\small Q}}(S_{2},S_{1}^{\prime})+\Pr_{{\small Q}}(S_{2},S_{2}%
^{\prime})=1.$ Now, acknowledging that there is no unique way in obtaining the
set (\ref{QprobabilitiesG}) from each players' strategies $(\mathbf{\hat{a}%
},\mathbf{\hat{b}})$, we propose to obtain this set as follows%

\begin{align}
\Pr_{{\small Q}}(S_{1},S_{1}^{\prime}) &  =\Pr_{{\small Q}}[(\mathbf{\hat{a}%
},m=+1),(\mathbf{\hat{b}},l=+1)],\text{ }\Pr_{{\small Q}}(S_{1},S_{2}^{\prime
})=\Pr_{{\small Q}}[(\mathbf{\hat{a}},m=+1),(\mathbf{\hat{b}}%
,l=-1)],\nonumber\\
\Pr_{{\small Q}}(S_{2},S_{1}^{\prime}) &  =\Pr_{{\small Q}}[(\mathbf{\hat{a}%
},m=-1),(\mathbf{\hat{b}},l=+1)],\text{ }\Pr_{{\small Q}}(S_{2},S_{2}^{\prime
})=\Pr_{{\small Q}}[(\mathbf{\hat{a}},m=-1),(\mathbf{\hat{b}}%
,l=-1)].\label{QprobsFromDirs}%
\end{align}
For instance, $\Pr_{{\small Q}}(S_{1},S_{2}^{\prime})$ is the quantum
probability that the polarization measurement along $\mathbf{\hat{a}}$ gives
the outcome $m=+1$ and polarization measurement along $\mathbf{\hat{b}}$ gives
the outcome $n=-1$.

The probabilities (\ref{QprobsFromDirs}) are obtained as%

\begin{align}
\Pr_{{\small Q}}(S_{1},S_{1}^{\prime}) &  =\left\vert \left\langle
\psi_{_{ini}}\mid(\mid\psi_{+1}\right\rangle _{\mathbf{\hat{a}}}%
\otimes\left\vert \psi_{+1}\right\rangle _{\mathbf{\hat{b}}})\right\vert
^{2}=\left\vert \left\langle \psi_{+1}^{\mathbf{\hat{a}}}\psi_{+1}%
^{\mathbf{\hat{b}}}\mid\psi_{\mathrm{ini}}\right\rangle \right\vert
^{2},\nonumber\\
\Pr_{{\small Q}}(S_{1},S_{2}^{\prime}) &  =\left\vert \left\langle
\psi_{_{ini}}\mid(\mid\psi_{+1}\right\rangle _{\mathbf{\hat{a}}}%
\otimes\left\vert \psi_{-1}\right\rangle _{\mathbf{\hat{b}}})\right\vert
^{2}=\left\vert \left\langle \psi_{+1}^{\mathbf{\hat{a}}}\psi_{-1}%
^{\mathbf{\hat{b}}}\mid\psi_{\mathrm{ini}}\right\rangle \right\vert
^{2},\nonumber\\
\Pr_{{\small Q}}(S_{2},S_{1}^{\prime}) &  =\left\vert \left\langle
\psi_{_{ini}}\mid(\mid\psi_{-1}\right\rangle _{\mathbf{\hat{a}}}%
\otimes\left\vert \psi_{+1}\right\rangle _{\mathbf{\hat{b}}})\right\vert
^{2}=\left\vert \left\langle \psi_{-1}^{\mathbf{\hat{a}}}\psi_{+1}%
^{\mathbf{\hat{b}}}\mid\psi_{\mathrm{ini}}\right\rangle \right\vert
^{2},\nonumber\\
\Pr_{{\small Q}}(S_{2},S_{2}^{\prime}) &  =\left\vert \left\langle
\psi_{_{ini}}\mid(\mid\psi_{-1}\right\rangle _{\mathbf{\hat{a}}}%
\otimes\left\vert \psi_{-1}\right\rangle _{\mathbf{\hat{b}}})\right\vert
^{2}=\left\vert \left\langle \psi_{-1}^{\mathbf{\hat{a}}}\psi_{-1}%
^{\mathbf{\hat{b}}}\mid\psi_{\mathrm{ini}}\right\rangle \right\vert
^{2},\label{Q_probabilities}%
\end{align}
and each players' payoff relations are then%

\begin{align}
\Pi_{\mathrm{A}}(\mathbf{\hat{a}},\mathbf{\hat{b}}) &  =\alpha\Pr_{{\small Q}%
}(S_{1},S_{1}^{\prime})+\beta\Pr_{{\small Q}}(S_{1},S_{2}^{\prime})+\gamma
\Pr_{{\small Q}}(S_{2},S_{1}^{\prime})+\delta\Pr_{{\small Q}}(S_{2}%
,S_{2}^{\prime}),\label{Payoff_Alice}\\
\Pi_{\mathrm{B}}(\mathbf{\hat{a}},\mathbf{\hat{b}}) &  =\alpha\Pr_{{\small Q}%
}(S_{1},S_{1}^{\prime})+\gamma\Pr_{{\small Q}}(S_{1},S_{2}^{\prime})+\beta
\Pr_{{\small Q}}(S_{2},S_{1}^{\prime})+\delta\Pr_{{\small Q}}(S_{2}%
,S_{2}^{\prime}).\label{Payoff_Bob}%
\end{align}
A directional pair $(\mathbf{\hat{a}}^{\ast},\mathbf{\hat{b}}^{\ast})$ is a NE
when the inequalities%

\begin{equation}
\Pi_{\mathrm{A}}(\mathbf{\hat{a}}^{\ast},\mathbf{\hat{b}}^{\ast}%
)-\Pi_{\mathrm{A}}(\mathbf{\hat{a}},\mathbf{\hat{b}}^{\ast})\geq0,\text{
\ \ }\Pi_{\mathrm{B}}(\mathbf{\hat{a}}^{\ast},\mathbf{\hat{b}}^{\ast}%
)-\Pi_{\mathrm{B}}(\mathbf{\hat{a}}^{\ast},\mathbf{\hat{b}})\geq0,
\end{equation}
are true for any directional choices $\mathbf{\hat{a}}$ and $\mathbf{\hat{b}}
$ by players \textrm{A} and \textrm{B}, respectively.

Given each player's strategies consisting of directional choices in three
dimensions, the classical mixed-strategy game is recoverede from the quantum
game if each player's directional choices consist of orientating their
respective unit vectors $\mathbf{\hat{a}}$ and $\mathbf{\hat{b}}$ along
specific trajectories on the surface of a unit sphere. When the players allow
their respective unit vectors $\mathbf{\hat{a}}$ and $\mathbf{\hat{b}}$ to be
orientated along directions beyond these trajectories, it results in obtaining
the quantum game.

\subsection{Orientating a unit vector considered as each player's strategy}

In an EPR setting, we note that with player \textrm{A}'s strategy
$\mathbf{\hat{a},}$ the polarization (or spin) measurement results in the
outcome $m=\pm1,$ and with the player \textrm{B}'s strategy $\mathbf{\hat{b}}$
the polarization measurement results in the outcome $n=\pm1$. We consider
Pauli's matrices $\sigma_{x}=\left(
\begin{array}
[c]{cc}%
0 & 1\\
1 & 0
\end{array}
\right)  ,$ $\sigma_{y}=\left(
\begin{array}
[c]{cc}%
0 & -i\\
i & 0
\end{array}
\right)  ,$ $\sigma_{z}=\left(
\begin{array}
[c]{cc}%
1 & 0\\
0 & -1
\end{array}
\right)  $ in the eigenbasis $\left\vert 0\right\rangle =\left(
\begin{array}
[c]{c}%
1\\
0
\end{array}
\right)  ,$ $\left\vert 1\right\rangle =\left(
\begin{array}
[c]{c}%
0\\
1
\end{array}
\right)  $:%

\begin{equation}
\sigma_{x}=\left\vert 0\right\rangle \left\langle 1\right\vert +\left\vert
1\right\rangle \left\langle 0\right\vert ,\text{ }\sigma_{y}=i(\left\vert
1\right\rangle \left\langle 0\right\vert -\left\vert 0\right\rangle
\left\langle 1\right\vert ),\text{ }\sigma_{z}=\left\vert 0\right\rangle
\left\langle 0\right\vert -\left\vert 1\right\rangle \left\langle 1\right\vert
,\label{DiagRep}%
\end{equation}
with $\mathbf{\sigma}=\sigma_{x}\hat{\imath}+\sigma_{y}\hat{\jmath}+\sigma
_{z}\hat{k}$ and $\mathbf{\hat{a}}=a_{x}\hat{\imath}+a_{y}\hat{\jmath}%
+a_{z}\hat{k},$ we have $\mathbf{\sigma}\cdot\mathbf{\hat{a}}=a_{x}\sigma
_{x}+a_{y}\sigma_{y}+a_{z}\sigma_{z},$ $\mathbf{\sigma}\cdot\mathbf{\hat{b}%
}=b_{x}\sigma_{x}+b_{y}\sigma_{y}+b_{z}\sigma_{z}$ that can be expressed in
the diagonal form as $\mathbf{\sigma}\cdot\mathbf{\hat{a}}=(a_{x}%
-ia_{y})\left\vert 0\right\rangle \left\langle 1\right\vert +(a_{x}%
+ia_{y})\left\vert 1\right\rangle \left\langle 0\right\vert +a_{z}(\left\vert
0\right\rangle \left\langle 0\right\vert -\left\vert 1\right\rangle
\left\langle 1\right\vert ).$ Let $\left\vert \psi\right\rangle =\alpha
\left\vert 0\right\rangle +\beta\left\vert 1\right\rangle $ with $\left\vert
\alpha\right\vert ^{2}+\left\vert \beta\right\vert ^{2}=1$ be the eigenstate
of $\mathbf{\sigma}\cdot\mathbf{\hat{a}}$ with the eigenvalue $k=\pm1$ i.e.
$(\mathbf{\sigma}\cdot\mathbf{\hat{a}})\left\vert \psi\right\rangle
=k\left\vert \psi\right\rangle $ or $(\mathbf{\sigma}\cdot\mathbf{\hat{a}%
})(\alpha\left\vert 0\right\rangle +\beta\left\vert 1\right\rangle
)=k(\alpha\left\vert 0\right\rangle +\beta\left\vert 1\right\rangle ),$ or
$(\mathbf{\sigma}\cdot\mathbf{\hat{a}})(\alpha\left\vert 0\right\rangle
+\beta\left\vert 1\right\rangle )=[\alpha a_{z}+\beta(a_{x}-ia_{y})]\left\vert
0\right\rangle +[\alpha(a_{x}+ia_{y})-\beta a_{z}]\left\vert 1\right\rangle
=k(\alpha\left\vert 0\right\rangle +\beta\left\vert 1\right\rangle )$ which
gives $\alpha a_{z}+\beta(a_{x}-ia_{y})=k\alpha,$ \ \ $\alpha(a_{x}%
+ia_{y})-\beta a_{z}=k\beta,$ and the normalized eigenstates for \textrm{A}
with eigenvalues $+1$ and $-1$, respectively, are%

\begin{equation}
\left\vert \psi_{+1}^{\mathbf{\hat{a}}}\right\rangle =\frac{1}{\sqrt{2}%
}{\huge [}\sqrt{1+a_{z}}\left\vert 0\right\rangle +\frac{a_{x}+ia_{y}}%
{\sqrt{1+a_{z}}}\left\vert 1\right\rangle {\huge ]},\text{ \ \ }\left\vert
\psi_{-1}^{\mathbf{\hat{a}}}\right\rangle =\frac{1}{\sqrt{2}}{\huge [}%
\sqrt{1-a_{z}}\left\vert 0\right\rangle -\frac{a_{x}+ia_{y}}{\sqrt{1-a_{z}}%
}\left\vert 1\right\rangle {\huge ]}.
\end{equation}
Likewise, the eigenstates for \textrm{B} with the eigenvalues $+1$ and $-1$,
respectively, are%

\begin{equation}
\left\vert \psi_{+1}^{\mathbf{\hat{b}}}\right\rangle =\frac{1}{\sqrt{2}%
}{\huge [}\sqrt{1+b_{z}}\left\vert 0\right\rangle +\frac{b_{x}+ib_{y}}%
{\sqrt{1+b_{z}}}\left\vert 1\right\rangle {\huge ]},\text{ \ \ }\left\vert
\psi_{-1}^{_{\mathbf{\hat{b}}}}\right\rangle =\frac{1}{\sqrt{2}}{\huge [}%
\sqrt{1-b_{z}}\left\vert 0\right\rangle -\frac{b_{x}+ib_{y}}{\sqrt{1-b_{z}}%
}\left\vert 1\right\rangle {\huge ]}.
\end{equation}
From these we then obtain the eigenstates:%

\begin{align}
\left\vert \psi_{+1}^{\mathbf{\hat{a}}}\psi_{+1}^{\mathbf{\hat{b}}%
}\right\rangle  &  =\frac{1}{2}{\huge [}\sqrt{(1+a_{z})(1+b_{z})}\left\vert
00\right\rangle +\sqrt{\frac{1+a_{z}}{1+b_{z}}}(b_{x}+ib_{y})\left\vert
01\right\rangle +\nonumber\\
&  \sqrt{\frac{1+b_{z}}{1+a_{z}}}(a_{x}+ia_{y})\left\vert 10\right\rangle
+\frac{(a_{x}+ia_{y})(b_{x}+ib_{y})}{\sqrt{(1+a_{z})(1+b_{z})}}\left\vert
11\right\rangle {\huge ],}\label{a_Eigen+1+1}\\
\left\vert \psi_{+1}^{\mathbf{\hat{a}}}\psi_{-1}^{\mathbf{\hat{b}}%
}\right\rangle  &  =\frac{1}{2}{\huge [}\sqrt{(1+a_{z})(1-b_{z})}\left\vert
00\right\rangle -\sqrt{\frac{1+a_{z}}{1-b_{z}}}(b_{x}+ib_{y})\left\vert
01\right\rangle +\nonumber\\
&  \sqrt{\frac{1-b_{z}}{1+a_{z}}}(a_{x}+ia_{y})\left\vert 10\right\rangle
-\frac{(a_{x}+ia_{y})(b_{x}+ib_{y})}{\sqrt{(1+a_{z})(1-b_{z})}}\left\vert
11\right\rangle {\huge ],}\label{a_Eigen+1-1}\\
\left\vert \psi_{-1}^{\mathbf{\hat{a}}}\psi_{+1}^{\mathbf{\hat{b}}%
}\right\rangle  &  =\frac{1}{2}{\huge [}\sqrt{(1-a_{z})(1+b_{z})}\left\vert
00\right\rangle +\sqrt{\frac{1-a_{z}}{1+b_{z}}}(b_{x}+ib_{y})\left\vert
01\right\rangle -\nonumber\\
&  \sqrt{\frac{1+b_{z}}{1-a_{z}}}(a_{x}+ia_{y})\left\vert 10\right\rangle
-\frac{(a_{x}+ia_{y})(b_{x}+ib_{y})}{\sqrt{(1-a_{z})(1+b_{z})}}\left\vert
11\right\rangle {\huge ],}\label{a_Eigen-1+1}\\
\left\vert \psi_{-1}^{\mathbf{\hat{a}}}\psi_{-1}^{\mathbf{\hat{b}}%
}\right\rangle  &  =\frac{1}{2}{\huge [}\sqrt{(1-a_{z})(1-b_{z})}\left\vert
00\right\rangle -\sqrt{\frac{1-a_{z}}{1-b_{z}}}(b_{x}+ib_{y})\left\vert
01\right\rangle -\nonumber\\
&  \sqrt{\frac{1-b_{z}}{1-a_{z}}}(a_{x}+ia_{y})\left\vert 10\right\rangle
+\frac{(a_{x}+ia_{y})(b_{x}+ib_{y})}{\sqrt{(1-a_{z})(1-b_{z})}}\left\vert
11\right\rangle {\huge ]}.\label{a_Eigen-1-1}%
\end{align}
For instance, the eigenstate (\ref{a_Eigen-1+1}) corresponds when player
\textrm{A}'s strategy consists of orientating her unit vector $\mathbf{\hat
{a}}$ in one specific spatial direction whereas player \textrm{B}'s strategy
consist of orientating his unit vector $\mathbf{\hat{b}}$ in other specific
spatial direction and the measurement in an EPR setting generates $-1$ on
\textrm{A}'s side and $+1$ on \textrm{B}'s side. Quantum probabilities
$\Pr_{{\small Q}}(S_{1},S_{1}^{\prime}),$ $\Pr_{{\small Q}}(S_{1}%
,S_{2}^{\prime}),$ $\Pr_{{\small Q}}(S_{2},S_{1}^{\prime}),$ and
$\Pr_{{\small Q}}(S_{2},S_{2}^{\prime})$ are determined from these eigenstates
using Eqs. (\ref{Q_probabilities}). That is, with the players' directional
choices $\mathbf{\hat{a}}$ and $\mathbf{\hat{b}}$, the new basis consisting of
the kets $\left\vert \psi_{+1}^{\mathbf{\hat{a}}}\psi_{+1}^{\mathbf{\hat{b}}%
}\right\rangle ,$ $\left\vert \psi_{+1}^{\mathbf{\hat{a}}}\psi_{-1}%
^{\mathbf{\hat{b}}}\right\rangle ,$ $\left\vert \psi_{-1}^{\mathbf{\hat{a}}%
}\psi_{+1}^{\mathbf{\hat{b}}}\right\rangle ,$ $\left\vert \psi_{-1}%
^{\mathbf{\hat{a}}}\psi_{-1}^{\mathbf{\hat{b}}}\right\rangle $ is prepared
onto which the initial state is then projected, during the quantum
measurement, to obtain the set of quantum probabilities.

Although the players' strategy sets consist of classical actions of rotating
their respective unit vectors in three dimensions, the considered game is
genuinely quantum mechanical because the player's payoff relations have an
underlying set of quantum mechanical probabilities. In particular, the players
have access to directional choices along which Bell's inequalities can be
violated. This indicates genuinely quantum mechanical character of this scheme.

In the following, we present the resulting quantum games when the initial
quantum states $\left\vert \psi_{\mathrm{ini}}\right\rangle $ are the product
state $\frac{1}{2}\left(  \left\vert 00\right\rangle +\left\vert
01\right\rangle +\left\vert 10\right\rangle +\left\vert 11\right\rangle
\right)  ,$ the maximally entangled state$\frac{1}{\sqrt{2}}\left(  \left\vert
00\right\rangle +i\left\vert 11\right\rangle \right)  ,$ and the entangled
state $\frac{1}{2}\left(  \left\vert 00\right\rangle +\left\vert
01\right\rangle -\left\vert 10\right\rangle +\left\vert 11\right\rangle
\right)  $.

\section{Game with the quantum state $\left\vert \psi_{\mathrm{ini}%
}\right\rangle =\frac{1}{2}\left(  \left\vert 00\right\rangle +\left\vert
01\right\rangle +\left\vert 10\right\rangle +\left\vert 11\right\rangle
\right)  $}

For this state we can write%

\begin{equation}
\left\vert \psi_{\mathrm{ini}}\right\rangle =\frac{1}{2}\left(  \left\vert
00\right\rangle +\left\vert 01\right\rangle +\left\vert 10\right\rangle
+\left\vert 11\right\rangle \right)  =\frac{(\left\vert 0\right\rangle
+\left\vert 1\right\rangle )_{\mathrm{A}}}{\sqrt{2}}\otimes\frac{(\left\vert
0\right\rangle +\left\vert 1\right\rangle )_{\mathrm{B}}}{\sqrt{2}},
\end{equation}
i.e. the state is a product state. For this state, we find%

\begin{gather}
\Pr(\mathbf{\hat{a}}_{+1},\mathbf{\hat{b}}_{+1})=\left\vert \left\langle
\psi_{+1}^{\mathbf{\hat{a}}}\psi_{+1}^{\mathbf{\hat{b}}}\mid\psi
_{\mathrm{ini}}\right\rangle \right\vert ^{2}=\nonumber\\
\frac{1}{16(1+a_{z})(1+b_{z})}{\Large \{}\left[  (1+a_{z})(1+b_{z}%
)+(1+a_{z})b_{x}+(1+b_{z})a_{x}+(a_{x}b_{x}-a_{y}b_{y})\right]  ^{2}%
+\nonumber\\
\left[  (1+a_{z})b_{y}+(1+b_{z})a_{y}+(a_{x}b_{y}+a_{y}b_{x})\right]
^{2}{\large \}},\\
\Pr(\mathbf{\hat{a}}_{+1},\mathbf{\hat{b}}_{-1})=\left\vert \left\langle
\psi_{+1}^{\mathbf{\hat{a}}}\psi_{-1}^{\mathbf{\hat{b}}}\mid\psi
_{\mathrm{ini}}\right\rangle \right\vert ^{2}=\nonumber\\
\frac{1}{16(1+a_{z})(1-b_{z})}{\Large \{}\left[  (1+a_{z})(1-b_{z}%
)-(1+a_{z})b_{x}+(1-b_{z})a_{x}-(a_{x}b_{x}-a_{y}b_{y})\right]  ^{2}%
+\nonumber\\
\left[  (1+a_{z})b_{y}-(1-b_{z})a_{y}+(a_{x}b_{y}+a_{y}b_{x})\right]
^{2}{\large \}},\\
\Pr(\mathbf{\hat{a}}_{-1},\mathbf{\hat{b}}_{+1})=\left\vert \left\langle
\psi_{-1}^{\mathbf{\hat{a}}}\psi_{+1}^{\mathbf{\hat{b}}}\mid\psi
_{\mathrm{ini}}\right\rangle \right\vert ^{2}=\nonumber\\
\frac{1}{16(1-a_{z})(1+b_{z})}{\Large \{}\left[  (1-a_{z})(1+b_{z}%
)+(1-a_{z})b_{x}-(1+b_{z})a_{x}-(a_{x}b_{x}-a_{y}b_{y})\right]  ^{2}%
+\nonumber\\
\left[  (1-a_{z})b_{y}-(1+b_{z})a_{y}-(a_{x}b_{y}+a_{y}b_{x})\right]
^{2}{\large \}},\\
\Pr(\mathbf{\hat{a}}_{-1},\mathbf{\hat{b}}_{-1})=\left\vert \left\langle
\psi_{-1}^{\mathbf{\hat{a}}}\psi_{-1}^{\mathbf{\hat{b}}}\mid\psi
_{\mathrm{ini}}\right\rangle \right\vert ^{2}=\nonumber\\
\frac{1}{16(1-a_{z})(1-b_{z})}{\Large \{}\left[  (1-a_{z})(1-b_{z}%
)-(1-a_{z})b_{x}-(1-b_{z})a_{x}+(a_{x}b_{x}-a_{y}b_{y})\right]  ^{2}%
+\nonumber\\
\left[  (1-a_{z})b_{y}+(1-b_{z})a_{y}-(a_{x}b_{y}+a_{y}b_{x})\right]
^{2}{\large \}}.
\end{gather}

The payoff to the players (\ref{Payoff_Alice}) can then be expressed as%

\begin{gather}
\Pi_{\mathrm{A},\mathrm{B}}(\mathbf{\hat{a}},\mathbf{\hat{b}})=\nonumber\\
\frac{1}{16(1+a_{z})}{\huge [}\frac{(\alpha,\alpha)}{(1+b_{z})}{\Large \{}%
\left[  (1+a_{z})(1+b_{z})+(1+a_{z})b_{x}+(1+b_{z})a_{x}+(a_{x}b_{x}%
-a_{y}b_{y})\right]  ^{2}+\nonumber\\
\left[  (1+a_{z})b_{y}+(1+b_{z})a_{y}+(a_{x}b_{y}+a_{y}b_{x})\right]
^{2}{\large \}}+\nonumber\\
\frac{(\beta,\gamma)}{(1-b_{z})}{\Large \{}\left[  (1+a_{z})(1-b_{z}%
)-(1+a_{z})b_{x}+(1-b_{z})a_{x}-(a_{x}b_{x}-a_{y}b_{y})\right]  ^{2}%
+\nonumber\\
\left[  (1+a_{z})b_{y}-(1-b_{z})a_{y}+(a_{x}b_{y}+a_{y}b_{x})\right]
^{2}{\large \}}{\huge ]}+\nonumber\\
\frac{1}{16(1-a_{z})}{\huge [}\frac{(\gamma,\beta)}{(1+b_{z})}{\Large \{}%
\left[  (1-a_{z})(1+b_{z})+(1-a_{z})b_{x}-(1+b_{z})a_{x}-(a_{x}b_{x}%
-a_{y}b_{y})\right]  ^{2}+\nonumber\\
\left[  (1-a_{z})b_{y}-(1+b_{z})a_{y}-(a_{x}b_{y}+a_{y}b_{x})\right]
^{2}{\large \}}+\nonumber\\
\frac{(\delta,\delta)}{(1-b_{z})}{\Large \{}\left[  (1-a_{z})(1-b_{z}%
)-(1-a_{z})b_{x}-(1-b_{z})a_{x}+(a_{x}b_{x}-a_{y}b_{y})\right]  ^{2}%
+\nonumber\\
\left[  (1-a_{z})b_{y}+(1-b_{z})a_{y}-(a_{x}b_{y}+a_{y}b_{x})\right]
^{2}{\large \}}{\huge ],}%
\end{gather}
To convert to polar coordinates we let $\mathbf{\hat{a}}=(\theta_{\mathrm{A}%
},\phi_{\mathrm{A}})$ and $\mathbf{\hat{b}}=(\theta_{\mathrm{B}}%
,\phi_{\mathrm{B}})$ with $\theta_{\mathrm{A}},\theta_{\mathrm{B}}\in
\lbrack0,\pi]$ and $\phi_{\mathrm{A}},\phi_{\mathrm{B}}\in\lbrack0,2\pi)$ and
have
\begin{align}
a_{x} &  =\sin\theta_{\mathrm{A}}\cos\phi_{\mathrm{A}},\text{ }b_{x}%
=\sin\theta_{\mathrm{B}}\cos\phi_{\mathrm{B}}\nonumber\\
a_{y} &  =\sin\theta_{\mathrm{A}}\sin\phi_{\mathrm{A}},\text{ }b_{y}%
=\sin\theta_{\mathrm{B}}\sin\phi_{\mathrm{B}}\nonumber\\
a_{z} &  =\cos\theta_{\mathrm{A}},\text{ }b_{z}=\cos\theta_{\mathrm{B}%
}.\label{Tranfs}%
\end{align}
This transformation reduces the independent variables $\hat{a}$ and $\hat{b}$
to $\theta_{A},\theta_{B},\phi_{A},$ and $\phi_{B}$ and players payoffs are
then expressed as%

\begin{gather}
\Pi_{A,B}(\hat{a},\hat{b})=\Pi_{A,B}(\theta_{A},\phi_{A};\theta_{B},\phi
_{B})=\nonumber\\
\frac{(\alpha,\alpha)}{16(1+\cos\theta_{A})(1+\cos\theta_{B})}{\Large \{}%
[(1+\cos\theta_{A})(1+\cos\theta_{B})+(1+\cos\theta_{A})\sin\theta_{B}\cos
\phi_{B}+\nonumber\\
(1+\cos\theta_{B})\sin\theta_{A}\cos\phi_{A}+\sin\theta_{A}\sin\theta_{B}%
\cos(\phi_{A}+\phi_{B})]^{2}+[(1+\cos\theta_{A})\sin\theta_{B}\sin\phi
_{B}+\nonumber\\
(1+\cos\theta_{B})\sin\theta_{A}\sin\phi_{A}+\sin\theta_{A}\sin\theta_{B}%
\sin(\phi_{A}+\phi_{B})]^{2}{\large \}}+\nonumber\\
\frac{(\beta,\gamma)}{16(1+\cos\theta_{A})(1-\cos\theta_{B})}{\Large \{}%
[(1+\cos\theta_{A})(1-\cos\theta_{B})-(1+\cos\theta_{A})\sin\theta_{B}\cos
\phi_{B}+\nonumber\\
(1-\cos\theta_{B})\sin\theta_{A}\cos\phi_{A}-\sin\theta_{A}\sin\theta_{B}%
\cos(\phi_{A}+\phi_{B})]^{2}+[(1+\cos\theta_{A})\sin\theta_{B}\sin\phi
_{B}-\nonumber\\
(1-\cos\theta_{B})\sin\theta_{A}\sin\phi_{A}+\sin\theta_{A}\sin\theta_{B}%
\sin(\phi_{A}+\phi_{B})]^{2}{\large \}}+\nonumber\\
\frac{(\gamma,\beta)}{16(1-\cos\theta_{A})(1+\cos\theta_{B})}{\Large \{}%
[(1-\cos\theta_{A})(1+\cos\theta_{B})+(1-\cos\theta_{A})\sin\theta_{B}\cos
\phi_{B}-\nonumber\\
(1+\cos\theta_{B})\sin\theta_{A}\cos\phi_{A}-\sin\theta_{A}\sin\theta_{B}%
\cos(\phi_{A}+\phi_{B})]^{2}+\nonumber\\
\lbrack(1-\cos\theta_{A})\sin\theta_{B}\sin\phi_{B}-(1+\cos\theta_{B}%
)\sin\theta_{A}\sin\phi_{A}-\sin\theta_{A}\sin\theta_{B}\sin(\phi_{A}+\phi
_{B})]^{2}{\large \}}+\nonumber\\
\frac{(\delta,\delta)}{16(1-\cos\theta_{A})(1-\cos\theta_{B})}{\Large \{}%
[(1-\cos\theta_{A})(1-\cos\theta_{B})-(1-\cos\theta_{A})\sin\theta_{B}\cos
\phi_{B}-\nonumber\\
(1-\cos\theta_{B})\sin\theta_{A}\cos\phi_{A}+\sin\theta_{A}\sin\theta_{B}%
\cos(\phi_{A}+\phi_{B})]^{2}+[(1-\cos\theta_{A})\sin\theta_{B}\sin\phi
_{B}+\nonumber\\
(1-\cos\theta_{B})\sin\theta_{A}\sin\phi_{A}-\sin\theta_{A}\sin\theta_{B}%
\sin(\phi_{A}+\phi_{B})]^{2}{\large \}}{\huge .}\label{Payoffs_gen}%
\end{gather}
The transformation (\ref{Tranfs}) reduces the independent variables
$\mathbf{\hat{a}}$ and $\mathbf{\hat{b}}$ to $\theta_{\mathrm{A}}%
,\theta_{\mathrm{B}},\phi_{\mathrm{A}},$ and $\phi_{\mathrm{B}}$ and players
payoffs are then expressed as%

\begin{gather}
\Pi_{\mathrm{A},\mathrm{B}}(\theta_{\mathrm{A}},\phi_{\mathrm{A}}%
;\theta_{\mathrm{B}},\phi_{\mathrm{B}})=\nonumber\\
\frac{1}{4}{\Large [}(\alpha,\alpha)(1+\sin\theta_{\mathrm{A}}\cos
\phi_{\mathrm{A}})(1+\sin\theta_{\mathrm{B}}\cos\phi_{\mathrm{B}}%
)+(\beta,\gamma)(1+\sin\theta_{\mathrm{A}}\cos\phi_{\mathrm{A}})(1-\sin
\theta_{\mathrm{B}}\cos\phi_{\mathrm{B}})+\nonumber\\
(\gamma,\beta)(1-\sin\theta_{\mathrm{A}}\cos\phi_{\mathrm{A}})(1+\sin
\theta_{\mathrm{B}}\cos\phi_{\mathrm{B}})+(\delta,\delta)(1-\sin
\theta_{\mathrm{A}}\cos\phi_{\mathrm{A}})(1-\sin\theta_{\mathrm{B}}\cos
\phi_{\mathrm{B}}){\Large ].}\label{QPayoffs}%
\end{gather}

Note that EWL used the notation $\phi_{\mathrm{A,B}}$ to describe one of the
two parameters in terms of which their (restricted) local unitary operators
are defined. In this paper, we have used notation $\phi_{\mathrm{A,B}}$ when
we change from Cartesian to spherical coordinates in accordance with Eqs.
(\ref{Tranfs}) i.e. our context is different. Appendix A and B detail the
simplification of the first and the second terms, respectively, in
Eq.~(\ref{Payoffs_gen}) to obtain Eq.~(\ref{QPayoffs}). Comparing Eq.
(\ref{QPayoffs}) with Eq. (\ref{Mixed_strategy_payoffs}), it is noticed that
when we take%

\begin{equation}
p=(1+\sin\theta_{\mathrm{A}}\cos\phi_{\mathrm{A}})/2,\text{ \ \ \ \ }%
q=(1+\sin\theta_{\mathrm{B}}\cos\phi_{\mathrm{B}})/2,\label{product_state_1}%
\end{equation}
and thus%

\begin{equation}
(1-p)=(1-\sin\theta_{\mathrm{A}}\cos\phi_{\mathrm{A}})/2,\text{ \ \ \ }%
(1-q)=(1-\sin\theta_{\mathrm{B}}\cos\phi_{\mathrm{B}}%
)/2,\label{product_state_2}%
\end{equation}
the quantum payoffs (\ref{QPayoffs}) are then reduced to players' classical
mixed strategy payoffs (\ref{Mixed_strategy_payoffs}).%

\begin{figure}[ptb]%
\centering
\includegraphics[
height=3.4791in,
width=4.6423in
]%
{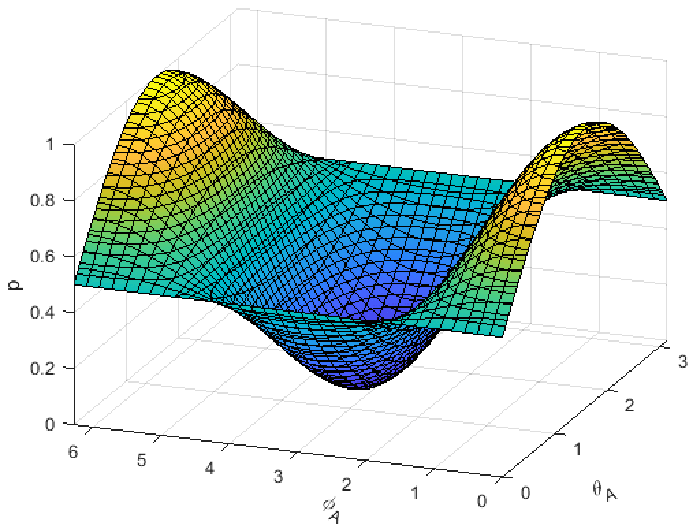}%
\caption{Alice's mixed strategy $p$ is plotted using Eq.
(\ref{product_state_1}) against $\theta_{A}$ and $\phi_{A}$ for the product
state $\left\vert \psi_{\mathrm{ini}}\right\rangle =\frac{1}{2}\left(
\left\vert 00\right\rangle +\left\vert 01\right\rangle +\left\vert
10\right\rangle +\left\vert 11\right\rangle \right)  .$}%
\end{figure}

This can be interpreted by stating that the quantum game considered here
results in the classical mixed strategy game (in which Alice plays the
strategy $p$ whereas Bob plays the strategy $q$) is obtained when the tips of
Alice's and Bob's unit vectors (representing their strategic choices) are
constrained to trajectories on a unit sphere that are defined by%

\begin{equation}
\sin\theta_{\mathrm{A}}\cos\phi_{\mathrm{A}}=2p-1,\text{ \ \ }\sin
\theta_{\mathrm{B}}\cos\phi_{\mathrm{B}}=2q-1,\label{p&q}%
\end{equation}
and the classical mixed strategy game is recovered by interpreting
$\frac{(1+\cos\phi_{\mathrm{A}})}{2}$ and $\frac{(1+\cos\phi_{\mathrm{B}})}%
{2}$ in these equations as the probabilities $p$ and $q$ in the mixed strategy
payoff relations (\ref{Mixed_strategy_payoffs}). Here $\frac{(1-\cos
\phi_{\mathrm{A}})}{2}$ and $\frac{(1-\cos\phi_{\mathrm{B}})}{2}$ are then
interpreted as $(1-p)$ and $(1-q)$ in (\ref{Mixed_strategy_payoffs}).%

\begin{figure}[ptb]%
\centering
\includegraphics[
height=3.5466in,
width=4.7331in
]%
{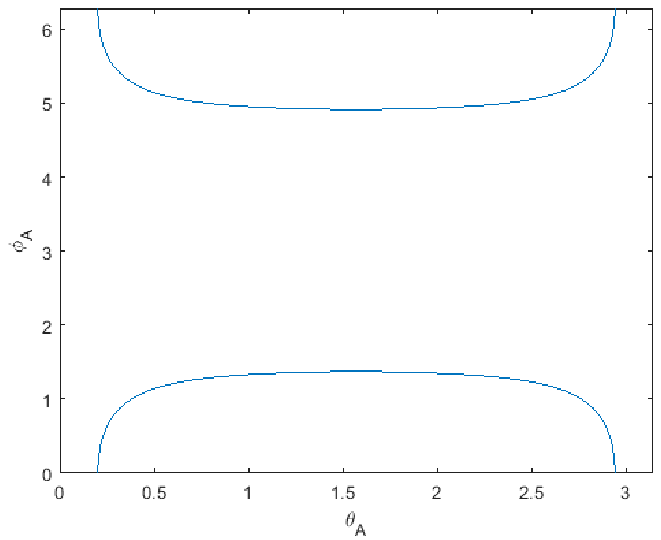}%
\caption{The plot of $\theta_{A} $ against $\phi_{A}$ for $p=0.6$ obtained
from the first Equation in (\ref{p&q}).}%
\end{figure}

Consider the Prisoners' Dilemma game, as defined by $\alpha=3,$ $\beta=0,$
$\gamma=5,$ $\delta=1$ in the game matrix (\ref{GameMatrix}), a quantized
version of which was considered in Ref. \cite{EWL}. The strategy pair $(p,q)$
is a NE in the classical game and therefore $(\theta_{\mathrm{A}}^{\ast}%
,\phi_{\mathrm{A}}^{\ast};\theta_{\mathrm{B}}^{\ast},\phi_{\mathrm{B}}^{\ast
})$ is a NE for which%

\begin{equation}
(1+\sin\theta_{\mathrm{A}}^{\ast}\cos\phi_{\mathrm{A}}^{\ast})=0=(1+\sin
\theta_{\mathrm{B}}^{\ast}\cos\phi_{\mathrm{B}}^{\ast}),
\end{equation}
and we obtain the NE of the game as%

\begin{equation}
(\theta_{\mathrm{A}}^{\ast},\phi_{\mathrm{A}}^{\ast};\theta_{\mathrm{B}}%
^{\ast},\phi_{\mathrm{B}}^{\ast})=(\pi/2,\pi;\pi/2,\pi),
\end{equation}
at which the players' payoffs are $\Pi_{\mathrm{A},\mathrm{B}}(\pi/2,\pi
;\pi/2,\pi)=1$. That is, playing the game with the state $\left\vert
\psi_{\mathrm{ini}}\right\rangle =\frac{1}{2}\left(  \left\vert
00\right\rangle +\left\vert 01\right\rangle +\left\vert 10\right\rangle
+\left\vert 11\right\rangle \right)  $ results in the classical mixed-strategy game.

\section{Game with the quantum state $\left\vert \psi_{\mathrm{ini}%
}\right\rangle =\frac{1}{\sqrt{2}}\left(  \left\vert 00\right\rangle
+i\left\vert 11\right\rangle \right)  $}

For the maximally entangled state $\left\vert \psi_{\mathrm{ini}}\right\rangle
=\frac{1}{\sqrt{2}}\left(  \left\vert 00\right\rangle +i\left\vert
11\right\rangle \right)  $ considered in Refs. \cite{EW,EWL}, following set of
quantum probabilities are obtained%

\begin{align}
\Pr(\mathbf{\hat{a}}_{+1},\mathbf{\hat{b}}_{+1}) &  =\left\vert \left\langle
\psi_{+1}^{\mathbf{\hat{a}}}\psi_{+1}^{\mathbf{\hat{b}}}\mid\psi
_{\mathrm{ini}}\right\rangle \right\vert ^{2}=\frac{1}{8}\left\vert
\sqrt{(1+a_{z})(1+b_{z})}+\frac{(a_{x}-ia_{y})(b_{x}-ib_{y})i}{\sqrt
{(1+a_{z})(1+b_{z})}}\right\vert ^{2},\nonumber\\
&  =\frac{1}{4}(1+a_{x}b_{y}+a_{y}b_{x}+a_{z}b_{z}),\nonumber\\
\Pr(\mathbf{\hat{a}}_{+1},\mathbf{\hat{b}}_{-1}) &  =\left\vert \left\langle
\psi_{+1}^{\mathbf{\hat{a}}}\psi_{-1}^{\mathbf{\hat{b}}}\mid\psi
_{\mathrm{ini}}\right\rangle \right\vert ^{2}=\frac{1}{8}\left\vert
\sqrt{(1+a_{z})(1-b_{z})}-\frac{(a_{x}-ia_{y})(b_{x}-ib_{y})i}{\sqrt
{(1+a_{z})(1-b_{z})}}\right\vert ^{2},\nonumber\\
&  =\frac{1}{4}(1-a_{x}b_{y}-a_{y}b_{x}-a_{z}b_{z}),\nonumber\\
\Pr(\mathbf{\hat{a}}_{-1},\mathbf{\hat{b}}_{+1}) &  =\left\vert \left\langle
\psi_{-1}^{\mathbf{\hat{a}}}\psi_{+1}^{\mathbf{\hat{b}}}\mid\psi
_{\mathrm{ini}}\right\rangle \right\vert ^{2}=\frac{1}{8}\left\vert
\sqrt{(1-a_{z})(1+b_{z})}-\frac{(a_{x}-ia_{y})(b_{x}-ib_{y})i}{\sqrt
{(1-a_{z})(1+b_{z})}}\right\vert ^{2},\nonumber\\
&  =\frac{1}{4}(1-a_{x}b_{y}-a_{y}b_{x}-a_{z}b_{z}),\nonumber\\
\Pr(\mathbf{\hat{a}}_{-1},\mathbf{\hat{b}}_{-1}) &  =\left\vert \left\langle
\psi_{-1}^{\mathbf{\hat{a}}}\psi_{-1}^{\mathbf{\hat{b}}}\mid\psi
_{\mathrm{ini}}\right\rangle \right\vert ^{2}=\frac{1}{8}\left\vert
\sqrt{(1-a_{z})(1-b_{z})}+\frac{(a_{x}-ia_{y})(b_{x}-ib_{y})i}{\sqrt
{(1-a_{z})(1-b_{z})}}\right\vert ^{2},\nonumber\\
&  =\frac{1}{4}(1+a_{x}b_{y}+a_{y}b_{x}+a_{z}b_{z}).\label{QProbabilities1}%
\end{align}
To express these in polar coordinates, we use Eqs. (\ref{Tranfs}) and the
quantum probabilities (\ref{QProbabilities1}) are%

\begin{align}
\Pr(\mathbf{\hat{a}}_{+1},\mathbf{\hat{b}}_{+1}) &  =\frac{1}{4}\{1+\sin
\theta_{\mathrm{A}}\sin\theta_{\mathrm{B}}\sin(\phi_{\mathrm{A}}%
+\phi_{\mathrm{B}})+\cos\theta_{\mathrm{A}}\cos\theta_{\mathrm{B}%
}\},\nonumber\\
\Pr(\mathbf{\hat{a}}_{+1},\mathbf{\hat{b}}_{-1}) &  =\frac{1}{4}\{1-\sin
\theta_{\mathrm{A}}\sin\theta_{\mathrm{B}}\sin(\phi_{\mathrm{A}}%
+\phi_{\mathrm{B}})-\cos\theta_{\mathrm{A}}\cos\theta_{\mathrm{B}%
}\},\nonumber\\
\Pr(\mathbf{\hat{a}}_{-1},\mathbf{\hat{b}}_{+1}) &  =\frac{1}{4}\{1-\sin
\theta_{\mathrm{A}}\sin\theta_{\mathrm{B}}\sin(\phi_{\mathrm{A}}%
+\phi_{\mathrm{B}})-\cos\theta_{\mathrm{A}}\cos\theta_{\mathrm{B}%
}\},\nonumber\\
\Pr(\mathbf{\hat{a}}_{-1},\mathbf{\hat{b}}_{-1}) &  =\frac{1}{4}\{1+\sin
\theta_{\mathrm{A}}\sin\theta_{\mathrm{B}}\sin(\phi_{\mathrm{A}}%
+\phi_{\mathrm{B}})+\cos\theta_{\mathrm{A}}\cos\theta_{\mathrm{B}%
}\}.\label{QProbabilities2}%
\end{align}
Players' payoffs are obtained as%

\begin{gather}
\Pi_{\mathrm{A}},_{\mathrm{B}}(\theta_{\mathrm{A}},\phi_{\mathrm{A}}%
;\theta_{\mathrm{B}},\phi_{\mathrm{B}})=\Pi(\theta_{\mathrm{A}},\phi
_{\mathrm{A}};\theta_{\mathrm{B}},\phi_{\mathrm{B}})=\nonumber\\
(\alpha,\alpha)\Pr(\mathbf{\hat{a}}_{+1},\mathbf{\hat{b}}_{+1})+(\beta
,\gamma)\Pr(\mathbf{\hat{a}}_{+1},\mathbf{\hat{b}}_{-1})+(\gamma,\beta
)\Pr(\mathbf{\hat{a}}_{-1},\mathbf{\hat{b}}_{+1})+(\delta,\delta
)\Pr(\mathbf{\hat{a}}_{-1},\mathbf{\hat{b}}_{-1})\nonumber\\
=\frac{1}{4}\{\Delta_{2}+\Delta_{1}[\sin\theta_{\mathrm{A}}\sin\theta
_{\mathrm{B}}\sin(\phi_{\mathrm{A}}+\phi_{\mathrm{B}})+\cos\theta_{\mathrm{A}%
}\cos\theta_{\mathrm{B}}]\},\label{Payoffs}%
\end{gather}
where%

\begin{equation}
\Delta_{1}=\alpha-\beta-\gamma+\delta\text{ and }\Delta_{2}=\alpha
+\beta+\gamma+\delta.
\end{equation}
%

\begin{figure}[ptb]%
\centering
\includegraphics[
height=3.4566in,
width=4.6112in
]%
{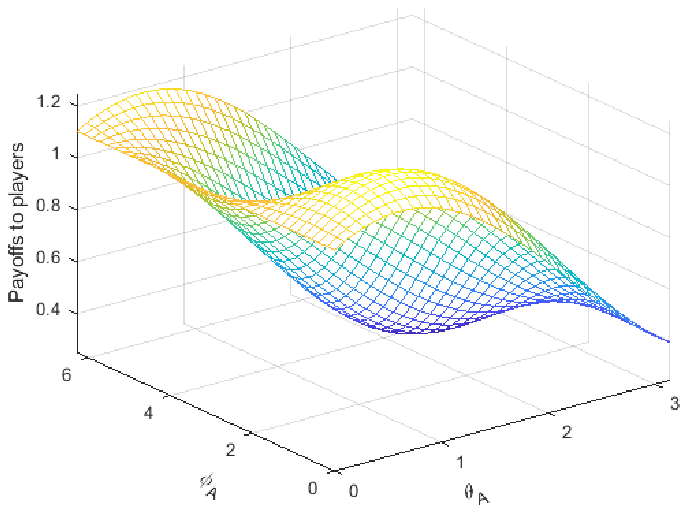}%
\caption{=2.}%
\end{figure}

We note that these payoffs cannot be reduced to the mixed strategy payoffs of
Eq. (\ref{Mixed_strategy_payoffs}). Stated alternatively, there do not exist
such trajectories for the tips of the players' unit vectors which if followed
would result in the mixed-strategy version of the classical game. To determine
the NE $(\theta_{\mathrm{A}}^{\ast},\phi_{\mathrm{A}}^{\ast};\theta
_{\mathrm{B}}^{\ast},\phi_{\mathrm{B}}^{\ast})$, we require%

\begin{gather}
\Pi(\theta_{\mathrm{A}}^{\ast},\phi_{\mathrm{A}}^{\ast};\theta_{\mathrm{B}%
}^{\ast},\phi_{\mathrm{B}}^{\ast})-\Pi(\theta_{\mathrm{A}},\phi_{\mathrm{A}%
}^{\ast};\theta_{\mathrm{B}}^{\ast},\phi_{\mathrm{B}}^{\ast})=(\theta
_{\mathrm{A}}^{\ast}-\theta_{\mathrm{A}})\frac{\partial\Pi}{\partial
\theta_{\mathrm{A}}}\mid_{\ast}\nonumber\\
=\frac{1}{4}\Delta_{1}[\cos\theta_{\mathrm{A}}^{\ast}\sin\theta_{\mathrm{B}%
}^{\ast}\sin(\phi_{\mathrm{A}}^{\ast}+\phi_{\mathrm{B}}^{\ast})-\sin
\theta_{\mathrm{A}}^{\ast}\cos\theta_{\mathrm{B}}^{\ast}](\theta_{\mathrm{A}%
}^{\ast}-\theta_{\mathrm{A}})\geq0,\nonumber\\
\Pi(\theta_{\mathrm{A}}^{\ast},\phi_{\mathrm{A}}^{\ast};\theta_{\mathrm{B}%
}^{\ast},\phi_{\mathrm{B}}^{\ast})-\Pi(\theta_{\mathrm{A}}^{\ast}%
,\phi_{\mathrm{A}}^{\ast};\theta_{\mathrm{B}},\phi_{\mathrm{B}}^{\ast
})=(\theta_{\mathrm{B}}^{\ast}-\theta_{\mathrm{B}})\frac{\partial\Pi}%
{\partial\theta_{\mathrm{B}}}\mid_{\ast}\nonumber\\
=\frac{1}{4}\Delta_{1}[\sin\theta_{\mathrm{A}}^{\ast}\cos\theta_{\mathrm{B}%
}^{\ast}\sin(\phi_{\mathrm{A}}^{\ast}+\phi_{\mathrm{B}}^{\ast})-\cos
\theta_{\mathrm{A}}^{\ast}\sin\theta_{\mathrm{B}}^{\ast}](\theta_{\mathrm{B}%
}^{\ast}-\theta_{\mathrm{B}})\geq0,\nonumber\\
\Pi(\theta_{\mathrm{A}}^{\ast},\phi_{\mathrm{A}}^{\ast};\theta_{\mathrm{B}%
}^{\ast},\phi_{\mathrm{B}}^{\ast})-\Pi(\theta_{\mathrm{A}}^{\ast}%
,\phi_{\mathrm{A}};\theta_{\mathrm{B}}^{\ast},\phi_{\mathrm{B}}^{\ast}%
)=(\phi_{\mathrm{A}}^{\ast}-\phi_{\mathrm{A}})\frac{\partial\Pi}{\partial
\phi_{\mathrm{A}}}\mid_{\ast}\nonumber\\
=\frac{1}{4}\Delta_{1}[\sin\theta_{\mathrm{A}}^{\ast}\sin\theta_{\mathrm{B}%
}^{\ast}\cos(\phi_{\mathrm{A}}^{\ast}+\phi_{\mathrm{B}}^{\ast})](\phi
_{\mathrm{A}}^{\ast}-\phi_{\mathrm{A}})\geq0,\nonumber\\
\Pi(\theta_{\mathrm{A}}^{\ast},\phi_{\mathrm{A}}^{\ast};\theta_{\mathrm{B}%
}^{\ast},\phi_{\mathrm{B}}^{\ast})-\Pi(\theta_{\mathrm{A}}^{\ast}%
,\phi_{\mathrm{A}}^{\ast};\theta_{\mathrm{B}}^{\ast},\phi_{\mathrm{B}}%
)=(\phi_{\mathrm{B}}^{\ast}-\phi_{\mathrm{B}})\frac{\partial\Pi}{\partial
\phi_{\mathrm{B}}}\mid_{\ast}\nonumber\\
=\frac{1}{4}\Delta_{1}[\sin\theta_{\mathrm{A}}^{\ast}\sin\theta_{\mathrm{B}%
}^{\ast}\cos(\phi_{\mathrm{A}}^{\ast}+\phi_{\mathrm{B}}^{\ast})](\phi
_{\mathrm{B}}^{\ast}-\phi_{\mathrm{B}})\geq0.\label{Nash_inequalities_1}%
\end{gather}

Now consider the case when only equalities are involved in the above
expressions i.e.%

\begin{gather}
\cos\theta_{\mathrm{A}}^{\ast}\sin\theta_{\mathrm{B}}^{\ast}\sin
(\phi_{\mathrm{A}}^{\ast}+\phi_{\mathrm{B}}^{\ast})-\sin\theta_{\mathrm{A}%
}^{\ast}\cos\theta_{\mathrm{B}}^{\ast}=0,\label{Eq_1}\\
\sin\theta_{\mathrm{A}}^{\ast}\cos\theta_{\mathrm{B}}^{\ast}\sin
(\phi_{\mathrm{A}}^{\ast}+\phi_{\mathrm{B}}^{\ast})-\cos\theta_{\mathrm{A}%
}^{\ast}\sin\theta_{\mathrm{B}}^{\ast}=0,\label{Eq_2}\\
\sin\theta_{\mathrm{A}}^{\ast}\sin\theta_{\mathrm{B}}^{\ast}\cos
(\phi_{\mathrm{A}}^{\ast}+\phi_{\mathrm{B}}^{\ast})=0.\label{Eq_3}%
\end{gather}

As $\theta_{\mathrm{A}},\theta_{\mathrm{B}}\in\lbrack0,\pi]$, these equations
would hold true when $\theta_{\mathrm{A}}^{\ast},\theta_{\mathrm{B}}^{\ast
}=0,\pi$ and for any $\phi_{\mathrm{A}},\phi_{\mathrm{B}}$. There both
players' payoffs are obtained from Eq.~(\ref{Payoffs}) as%

\begin{equation}
\Pi_{\mathrm{A},\mathrm{B}}(\theta_{\mathrm{A}}^{\ast},\phi_{\mathrm{A}%
};\theta_{\mathrm{B}}^{\ast},\phi_{\mathrm{B}})=\frac{1}{4}\{\Delta_{2}%
+\Delta_{1}\cos\theta_{\mathrm{A}}^{\ast}\cos\theta_{\mathrm{B}}^{\ast
}\}=\frac{1}{2}(\alpha+\delta),\text{ }\frac{1}{2}(\beta+\gamma).
\end{equation}
Alternatively, Eq.~(\ref{Eq_3}) holds when $\cos(\phi_{\mathrm{A}}^{\ast}%
+\phi_{\mathrm{B}}^{\ast})=0$ i.e. $\sin(\phi_{\mathrm{A}}^{\ast}%
+\phi_{\mathrm{B}}^{\ast})=\pm1.$

We note that for $\sin(\phi_{\mathrm{A}}^{\ast}+\phi_{\mathrm{B}}^{\ast})=+1$,
Eqs.~(\ref{Eq_1},~\ref{Eq_2}) give $\sin(\theta_{\mathrm{A}}^{\ast}%
-\theta_{\mathrm{B}}^{\ast})=0$ or $\theta_{\mathrm{A}}^{\ast}-\theta
_{\mathrm{B}}^{\ast}=0,$ $\pm\pi$. For this NE, both players' payoffs are
obtained from Eq.~(\ref{Payoffs}) as%

\begin{equation}
\Pi_{\mathrm{A},\mathrm{B}}(\theta_{\mathrm{A}}^{\ast},\phi_{\mathrm{A}}%
^{\ast};\theta_{\mathrm{B}}^{\ast},\phi_{\mathrm{B}}^{\ast})=\frac{1}%
{4}\{\Delta_{2}+\Delta_{1}\cos(\theta_{\mathrm{A}}^{\ast}-\theta_{\mathrm{B}%
}^{\ast})\}=\frac{1}{2}(\alpha+\delta),\text{ }\frac{1}{2}(\beta+\gamma).
\end{equation}
However, for $\sin(\phi_{\mathrm{A}}^{\ast}+\phi_{\mathrm{B}}^{\ast})=-1 $,
Eqs.~(\ref{Eq_1},~\ref{Eq_2}) give $\sin(\theta_{\mathrm{A}}^{\ast}%
+\theta_{\mathrm{B}}^{\ast})=0$ or $\theta_{\mathrm{A}}^{\ast}+\theta
_{\mathrm{B}}^{\ast}=0,$ $\pi,2\pi.$ For this NE, both players' payoffs are
then obtained from Eq.~(\ref{Payoffs}) as%

\begin{equation}
\Pi_{\mathrm{A},\mathrm{B}}(\theta_{\mathrm{A}}^{\ast},\phi_{\mathrm{A}}%
^{\ast};\theta_{\mathrm{B}}^{\ast},\phi_{\mathrm{B}}^{\ast})=\frac{1}%
{4}\{\Delta_{2}+\Delta_{1}\cos(\theta_{\mathrm{A}}^{\ast}+\theta_{\mathrm{B}%
}^{\ast})\}=\frac{1}{2}(\alpha+\delta),\text{ }\frac{1}{2}(\beta
+\gamma).\label{NE_solutions}%
\end{equation}
Therefore, for all these equilibria, both players' payoffs are same i.e.
either $\frac{1}{2}(\alpha+\delta)$ or $\frac{1}{2}(\beta+\gamma)$. We also
note that for the edges located at%

\begin{align}
&  (0,0;0,0),(0,0;0,2\pi),(0,2\pi;0,0),(0,2\pi;0,2\pi);\nonumber\\
&  (0,0;\pi,0),(0,0;\pi,2\pi),(0,2\pi;\pi,0),(0,2\pi;\pi,2\pi);\nonumber\\
&  (\pi,0;0,0),(\pi,0;0,2\pi),(\pi,2\pi;0,0),(\pi,2\pi;0,2\pi);\nonumber\\
&  (\pi,0;\pi,0),(\pi,0;\pi,2\pi),(\pi,2\pi;\pi,0),(\pi,2\pi;\pi,2\pi),
\end{align}
and we have $\theta_{\mathrm{A}}^{\ast},\theta_{\mathrm{B}}^{\ast}=0$ or $\pi$
and therefore $\sin\theta_{\mathrm{A}}^{\ast}=0=\sin\theta_{\mathrm{B}}^{\ast
}$. That is, Eqs. (\ref{Eq_1},~\ref{Eq_2},~\ref{Eq_3}) are true for all these
edges and both players' payoffs at these are the same as given by
Eq.~(\ref{NE_solutions}).

\section{Game with the quantum state $\left\vert \psi_{\mathrm{ini}%
}\right\rangle =\frac{1}{2}\left(  \left\vert 00\right\rangle +\left\vert
01\right\rangle -\left\vert 10\right\rangle +\left\vert 11\right\rangle
\right)  $}

This is an entangled state for which we find%

\begin{gather}
\Pr(\mathbf{\hat{a}}_{+1},\mathbf{\hat{b}}_{+1})=\left\vert \left\langle
\psi_{+1}^{\mathbf{\hat{a}}}\psi_{+1}^{\mathbf{\hat{b}}}\mid\psi
_{\mathrm{ini}}\right\rangle \right\vert ^{2}=\nonumber\\
\frac{1}{16(1+a_{z})(1+b_{z})}{\Large \{}\left[  (1+a_{z})(1+b_{z}%
)+(1+a_{z})b_{x}-(1+b_{z})a_{x}+(a_{x}b_{x}-a_{y}b_{y})\right]  ^{2}%
+\nonumber\\
\left[  (1+a_{z})b_{y}-(1+b_{z})a_{y}+(a_{x}b_{y}+a_{y}b_{x})\right]
^{2}{\large \}},\nonumber\\
\Pr(\mathbf{\hat{a}}_{+1},\mathbf{\hat{b}}_{-1})=\left\vert \left\langle
\psi_{+1}^{\mathbf{\hat{a}}}\psi_{-1}^{\mathbf{\hat{b}}}\mid\psi
_{\mathrm{ini}}\right\rangle \right\vert ^{2}=\nonumber\\
\frac{1}{16(1+a_{z})(1-b_{z})}{\Large \{}\left[  (1+a_{z})(1-b_{z}%
)-(1+a_{z})b_{x}-(1-b_{z})a_{x}-(a_{x}b_{x}-a_{y}b_{y})\right]  ^{2}%
+\nonumber\\
\left[  (1+a_{z})b_{y}+(1-b_{z})a_{y}+(a_{x}b_{y}+a_{y}b_{x})\right]
^{2}{\large \}},\nonumber\\
\Pr(\mathbf{\hat{a}}_{-1},\mathbf{\hat{b}}_{+1})=\left\vert \left\langle
\psi_{-1}^{\mathbf{\hat{a}}}\psi_{+1}^{\mathbf{\hat{b}}}\mid\psi
_{\mathrm{ini}}\right\rangle \right\vert ^{2}=\nonumber\\
\frac{1}{16(1-a_{z})(1+b_{z})}{\Large \{}\left[  (1-a_{z})(1+b_{z}%
)+(1-a_{z})b_{x}+(1+b_{z})a_{x}-(a_{x}b_{x}-a_{y}b_{y})\right]  ^{2}%
+\nonumber\\
\left[  (1-a_{z})b_{y}+(1+b_{z})a_{y}-(a_{x}b_{y}+a_{y}b_{x})\right]
^{2}{\large \}},\nonumber\\
\Pr(\mathbf{\hat{a}}_{-1},\mathbf{\hat{b}}_{-1})=\left\vert \left\langle
\psi_{-1}^{\mathbf{\hat{a}}}\psi_{-1}^{\mathbf{\hat{b}}}\mid\psi
_{\mathrm{ini}}\right\rangle \right\vert ^{2}=\nonumber\\
\frac{1}{16(1-a_{z})(1-b_{z})}{\Large \{}\left[  (1-a_{z})(1-b_{z}%
)-(1-a_{z})b_{x}+(1-b_{z})a_{x}+(a_{x}b_{x}-a_{y}b_{y})\right]  ^{2}%
+\nonumber\\
\left[  (1-a_{z})b_{y}-(1-b_{z})a_{y}-(a_{x}b_{y}+a_{y}b_{x})\right]
^{2}{\large \}}.
\end{gather}

The payoff to the players can now be expressed as%

\begin{gather}
\Pi_{A,B}(\hat{a},\hat{b})=\nonumber\\
\frac{1}{16(1+a_{z})}{\huge [}\frac{(\alpha,\alpha)}{(1+b_{z})}{\Large \{}%
\left[  (1+a_{z})(1+b_{z})+(1+a_{z})b_{x}-(1+b_{z})a_{x}+(a_{x}b_{x}%
-a_{y}b_{y})\right]  ^{2}+\nonumber\\
\left[  (1+a_{z})b_{y}-(1+b_{z})a_{y}+(a_{x}b_{y}+a_{y}b_{x})\right]
^{2}{\large \}}+\nonumber\\
\frac{(\beta,\gamma)}{(1-b_{z})}{\Large \{}\left[  (1+a_{z})(1-b_{z}%
)-(1+a_{z})b_{x}-(1-b_{z})a_{x}-(a_{x}b_{x}-a_{y}b_{y})\right]  ^{2}%
+\nonumber\\
\left[  (1+a_{z})b_{y}+(1-b_{z})a_{y}+(a_{x}b_{y}+a_{y}b_{x})\right]
^{2}{\large \}}{\huge ]}+\nonumber\\
\frac{1}{16(1-a_{z})}{\huge [}\frac{(\gamma,\beta)}{(1+b_{z})}{\Large \{}%
\left[  (1-a_{z})(1+b_{z})+(1-a_{z})b_{x}+(1+b_{z})a_{x}-(a_{x}b_{x}%
-a_{y}b_{y})\right]  ^{2}+\nonumber\\
\left[  (1-a_{z})b_{y}+(1+b_{z})a_{y}-(a_{x}b_{y}+a_{y}b_{x})\right]
^{2}{\large \}}+\nonumber\\
\frac{(\delta,\delta)}{(1-b_{z})}{\Large \{}\left[  (1-a_{z})(1-b_{z}%
)-(1-a_{z})b_{x}+(1-b_{z})a_{x}+(a_{x}b_{x}-a_{y}b_{y})\right]  ^{2}%
+\nonumber\\
\left[  (1-a_{z})b_{y}-(1-b_{z})a_{y}-(a_{x}b_{y}+a_{y}b_{x})\right]
^{2}{\large \}}{\huge ],}%
\end{gather}

The above transformation reduces the independent variables $\hat{a}$ and
$\hat{b}$ to $\theta_{A},\theta_{B},\phi_{A},$and $\phi_{B}$ and players
payoffs are then expressed as%

\begin{gather}
\Pi_{A,B}(\hat{a},\hat{b})=\Pi_{A,B}(\theta_{A},\phi_{A};\theta_{B},\phi
_{B})=\nonumber\\
\frac{(\alpha,\alpha)}{16(1+\cos\theta_{A})(1+\cos\theta_{B})}{\Large \{}%
[(1+\cos\theta_{A})(1+\cos\theta_{B})+(1+\cos\theta_{A})\sin\theta_{B}\cos
\phi_{B}-\nonumber\\
(1+\cos\theta_{B})\sin\theta_{A}\cos\phi_{A}+\sin\theta_{A}\sin\theta_{B}%
\cos(\phi_{A}+\phi_{B})]^{2}+[(1+\cos\theta_{A})\sin\theta_{B}\sin\phi
_{B}-\nonumber\\
(1+\cos\theta_{B})\sin\theta_{A}\sin\phi_{A}+\sin\theta_{A}\sin\theta_{B}%
\sin(\phi_{A}+\phi_{B})]^{2}{\large \}}+\nonumber\\
\frac{(\beta,\gamma)}{16(1+\cos\theta_{A})(1-\cos\theta_{B})}{\Large \{}%
[(1+\cos\theta_{A})(1-\cos\theta_{B})-(1+\cos\theta_{A})\sin\theta_{B}\cos
\phi_{B}-\nonumber\\
(1-\cos\theta_{B})\sin\theta_{A}\cos\phi_{A}-\sin\theta_{A}\sin\theta_{B}%
\cos(\phi_{A}+\phi_{B})]^{2}+[(1+\cos\theta_{A})\sin\theta_{B}\sin\phi
_{B}+\nonumber\\
(1-\cos\theta_{B})\sin\theta_{A}\sin\phi_{A}+\sin\theta_{A}\sin\theta_{B}%
\sin(\phi_{A}+\phi_{B})]^{2}{\large \}}+\nonumber\\
\frac{(\gamma,\beta)}{16(1-\cos\theta_{A})(1+\cos\theta_{B})}{\Large \{}%
[(1-\cos\theta_{A})(1+\cos\theta_{B})+(1-\cos\theta_{A})\sin\theta_{B}\cos
\phi_{B}+\nonumber\\
(1+\cos\theta_{B})\sin\theta_{A}\cos\phi_{A}-\sin\theta_{A}\sin\theta_{B}%
\cos(\phi_{A}+\phi_{B})]^{2}+\nonumber\\
\lbrack(1-\cos\theta_{A})\sin\theta_{B}\sin\phi_{B}+(1+\cos\theta_{B}%
)\sin\theta_{A}\sin\phi_{A}-\sin\theta_{A}\sin\theta_{B}\sin(\phi_{A}+\phi
_{B})]^{2}{\large \}}+\nonumber\\
\frac{(\delta,\delta)}{16(1-\cos\theta_{A})(1-\cos\theta_{B})}{\Large \{}%
[(1-\cos\theta_{A})(1-\cos\theta_{B})-(1-\cos\theta_{A})\sin\theta_{B}\cos
\phi_{B}+\nonumber\\
(1-\cos\theta_{B})\sin\theta_{A}\cos\phi_{A}+\sin\theta_{A}\sin\theta_{B}%
\cos(\phi_{A}+\phi_{B})]^{2}+[(1-\cos\theta_{A})\sin\theta_{B}\sin\phi
_{B}-\nonumber\\
(1-\cos\theta_{B})\sin\theta_{A}\sin\phi_{A}-\sin\theta_{A}\sin\theta_{B}%
\sin(\phi_{A}+\phi_{B})]^{2}{\large \}},\label{Payoff_asym}%
\end{gather}
which can be simplified to%

\begin{gather}
\Pi_{A,B}(\hat{a},\hat{b})=\Pi_{A,B}(\theta_{A},\phi_{A};\theta_{B},\phi
_{B})=\nonumber\\
(\alpha,\alpha)\Pr(\hat{a}_{+1},\hat{b}_{+1})+(\beta,\gamma)\Pr(\hat{a}%
_{+1},\hat{b}_{-1})+(\gamma,\beta)\Pr(\hat{a}_{-1},\hat{b}_{+1})+(\delta
,\delta)\Pr(\hat{a}_{-1},\hat{b}_{-1})\nonumber\\
=\frac{(\alpha,\alpha)}{4}\{1-\sin\theta_{A}\sin\theta_{B}\sin\phi_{A}\sin
\phi_{B}-\sin\theta_{A}\cos\theta_{B}\cos\phi_{A}+\cos\theta_{A}\sin\theta
_{B}\cos\phi_{B}\}+\nonumber\\
\frac{(\beta,\gamma)}{4}[1+\sin\theta_{A}\sin\theta_{B}\sin\phi_{A}\sin
\phi_{B}+\sin\theta_{A}\cos\theta_{B}\cos\phi_{A}-\cos\theta_{A}\sin\theta
_{B}\cos\phi_{B}]+\nonumber\\
\frac{(\gamma,\beta)}{4}(1+\sin\theta_{A}\sin\theta_{B}\sin\phi_{A}\sin
\phi_{B}+\sin\theta_{A}\cos\theta_{B}\cos\phi_{A}-\cos\theta_{A}\sin\theta
_{B}\cos\phi_{B})+\nonumber\\
\frac{(\delta,\delta)}{4}(1-\sin\theta_{A}\sin\theta_{B}\sin\phi_{A}\sin
\phi_{B}-\sin\theta_{A}\cos\theta_{B}\cos\phi_{A}+\cos\theta_{A}\sin\theta
_{B}\cos\phi_{B}).\label{Payoffs_asym_1}%
\end{gather}

As an example, Appendix C details the simplification of third term in the
payoff relations (\ref{Payoff_asym}) to obtain the third term of
(\ref{Payoffs_asym_1}). The transformations (\ref{Tranfs}) reduce the
independent variables $\mathbf{\hat{a}}$ and $\mathbf{\hat{b}}$ to
$\theta_{\mathrm{A}},\theta_{\mathrm{B}},\phi_{\mathrm{A}},$and $\phi
_{\mathrm{B}}$ and players payoffs can be expressed as%

\begin{gather}
\Pi_{\mathrm{A},\mathrm{B}}(\theta_{\mathrm{A}},\phi_{\mathrm{A}}%
;\theta_{\mathrm{B}},\phi_{\mathrm{B}})=\nonumber\\
=\frac{1}{4}\{\Delta_{2}-\Delta_{1}[\sin\theta_{\mathrm{A}}\sin\theta
_{\mathrm{B}}\sin\phi_{\mathrm{A}}\sin\phi_{\mathrm{B}}+\sin\theta
_{\mathrm{A}}\cos\theta_{\mathrm{B}}\cos\phi_{\mathrm{A}}-\cos\theta
_{\mathrm{A}}\sin\theta_{\mathrm{B}}\cos\phi_{\mathrm{B}}]\}.\label{Payoffs_3}%
\end{gather}
We note that as was the case for the state $\frac{1}{\sqrt{2}}\left(
\left\vert 00\right\rangle +i\left\vert 11\right\rangle \right)  $ these
payoffs cannot be reduced to the classical mixed strategy payoffs in the game.
In other words, there do not exist such trajectories for the tips of each
players' unit vectors which if followed can result in the classical
mixed-strategy game. To determine the NE $(\theta_{\mathrm{A}}^{\ast}%
,\phi_{\mathrm{A}}^{\ast};\theta_{\mathrm{B}}^{\ast},\phi_{\mathrm{B}}^{\ast
})$ we require%

\begin{gather}
\Pi(\theta_{\mathrm{A}}^{\ast},\phi_{\mathrm{A}}^{\ast};\theta_{\mathrm{B}%
}^{\ast},\phi_{\mathrm{B}}^{\ast})-\Pi(\theta_{\mathrm{A}},\phi_{\mathrm{A}%
}^{\ast};\theta_{\mathrm{B}}^{\ast},\phi_{\mathrm{B}}^{\ast})=(\theta
_{\mathrm{A}}^{\ast}-\theta_{\mathrm{A}})\frac{\partial\Pi}{\partial
\theta_{\mathrm{A}}}\mid_{\ast}\nonumber\\
=-\frac{1}{4}\Delta_{1}[\cos\theta_{\mathrm{A}}^{\ast}\sin\theta_{\mathrm{B}%
}^{\ast}\sin\phi_{\mathrm{A}}^{\ast}\sin\phi_{\mathrm{B}}^{\ast}+\cos
\theta_{\mathrm{A}}^{\ast}\cos\theta_{\mathrm{B}}^{\ast}\cos\phi_{\mathrm{A}%
}^{\ast}+\sin\theta_{\mathrm{A}}^{\ast}\sin\theta_{\mathrm{B}}^{\ast}\cos
\phi_{\mathrm{B}}^{\ast}](\theta_{\mathrm{A}}^{\ast}-\theta_{\mathrm{A}}%
)\geq0,\nonumber\\
\Pi(\theta_{\mathrm{A}}^{\ast},\phi_{\mathrm{A}}^{\ast};\theta_{\mathrm{B}%
}^{\ast},\phi_{\mathrm{B}}^{\ast})-\Pi(\theta_{\mathrm{A}}^{\ast}%
,\phi_{\mathrm{A}}^{\ast};\theta_{\mathrm{B}},\phi_{\mathrm{B}}^{\ast
})=(\theta_{\mathrm{B}}^{\ast}-\theta_{\mathrm{B}})\frac{\partial\Pi}%
{\partial\theta_{\mathrm{B}}}\mid_{\ast}\nonumber\\
=-\frac{1}{4}\Delta_{1}[\sin\theta_{\mathrm{A}}^{\ast}\cos\theta_{\mathrm{B}%
}^{\ast}\sin\phi_{\mathrm{A}}^{\ast}\sin\phi_{\mathrm{B}}^{\ast}-\sin
\theta_{\mathrm{A}}^{\ast}\sin\theta_{\mathrm{B}}^{\ast}\cos\phi_{\mathrm{A}%
}^{\ast}-\cos\theta_{\mathrm{A}}^{\ast}\cos\theta_{\mathrm{B}}^{\ast}\cos
\phi_{\mathrm{B}}^{\ast}](\theta_{\mathrm{B}}^{\ast}-\theta_{\mathrm{B}}%
)\geq0,\nonumber\\
\Pi(\theta_{\mathrm{A}}^{\ast},\phi_{\mathrm{A}}^{\ast};\theta_{\mathrm{B}%
}^{\ast},\phi_{\mathrm{B}}^{\ast})-\Pi(\theta_{\mathrm{A}}^{\ast}%
,\phi_{\mathrm{A}};\theta_{\mathrm{B}}^{\ast},\phi_{\mathrm{B}}^{\ast}%
)=(\phi_{\mathrm{A}}^{\ast}-\phi_{\mathrm{A}})\frac{\partial\Pi}{\partial
\phi_{\mathrm{A}}}\mid_{\ast}\nonumber\\
=-\frac{1}{4}\Delta_{1}[\sin\theta_{\mathrm{A}}^{\ast}\sin\theta_{\mathrm{B}%
}^{\ast}\cos\phi_{\mathrm{A}}^{\ast}\sin\phi_{\mathrm{B}}^{\ast}-\sin
\theta_{\mathrm{A}}^{\ast}\cos\theta_{\mathrm{B}}^{\ast}\sin\phi_{\mathrm{A}%
}^{\ast}](\phi_{\mathrm{A}}^{\ast}-\phi_{\mathrm{A}})\geq0,\nonumber\\
\Pi(\theta_{\mathrm{A}}^{\ast},\phi_{\mathrm{A}}^{\ast};\theta_{\mathrm{B}%
}^{\ast},\phi_{\mathrm{B}}^{\ast})-\Pi(\theta_{\mathrm{A}}^{\ast}%
,\phi_{\mathrm{A}}^{\ast};\theta_{\mathrm{B}}^{\ast},\phi_{\mathrm{B}}%
)=(\phi_{\mathrm{B}}^{\ast}-\phi_{\mathrm{B}})\frac{\partial\Pi}{\partial
\phi_{\mathrm{B}}}\mid_{\ast}\nonumber\\
=-\frac{1}{4}\Delta_{1}[\sin\theta_{\mathrm{A}}^{\ast}\sin\theta_{\mathrm{B}%
}^{\ast}\sin\phi_{\mathrm{A}}^{\ast}\cos\phi_{\mathrm{B}}^{\ast}+\cos
\theta_{\mathrm{A}}^{\ast}\sin\theta_{\mathrm{B}}^{\ast}\sin\phi_{\mathrm{B}%
}^{\ast}](\phi_{\mathrm{B}}^{\ast}-\phi_{\mathrm{B}})\geq0.
\end{gather}
We firstly consider the case when only equalities are involved in the above
expressions i.e.%

\begin{gather}
\sin\theta_{\mathrm{B}}^{\ast}(\cos\theta_{\mathrm{A}}^{\ast}\sin
\phi_{\mathrm{A}}^{\ast}\sin\phi_{\mathrm{B}}^{\ast}+\sin\theta_{\mathrm{A}%
}^{\ast}\cos\phi_{\mathrm{B}}^{\ast})+\cos\theta_{\mathrm{A}}^{\ast}\cos
\theta_{\mathrm{B}}^{\ast}\cos\phi_{\mathrm{A}}^{\ast}=0,\nonumber\\
\sin\theta_{\mathrm{A}}^{\ast}(\cos\theta_{\mathrm{B}}^{\ast}\sin
\phi_{\mathrm{A}}^{\ast}\sin\phi_{\mathrm{B}}^{\ast}-\sin\theta_{\mathrm{B}%
}^{\ast}\cos\phi_{\mathrm{A}}^{\ast})-\cos\theta_{\mathrm{A}}^{\ast}\cos
\theta_{\mathrm{B}}^{\ast}\cos\phi_{\mathrm{B}}^{\ast}=0,\nonumber\\
\sin\theta_{\mathrm{A}}^{\ast}(\sin\theta_{\mathrm{B}}^{\ast}\cos
\phi_{\mathrm{A}}^{\ast}\sin\phi_{\mathrm{B}}^{\ast}-\cos\theta_{\mathrm{B}%
}^{\ast}\sin\phi_{\mathrm{A}}^{\ast})=0,\nonumber\\
\sin\theta_{\mathrm{B}}^{\ast}(\sin\theta_{\mathrm{A}}^{\ast}\sin
\phi_{\mathrm{A}}^{\ast}\cos\phi_{\mathrm{B}}^{\ast}+\cos\theta_{\mathrm{A}%
}^{\ast}\sin\phi_{\mathrm{B}}^{\ast})=0,\label{Non_edge_3}%
\end{gather}
where $\theta_{\mathrm{A}},\theta_{\mathrm{B}}\in\lbrack0,\pi]$ and
$\phi_{\mathrm{A}},\phi_{\mathrm{B}}\in\lbrack0,2\pi).$ Now we consider the
following cases:

\subsection{Case $\sin\theta_{\mathrm{A}}^{\ast}=0=\sin\theta_{\mathrm{B}%
}^{\ast}$}

For $\sin\theta_{\mathrm{A}}^{\ast}=0=\sin\theta_{\mathrm{B}}^{\ast}$ we also
have $\cos\theta_{\mathrm{A}}^{\ast}=\pm1$ and $\cos\theta_{\mathrm{B}}^{\ast
}=\pm1$ and this results the first two equation in (\ref{Non_edge_3}) to give
$\pm\cos\phi_{\mathrm{A}}^{\ast}=0$ and $\pm\cos\phi_{\mathrm{B}}^{\ast}=0$.
This gives%

\begin{equation}
\theta_{\mathrm{A}}^{\ast}=0,\pi;\text{ \ \ \ }\theta_{\mathrm{B}}^{\ast
}=0,\pi;\text{ \ \ \ }\phi_{\mathrm{A}}^{\ast}=\pi/2,3\pi/2;\text{ and\ }%
\phi_{\mathrm{B}}^{\ast}=\pi/2,3\pi/2,
\end{equation}
which result in the set of solutions for $(\theta_{\mathrm{A}}^{\ast}%
,\phi_{\mathrm{A}}^{\ast};\theta_{\mathrm{B}}^{\ast},\phi_{\mathrm{B}}^{\ast
})$ as%

\begin{align}
&  (0,\pi/2;0,\pi/2),(0,\pi/2;0,3\pi/2),(0,3\pi/2;0,\pi/2),(0,3\pi
/2;0,3\pi/2),\nonumber\\
&  (0,\pi/2;\pi,\pi/2),(0,\pi/2;\pi,3\pi/2),(0,3\pi/2;\pi,\pi/2),(0,3\pi
/2;\pi,3\pi/2),\nonumber\\
&  (\pi,\pi/2;0,\pi/2),(\pi,\pi/2;0,3\pi/2),(\pi,3\pi/2;0,\pi/2),(\pi
,3\pi/2;0,3\pi/2),\nonumber\\
&  (\pi,\pi/2;\pi,\pi/2),(\pi,\pi/2;\pi,3\pi/2),(\pi,3\pi/2;\pi,\pi
/2),(\pi,3\pi/2;\pi,3\pi/2),
\end{align}
and the players' payoffs at these Nash equilibria are obtained from
Eq.~(\ref{Payoffs_3}) as%

\begin{equation}
\Pi_{\mathrm{A},\mathrm{B}}(\theta_{\mathrm{A}}^{\ast},\phi_{\mathrm{A}}%
^{\ast};\theta_{\mathrm{B}}^{\ast},\phi_{\mathrm{B}}^{\ast})=\frac{1}{4}%
\Delta_{2}=\frac{1}{4}(\alpha+\beta+\gamma+\delta).
\end{equation}

\subsection{Case $\sin\theta_{\mathrm{A}}^{\ast}\neq0$ and $\sin
\theta_{\mathrm{B}}^{\ast}\neq0$}

When $\sin\theta_{\mathrm{A}}^{\ast}\neq0$ and $\sin\theta_{\mathrm{B}}^{\ast
}\neq0$, we have from the last two equations in (\ref{Non_edge_3})%

\begin{equation}
\sin\theta_{\mathrm{B}}^{\ast}\cos\phi_{\mathrm{A}}^{\ast}\sin\phi
_{\mathrm{B}}^{\ast}-\cos\theta_{\mathrm{B}}^{\ast}\sin\phi_{\mathrm{A}}%
^{\ast}=0,\text{ }\sin\theta_{\mathrm{A}}^{\ast}\sin\phi_{\mathrm{A}}^{\ast
}\cos\phi_{\mathrm{B}}^{\ast}+\cos\theta_{\mathrm{A}}^{\ast}\sin
\phi_{\mathrm{B}}^{\ast}=0,
\end{equation}
that can be expressed as%

\begin{align}
\cos\phi_{\mathrm{A}}^{\ast}\sin\phi_{\mathrm{B}}^{\ast}-\cot\theta
_{\mathrm{B}}^{\ast}\sin\phi_{\mathrm{A}}^{\ast} &  =0,\label{c}\\
\sin\phi_{\mathrm{A}}^{\ast}\cos\phi_{\mathrm{B}}^{\ast}+\cot\theta
_{\mathrm{A}}^{\ast}\sin\phi_{\mathrm{B}}^{\ast} &  =0.\label{d}%
\end{align}
Now, the first two equations in (\ref{Non_edge_3}) are%

\begin{align}
\sin\theta_{\mathrm{B}}^{\ast}(\cos\theta_{\mathrm{A}}^{\ast}\sin
\phi_{\mathrm{B}}^{\ast}\sin\phi_{\mathrm{A}}^{\ast}+\sin\theta_{\mathrm{A}%
}^{\ast}\cos\phi_{\mathrm{B}}^{\ast})+\cos\theta_{\mathrm{A}}^{\ast}\cos
\theta_{\mathrm{B}}^{\ast}\cos\phi_{\mathrm{A}}^{\ast} &  =0,\label{first}\\
\sin\theta_{\mathrm{A}}^{\ast}(\cos\theta_{\mathrm{B}}^{\ast}\sin
\phi_{\mathrm{A}}^{\ast}\sin\phi_{\mathrm{B}}^{\ast}-\sin\theta_{\mathrm{B}%
}^{\ast}\cos\phi_{\mathrm{A}}^{\ast})-\cos\theta_{\mathrm{A}}^{\ast}\cos
\theta_{\mathrm{B}}^{\ast}\cos\phi_{\mathrm{B}}^{\ast} &  =0,\label{second}%
\end{align}
and given that $\sin\theta_{\mathrm{A}}^{\ast}\neq0$ and $\sin\theta
_{\mathrm{B}}^{\ast}\neq0$, we divide Eq.~(\ref{first}) with $\sin
\theta_{\mathrm{B}}^{\ast}$ and Eq.~(\ref{second}) by $\sin\theta_{\mathrm{A}%
}^{\ast}$ to obtain%

\begin{align}
\cos\theta_{\mathrm{A}}^{\ast}(\sin\phi_{\mathrm{B}}^{\ast}\sin\phi
_{\mathrm{A}}^{\ast}+\cot\theta_{\mathrm{B}}^{\ast}\cos\phi_{\mathrm{A}}%
^{\ast})+\sin\theta_{\mathrm{A}}^{\ast}\cos\phi_{\mathrm{B}}^{\ast} &
=0,\label{first_}\\
\cos\theta_{\mathrm{B}}^{\ast}(\sin\phi_{\mathrm{A}}^{\ast}\sin\phi
_{\mathrm{B}}^{\ast}-\cot\theta_{\mathrm{A}}^{\ast}\cos\phi_{\mathrm{B}}%
^{\ast})-\sin\theta_{\mathrm{B}}^{\ast}\cos\phi_{\mathrm{A}}^{\ast} &
=0.\label{second_}%
\end{align}
Now divide Eq.~(\ref{first_}) by $\sin\theta_{\mathrm{A}}^{\ast}$ and divide
Eq.~(\ref{second_}) by $\sin\theta_{\mathrm{B}}^{\ast}$ to obtain%

\begin{align}
\cot\theta_{\mathrm{A}}^{\ast}(\sin\phi_{\mathrm{B}}^{\ast}\sin\phi
_{\mathrm{A}}^{\ast}+\cot\theta_{\mathrm{B}}^{\ast}\cos\phi_{\mathrm{A}}%
^{\ast})+\cos\phi_{\mathrm{B}}^{\ast} &  =0,\label{a}\\
\cot\theta_{\mathrm{B}}^{\ast}(\sin\phi_{\mathrm{A}}^{\ast}\sin\phi
_{\mathrm{B}}^{\ast}-\cot\theta_{\mathrm{A}}^{\ast}\cos\phi_{\mathrm{B}}%
^{\ast})-\cos\phi_{\mathrm{A}}^{\ast} &  =0.\label{b}%
\end{align}
As Eqs.~(\ref{a},~\ref{b}) are to be considered along with Eqs.~(\ref{c}%
,~\ref{d}), we rewrite (\ref{a},~\ref{b}) as%

\begin{align}
(\cot\theta_{\mathrm{A}}^{\ast}\sin\phi_{\mathrm{B}}^{\ast})\sin
\phi_{\mathrm{A}}^{\ast}+\cot\theta_{\mathrm{A}}^{\ast}\cot\theta_{\mathrm{B}%
}^{\ast}\cos\phi_{\mathrm{A}}^{\ast}+\cos\phi_{\mathrm{B}}^{\ast} &
=0,\label{e}\\
(\cot\theta_{\mathrm{B}}^{\ast}\sin\phi_{\mathrm{A}}^{\ast})\sin
\phi_{\mathrm{B}}^{\ast}-\cot\theta_{\mathrm{A}}^{\ast}\cot\theta_{\mathrm{B}%
}^{\ast}\cos\phi_{\mathrm{B}}^{\ast}-\cos\phi_{\mathrm{A}}^{\ast} &
=0,\label{f}%
\end{align}
and substitute from (\ref{c}, \ref{d}) to (\ref{e}, \ref{f}) to obtain%

\begin{align}
(\cos\phi_{\mathrm{A}}^{\ast}\cos\phi_{\mathrm{B}}^{\ast}+\cot\theta
_{\mathrm{A}}^{\ast}\cot\theta_{\mathrm{B}}^{\ast})\cos\phi_{\mathrm{A}}%
^{\ast} &  =0,\label{a_}\\
(\cos\phi_{\mathrm{A}}^{\ast}\cos\phi_{\mathrm{B}}^{\ast}+\cot\theta
_{\mathrm{A}}^{\ast}\cot\theta_{\mathrm{B}}^{\ast})\cos\phi_{\mathrm{B}}%
^{\ast} &  =0.\label{b_}%
\end{align}
The above solution of Eqs.~(\ref{a_},~\ref{b_},~\ref{c}, \ref{d}) are obtained
under the requirement that $\sin\theta_{\mathrm{A}}^{\ast}\neq0$ and
$\sin\theta_{\mathrm{B}}^{\ast}\neq0.$ This leads us to consider the following cases:

\subsubsection{Case $\sin\theta_{\mathrm{A}}^{\ast}\neq0$, $\sin
\theta_{\mathrm{B}}^{\ast}\neq0$ and $\cos\phi_{\mathrm{A}}^{\ast}=0=\cos
\phi_{\mathrm{B}}^{\ast}$}

In this case we have a solution for which $\sin\phi_{\mathrm{A}}^{\ast}=\pm1$
and $\sin\phi_{\mathrm{B}}^{\ast}=\pm1$ and from Eqs. (\ref{c},~\ref{d}) we
then have $\cot\theta_{\mathrm{A}}^{\ast}=0=\cot\theta_{\mathrm{B}}^{\ast}$
i.e. $\cos\theta_{\mathrm{A}}^{\ast}=0=\cos\theta_{\mathrm{B}}^{\ast}$ and
therefore $\sin\theta_{\mathrm{A}}^{\ast}=\pm1$ and $\sin\theta_{\mathrm{B}%
}^{\ast}=\pm1.$ Players' payoffs at these Nash equilibria are then obtained
from Eq.~(\ref{Payoffs_3}) as%

\begin{equation}
\Pi_{\mathrm{A},\mathrm{B}}(\theta_{\mathrm{A}}^{\ast},\phi_{\mathrm{A}}%
^{\ast};\theta_{\mathrm{B}}^{\ast},\phi_{\mathrm{B}}^{\ast})=\frac{1}%
{4}(\Delta_{2}\pm\Delta_{1})=\frac{1}{2}(\alpha+\delta),\text{ }\frac{1}%
{2}(\beta+\gamma).
\end{equation}

\subsubsection{Case $\sin\theta_{\mathrm{A}}^{\ast}\neq0$, $\sin
\theta_{\mathrm{B}}^{\ast}\neq0$ and $\sin\phi_{\mathrm{A}}^{\ast}=0=\sin
\phi_{\mathrm{B}}^{\ast}$}

In this case we have a solution for which $\cos\phi_{\mathrm{A}}^{\ast}=\pm1$
and $\cos\phi_{\mathrm{B}}^{\ast}=\pm1$ and from (\ref{a_},~\ref{b_}) we then have%

\begin{equation}
\pm(\pm1+\cot\theta_{\mathrm{A}}^{\ast}\cot\theta_{\mathrm{B}}^{\ast})=0,
\end{equation}
whereas (\ref{c},~\ref{d}) hold true. That is when $\cot\theta_{\mathrm{A}%
}^{\ast}\cot\theta_{\mathrm{B}}^{\ast}=\pm1$ or when $\cot\theta_{\mathrm{A}%
}^{\ast}=\pm1$ and $\cot\theta_{\mathrm{B}}^{\ast}=\pm1$ i.e.%

\begin{equation}
\sin\phi_{\mathrm{A}}^{\ast}=0=\sin\phi_{\mathrm{B}}^{\ast},\text{ \ }%
\cot\theta_{\mathrm{A}}^{\ast}=\pm1\text{ and }\cot\theta_{\mathrm{B}}^{\ast
}=\pm1.
\end{equation}
As $\theta_{\mathrm{A}},\theta_{\mathrm{B}}\in\lbrack0,\pi]$ we have
$\cos\theta_{\mathrm{A}}^{\ast}=\pm\frac{1}{\sqrt{2}}$, $\sin\theta
_{\mathrm{A}}^{\ast}=\frac{1}{\sqrt{2}}$ and $\cos\theta_{\mathrm{B}}^{\ast
}=\pm\frac{1}{\sqrt{2}}$, $\sin\theta_{\mathrm{B}}^{\ast}=\frac{1}{\sqrt{2}}.$
Therefore $\sin\theta_{\mathrm{A}}^{\ast}\cos\theta_{\mathrm{B}}^{\ast}%
=\pm\frac{1}{2}$ and $\cos\theta_{\mathrm{A}}^{\ast}\sin\theta_{\mathrm{B}%
}^{\ast}=\pm\frac{1}{2}.$ Also, then we have $\cos\phi_{\mathrm{A}}^{\ast}%
=\pm1$ and $\cos\phi_{\mathrm{B}}^{\ast}=\pm1.$ This yields%

\begin{gather}
\Pi_{\mathrm{A},\mathrm{B}}(\theta_{\mathrm{A}}^{\ast},\phi_{\mathrm{A}}%
^{\ast};\theta_{\mathrm{B}}^{\ast},\phi_{\mathrm{B}}^{\ast})=\nonumber\\
=\frac{1}{4}\{\Delta_{2}-\Delta_{1}[\sin\theta_{\mathrm{A}}^{\ast}\cos
\theta_{\mathrm{B}}^{\ast}\cos\phi_{\mathrm{A}}^{\ast}-\cos\theta_{\mathrm{A}%
}^{\ast}\sin\theta_{\mathrm{B}}^{\ast}\cos\phi_{\mathrm{B}}^{\ast
}]\},\nonumber\\
=\frac{1}{4}\{\Delta_{2}-\Delta_{1}[\pm(\pm\frac{1}{2})\pm(\pm\frac{1}%
{2})]\},\nonumber\\
=\frac{1}{4}\{\Delta_{2}-\frac{1}{2}\Delta_{1}[\pm1\pm1]\}=\frac{1}{4}%
\{\Delta_{2}-\frac{1}{2}\Delta_{1}(2,-2,0)\},\nonumber\\
=\frac{1}{4}(\Delta_{2}\pm\Delta_{1}),\text{ }\frac{1}{4}\Delta_{2}%
,\nonumber\\
=\frac{1}{2}(\alpha+\delta),\text{ }\frac{1}{2}(\beta+\gamma),\text{ }\frac
{1}{4}(\alpha+\beta+\gamma+\delta).
\end{gather}

\subsubsection{Case $\sin\theta_{\mathrm{A}}^{\ast}\neq0$, $\sin
\theta_{\mathrm{B}}^{\ast}\neq0$ and $\cos\phi_{\mathrm{A}}^{\ast}\neq0$ and
$\cos\phi_{\mathrm{B}}^{\ast}\neq0$}

Referring to (\ref{a_},~\ref{b_}) we then have%

\begin{equation}
\cos\phi_{\mathrm{A}}^{\ast}\cos\phi_{\mathrm{B}}^{\ast}+\cot\theta
_{\mathrm{A}}^{\ast}\cot\theta_{\mathrm{B}}^{\ast}=0,\label{e_}%
\end{equation}
which must hold true along with Eqs.~(\ref{c},~\ref{d}). That is, the problem
then is to find a solution for $(\theta_{\mathrm{A}}^{\ast},\phi_{\mathrm{A}%
}^{\ast};\theta_{\mathrm{B}}^{\ast},\phi_{\mathrm{B}}^{\ast})$ from
Eqs.~(\ref{c},~\ref{d},~\ref{e_}). Eqs.~(\ref{c},~\ref{d}) can be written as%

\begin{equation}
\cos\phi_{\mathrm{A}}^{\ast}\sin\phi_{\mathrm{B}}^{\ast}=\cot\theta
_{\mathrm{B}}^{\ast}\sin\phi_{\mathrm{A}}^{\ast},\text{ \ \ }\sin
\phi_{\mathrm{A}}^{\ast}\cos\phi_{\mathrm{B}}^{\ast}=-\cot\theta_{\mathrm{A}%
}^{\ast}\sin\phi_{\mathrm{B}}^{\ast},
\end{equation}
and on multiplying the sides together we obtain%

\[
\cos\phi_{\mathrm{A}}^{\ast}\cos\phi_{\mathrm{B}}^{\ast}\sin\phi_{\mathrm{A}%
}^{\ast}\sin\phi_{\mathrm{B}}^{\ast}=-\sin\phi_{\mathrm{A}}^{\ast}\sin
\phi_{\mathrm{B}}^{\ast}\cot\theta_{\mathrm{A}}^{\ast}\cot\theta_{\mathrm{B}%
}^{\ast},
\]
from which Eq.~(\ref{e_}) follows as given below%

\begin{equation}
\cos\phi_{\mathrm{A}}^{\ast}\cos\phi_{\mathrm{B}}^{\ast}+\cot\theta
_{\mathrm{A}}^{\ast}\cot\theta_{\mathrm{B}}^{\ast}=0.
\end{equation}
As Eq.~(\ref{e_}) follows from (\ref{c},~\ref{d}), it is not required to
consider Eq.~(\ref{e_}) and can rewrite Eqs.~(\ref{c},~\ref{d}) as%

\begin{equation}
\cos\phi_{\mathrm{A}}^{\ast}\sin\phi_{\mathrm{B}}^{\ast}-\cot\theta
_{\mathrm{B}}^{\ast}\sin\phi_{\mathrm{A}}^{\ast}=0,\text{ \ \ }\sin
\phi_{\mathrm{A}}^{\ast}\cos\phi_{\mathrm{B}}^{\ast}+\cot\theta_{\mathrm{A}%
}^{\ast}\sin\phi_{\mathrm{B}}^{\ast}=0.
\end{equation}
When $\sin\phi_{\mathrm{A}}^{\ast}\neq0$ and $\sin\phi_{\mathrm{B}}^{\ast}%
\neq0$ then the above equations can be written as%

\begin{equation}
\cot\phi_{\mathrm{A}}^{\ast}\sin\phi_{\mathrm{B}}^{\ast}=\cot\theta
_{\mathrm{B}}^{\ast},\text{ \ \ }-\sin\phi_{\mathrm{A}}^{\ast}\cot
\phi_{\mathrm{B}}^{\ast}=\cot\theta_{\mathrm{A}}^{\ast}.\label{cots}%
\end{equation}
Note that substituting from Eqs.~(\ref{cots}) into Eqs.~(\ref{a},~\ref{b}) and
Eqs.~(\ref{c},~\ref{d}) changes them to identities. From (\ref{cots}) we obtain%

\begin{equation}
\theta_{\mathrm{A}}^{\ast}=\operatorname{arccot}(-\sin\phi_{\mathrm{A}}^{\ast
}\cot\phi_{\mathrm{B}}^{\ast}),\text{ \ \ }\theta_{\mathrm{B}}^{\ast
}=\operatorname{arccot}(\cot\phi_{\mathrm{A}}^{\ast}\sin\phi_{\mathrm{B}%
}^{\ast}).\label{arccots}%
\end{equation}
As $\sin\theta_{\mathrm{A}}^{\ast}\neq0$, $\sin\theta_{\mathrm{B}}^{\ast}%
\neq0$ and $\cos\phi_{\mathrm{A}}^{\ast}\neq0$, $\cos\phi_{\mathrm{B}}^{\ast
}\neq0,$ the players' payoffs are obtained as%

\begin{gather}
\Pi_{\mathrm{A},\mathrm{B}}(\theta_{\mathrm{A}}^{\ast},\phi_{\mathrm{A}}%
^{\ast};\theta_{\mathrm{B}}^{\ast},\phi_{\mathrm{B}}^{\ast})=\nonumber\\
=\frac{1}{4}\{\Delta_{2}-\Delta_{1}\sin\theta_{\mathrm{A}}^{\ast}\sin
\theta_{\mathrm{B}}^{\ast}[\sin\phi_{\mathrm{A}}^{\ast}\sin\phi_{\mathrm{B}%
}^{\ast}+\cot\theta_{\mathrm{B}}^{\ast}\cos\phi_{\mathrm{A}}^{\ast}-\cot
\theta_{\mathrm{A}}^{\ast}\cos\phi_{\mathrm{B}}^{\ast}]\},\label{4_}%
\end{gather}
and by substituting from Eqs.~(\ref{cots}) to Eq.~(\ref{4_}) we obtain%

\begin{gather}
\Pi_{\mathrm{A},\mathrm{B}}(\theta_{\mathrm{A}}^{\ast},\phi_{\mathrm{A}}%
^{\ast};\theta_{\mathrm{B}}^{\ast},\phi_{\mathrm{B}}^{\ast})=\Pi
_{\mathrm{A},\mathrm{B}}(\phi_{\mathrm{A}}^{\ast};\phi_{\mathrm{B}}^{\ast
})\nonumber\\
=\frac{1}{4}\{\Delta_{2}-\Delta_{1}\sin[\operatorname{arccot}(-\sin
\phi_{\mathrm{A}}^{\ast}\cot\phi_{\mathrm{B}}^{\ast})]\sin
[\operatorname{arccot}(\cot\phi_{\mathrm{A}}^{\ast}\sin\phi_{\mathrm{B}}%
^{\ast})]\times\nonumber\\
\lbrack\sin\phi_{\mathrm{A}}^{\ast}\sin\phi_{\mathrm{B}}^{\ast}+\cot
\phi_{\mathrm{A}}^{\ast}\cos\phi_{\mathrm{A}}^{\ast}\sin\phi_{\mathrm{B}%
}^{\ast}+\sin\phi_{\mathrm{A}}^{\ast}\cos\phi_{\mathrm{B}}^{\ast}\cot
\phi_{\mathrm{B}}^{\ast}]\}.\label{5_}%
\end{gather}

An example, consider the case when $\phi_{\mathrm{A}}^{\ast}=\pi/4$ and
$\phi_{\mathrm{B}}^{\ast}=3\pi/4$ for which $\theta_{\mathrm{A}}^{\ast
}=0.95532=\theta_{\mathrm{B}}^{\ast}.$ As the pair $(\theta_{\mathrm{A}}%
^{\ast},\theta_{\mathrm{B}}^{\ast})$ can be determined from a pair
$(\phi_{\mathrm{A}}^{\ast},\phi_{\mathrm{B}}^{\ast})$ that is arbitrarily
chosen, there exist an infinite set of Nash equilibria. With $(\phi
_{\mathrm{A}}^{\ast},\phi_{\mathrm{B}}^{\ast})\in\lbrack0,2\pi)$ the players'
payoffs at all these equilibria can be plotted as below with $\phi
_{\mathrm{A}}^{\ast}$ and $\phi_{\mathrm{B}}^{\ast}$ taken as independent coordinates.%

\begin{figure}[ptb]%
\centering
\includegraphics[
height=3.845in,
width=5.1309in
]%
{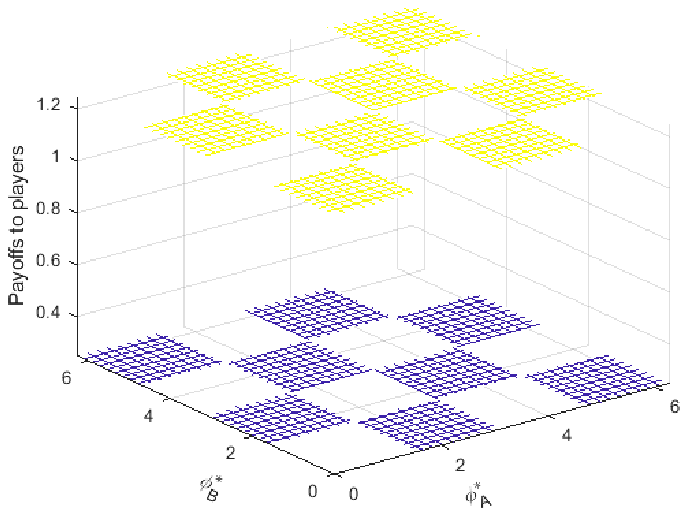}%
\caption{An infinite number of Nash equilbria exist when the game is played
with the state $\left\vert \psi_{ini}\right\rangle =\frac{1}{2}\left(
\left\vert 00\right\rangle +\left\vert 01\right\rangle -\left\vert
10\right\rangle +\left\vert 11\right\rangle \right)  .$ Players' payoffs at
these equilibria $\Pi_{A,B}(\phi_{A}^{\ast};\phi_{B}^{\ast})$ and given in Eq.
(\ref{5_}) are plotted for $\Delta_{2}=3$ and $\Delta_{1}=2$ againt variables
$\phi_{A}^{\ast},\phi_{B}^{\ast}\in\lbrack0,2\pi)$ considered independent. The
$\phi_{A}^{\ast},\phi_{B}^{\ast}$ plane is found to be divided into
rectangular patches with respect to the variation of players' payoffs. Angles
$\theta_{A}^{\ast},\theta_{B}^{\ast}$ that correspond to $\phi_{A}^{\ast}%
,\phi_{B}^{\ast}$ are determined from Eq. (\ref{arccots}).}%
\end{figure}

The above plot in a different range of values for $\phi_{\mathrm{A}}^{\ast
},\phi_{\mathrm{B}}^{\ast}$ is given below.%

\begin{figure}[ptb]%
\centering
\includegraphics[
height=3.7533in,
width=5.009in
]%
{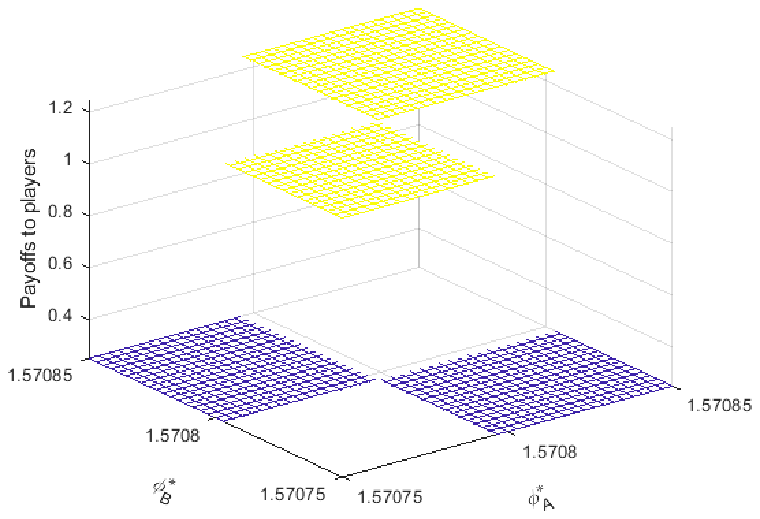}%
\caption{Players' payoffs $\Pi_{A,B}(\phi_{A}^{\ast};\phi_{B}^{\ast})$ for the
state $\left\vert \psi_{\mathrm{ini}}\right\rangle =\frac{1}{2}\left(
\left\vert 00\right\rangle +\left\vert 01\right\rangle -\left\vert
10\right\rangle +\left\vert 11\right\rangle \right)  .$as given in Eq.
(\ref{5_}) are plotted for $\Delta_{2}=3$ and $\Delta_{1}=2$ againt the
variables $\phi_{A}^{\ast},\phi_{B}^{\ast}$ in a different range.}%
\end{figure}

\section{Players' directional choices and the violation of Bell-CHSH
inequality}

The proposed setup for playing a two-player quantum game uses the setting of
an EPR type experiment. Consider such an experiment that is designed to test
the Bell-CHSH inequality \cite{Peres} in which two correlated particles $1$
and $2$ fly apart in opposite directions from some common source.
Subsequently, each of the particles enters its own measuring apparatus which
can measure either along $\mathbf{\hat{a}}$ or $\mathbf{\hat{a}}^{\prime}$ for
particle $1$ and $\mathbf{\hat{b}}$ or $\mathbf{\hat{b}}^{\prime}$ for
particle $2$. The possible values of these variables may be taken to be $+1$
and $-1$ and the source emits a very large number of particle pairs. We let%

\begin{equation}
\mathbf{\hat{a}}=(\theta_{\mathrm{A}},\phi_{\mathrm{A}})\text{, }%
\mathbf{\hat{a}}^{\prime}=(\theta_{\mathrm{A}}^{\prime},\phi_{\mathrm{A}%
}^{\prime}),\text{ }\mathbf{\hat{b}}=(\theta_{\mathrm{B}},\phi_{\mathrm{B}%
}),\text{ }\mathbf{\hat{b}}^{\prime}=(\theta_{\mathrm{B}}^{\prime}%
,\phi_{\mathrm{B}}^{\prime}),
\end{equation}
where $\theta_{\mathrm{A}},\theta_{\mathrm{B}},\theta_{\mathrm{A}}^{\prime
},\theta_{\mathrm{B}}^{\prime}\in\lbrack0,\pi]$ and $\phi_{\mathrm{A}}%
,\phi_{\mathrm{B}},\phi_{\mathrm{A}}^{\prime},\phi_{\mathrm{B}}^{\prime}%
\in\lbrack0,2\pi)$. Bell-CHSH inequality can be written as $\left\vert
\Lambda\right\vert \leq2$ where%

\begin{align}
\Lambda & =2[\Pr(\mathbf{\hat{a}}_{+1},\mathbf{\hat{b}}_{+1})+\Pr
(\mathbf{\hat{a}}_{-1},\mathbf{\hat{b}}_{-1})+\Pr(\mathbf{\hat{a}}%
_{+1},\mathbf{\hat{b}}_{+1}^{\prime})+\Pr(\mathbf{\hat{a}}_{-1},\mathbf{\hat
{b}}_{-1}^{\prime})+\nonumber\\
& \Pr(\mathbf{\hat{a}}_{+1}^{\prime},\mathbf{\hat{b}}_{+1})+\Pr(\mathbf{\hat
{a}}_{-1}^{\prime},\mathbf{\hat{b}}_{-1})+\Pr(\mathbf{\hat{a}}_{+1}^{\prime
},\mathbf{\hat{b}}_{-1}^{\prime})+\Pr(\mathbf{\hat{a}}_{-1}^{\prime
},\mathbf{\hat{b}}_{+1}^{\prime})-2]
\end{align}
Now, for the state $\left\vert \psi_{\mathrm{ini}}\right\rangle =\frac
{1}{\sqrt{2}}\left(  \left\vert 00\right\rangle +i\left\vert 11\right\rangle
\right)  $, considered above, we have%

\begin{align}
\Pr(\mathbf{\hat{a}}_{+1},\mathbf{\hat{b}}_{+1}) &  =\frac{1}{4}\{1+\sin
\theta_{\mathrm{A}}\sin\theta_{\mathrm{B}}\sin(\phi_{\mathrm{A}}%
+\phi_{\mathrm{B}})+\cos\theta_{\mathrm{A}}\cos\theta_{\mathrm{B}%
}\},\nonumber\\
\Pr(\mathbf{\hat{a}}_{-1},\mathbf{\hat{b}}_{-1}) &  =\frac{1}{4}\{1+\sin
\theta_{\mathrm{A}}\sin\theta_{\mathrm{B}}\sin(\phi_{\mathrm{A}}%
+\phi_{\mathrm{B}})+\cos\theta_{\mathrm{A}}\cos\theta_{\mathrm{B}%
}\},\nonumber\\
&  ....\nonumber\\
\Pr(\mathbf{\hat{a}}_{-1}^{\prime},\mathbf{\hat{b}}_{+1}^{\prime}) &
=\frac{1}{4}\{1-\sin\theta_{\mathrm{A}}^{\prime}\sin\theta_{\mathrm{B}%
}^{\prime}\sin(\phi_{\mathrm{A}}^{\prime}+\phi_{\mathrm{B}}^{\prime}%
)-\cos\theta_{\mathrm{A}}^{\prime}\cos\theta_{\mathrm{B}}^{\prime}\},
\end{align}
and we obtain%

\begin{align}
\Lambda & =\sin\theta_{\mathrm{A}}\sin\theta_{\mathrm{B}}\sin(\phi
_{\mathrm{A}}+\phi_{\mathrm{B}})+\sin\theta_{\mathrm{A}}\sin\theta
_{\mathrm{B}}^{\prime}\sin(\phi_{\mathrm{A}}+\phi_{\mathrm{B}}^{\prime}%
)+\sin\theta_{\mathrm{A}}^{\prime}\sin\theta_{\mathrm{B}}\sin(\phi
_{\mathrm{A}}^{\prime}+\phi_{\mathrm{B}})\nonumber\\
& -\sin\theta_{\mathrm{A}}^{\prime}\sin\theta_{\mathrm{B}}^{\prime}\sin
(\phi_{\mathrm{A}}^{\prime}+\phi_{\mathrm{B}}^{\prime})+\cos\theta
_{\mathrm{A}}\cos\theta_{\mathrm{B}}+\cos\theta_{\mathrm{A}}\cos
\theta_{\mathrm{B}}^{\prime}+\cos\theta_{\mathrm{A}}^{\prime}\cos
\theta_{\mathrm{B}}-\cos\theta_{\mathrm{A}}^{\prime}\cos\theta_{\mathrm{B}%
}^{\prime},\nonumber\\
&
\end{align}
that can be expressed as%

\begin{align}
\Lambda & =\sin\theta_{\mathrm{A}}[\sin\theta_{\mathrm{B}}\sin(\phi
_{\mathrm{A}}+\phi_{\mathrm{B}})+\sin\theta_{\mathrm{B}}^{\prime}\sin
(\phi_{\mathrm{A}}+\phi_{\mathrm{B}}^{\prime})]+\nonumber\\
& \sin\theta_{\mathrm{A}}^{\prime}[\sin\theta_{\mathrm{B}}\sin(\phi
_{\mathrm{A}}^{\prime}+\phi_{\mathrm{B}})-\sin\theta_{\mathrm{B}}^{\prime}%
\sin(\phi_{\mathrm{A}}^{\prime}+\phi_{\mathrm{B}}^{\prime})]+\nonumber\\
& \cos\theta_{\mathrm{A}}(\cos\theta_{\mathrm{B}}+\cos\theta_{\mathrm{B}%
}^{\prime})+\cos\theta_{\mathrm{A}}^{\prime}(\cos\theta_{\mathrm{B}}%
-\cos\theta_{\mathrm{B}}^{\prime}).\label{Discrim}%
\end{align}
We take, for instance, $\phi_{\mathrm{A}}=\phi_{\mathrm{B}}=\phi_{\mathrm{A}%
}^{\prime}=\phi_{\mathrm{B}}^{\prime}=\pi/4$ and this reduces (\ref{Discrim}) to%

\begin{align}
\Lambda & =\sin\theta_{\mathrm{A}}(\sin\theta_{\mathrm{B}}+\sin\theta
_{\mathrm{B}}^{\prime})+\sin\theta_{\mathrm{A}}^{\prime}(\sin\theta
_{\mathrm{B}}-\sin\theta_{\mathrm{B}}^{\prime})+\nonumber\\
& \cos\theta_{\mathrm{A}}(\cos\theta_{\mathrm{B}}+\cos\theta_{\mathrm{B}%
}^{\prime})+\cos\theta_{\mathrm{A}}^{\prime}(\cos\theta_{\mathrm{B}}%
-\cos\theta_{\mathrm{B}}^{\prime}).
\end{align}
Now, consider the case when $\theta_{\mathrm{A}}=\pi/4,$ $\theta_{\mathrm{A}%
}^{\prime}=3\pi/4,$ $\theta_{\mathrm{B}}=\pi/2,$ $\theta_{\mathrm{B}}^{\prime
}=\pi/4$ and we obtain $\Lambda=1+\sqrt{2}\geq2$ and Bell's inequality is
violated. For the state $\left\vert \psi_{\mathrm{ini}}\right\rangle =\frac
{1}{\sqrt{2}}\left(  \left\vert 00\right\rangle +i\left\vert 11\right\rangle
\right)  ,$ and with $\phi_{\mathrm{A}}=\phi_{\mathrm{B}}=\phi_{\mathrm{A}%
}^{\prime}=\phi_{\mathrm{B}}^{\prime}=\pi/4, $ the players' payoffs
(\ref{Payoffs}) are then obtained as%

\begin{equation}
\Pi_{\mathrm{A}},_{\mathrm{B}}(\pi/4,\pi/4;\pi/2,\pi/4)=\frac{1}{4}%
\{\alpha(1+1/\sqrt{2})+\beta(1-1/\sqrt{2})+\gamma(1-1/\sqrt{2})+\delta
(1+1/\sqrt{2})\}.
\end{equation}
To know whether these players' payoffs in the quantum game can be embedded
within the classical game,we refer to the players' payoffs
(\ref{Mixed_strategy_payoffs}) in the mixed strategy game. We require
$\Pi_{\mathrm{A}}(p,q)=\Pi_{\mathrm{B}}(p,q)$ in accordance with the players'
payoff relations (\ref{Payoffs}) in the quantum game. This results in
$\beta=\gamma$ and the players' payoffs in the mixed-strategy classical game
(\ref{Mixed_strategy_payoffs}) become%

\begin{equation}
\Pi_{\text{\textrm{A,B}}}(p,q)=\alpha pq+\beta(p+q-2pq)+\delta
(1-p)(1-q).\label{ClassPayoffs}%
\end{equation}
The players' payoffs in the quantum game for the directional choice
$(\pi/4,\pi/4;\pi/2,\pi/4)$, and at which Bell's inequalities are violated, are%

\begin{equation}
\Pi_{\mathrm{A}},_{\mathrm{B}}(\pi/4,\pi/4;\pi/2,\pi/4)=\frac{1}{4}%
\{\alpha(1+1/\sqrt{2})+\beta(2-\sqrt{2})+\delta(1+1/\sqrt{2}%
)\}.\label{Qpayoffs}%
\end{equation}
Comparing (\ref{ClassPayoffs}) with (\ref{Qpayoffs}) gives%

\begin{equation}
pq=\frac{1}{4}(1+1/\sqrt{2}),\text{ }p+q-2pq=\frac{1}{4}(2-\sqrt{2}),\text{
}(1-p)(1-q)=\frac{1}{4}(1+1/\sqrt{2}),
\end{equation}
and from which we obtain $p+q=1$ and $q=\frac{1\pm\sqrt{-1/\sqrt{2}}}{2},$
showing that for the directional choice $(\pi/4,\pi/4;\pi/2,\pi/4)$ on behalf
of the two players, and at which the players' payoffs are given by
(\ref{Qpayoffs}), the players' payoffs in the quantum game have no mapping
within the classical mixed-strategy game.

\section{Discussion}

This paper presents a quantization scheme for playing two-player games in
which each player's strategy consist of orientating a unit vector in three
dimensions. In the usual approach, a Nash equilibrium is a pair of unitary
operators $(\hat{U}_{\mathrm{A}}^{\ast},\hat{U}_{\mathrm{B}}^{\ast})$ defined
by the inequalities (\ref{NashInequalities}). For the given initial quantum
state $\left\vert \psi_{\mathrm{ini}}\right\rangle $, the proposed quantum
game uses an EPR setting in which player \textrm{A}'s and player \textrm{B}'s
strategies consist of orientating the unit vector $\mathbf{\hat{a}}$ and
$\mathbf{\hat{b}}$, respectively. The polarization (or spin) measurements in
an EPR setting result in the outcome $m=\pm1$ along $\mathbf{\hat{a}}$ and
$n=\pm1$ along $\mathbf{\hat{b}}$. The players' payoff relations in the
considered scheme involves a set of quantum probabilities that are obtained,
according to the Eqs. (\ref{Q_probabilities}) from each player's strategies,
entries of the matrix of the game, and the initial quantum state $\left\vert
\psi_{\mathrm{ini}}\right\rangle $. The payoff relations in the quantum game
are defined in terms of this set as described by Eqs.~(\ref{Payoff_Alice}%
,~\ref{Payoff_Bob}). That is, the set of underlying quantum probabilities are
generated by each player's strategies---consisting of the players' directional
choices--- along with the initial quantum state $\left\vert \psi
_{\mathrm{ini}}\right\rangle $.

With directional choices as player's strategies, the NE in the quantum game
consists of a pair of unit vectors $(\mathbf{\hat{a}}^{\ast},\mathbf{\hat{b}%
}^{\ast})$ in three dimensional space. Also, the classical mixed strategy game
is recovered---for certain initial states $\left\vert \psi_{\mathrm{ini}%
}\right\rangle $---when each player's directional choices $\mathbf{\hat{a}}$
and $\mathbf{\hat{b}}$ follow the assigned trajectories in space.

The scheme is analyzed for three initial states $\left\vert \psi
_{\mathrm{ini}}\right\rangle $. We show that playing the game with the quantum
state $\left\vert \psi_{\mathrm{ini}}\right\rangle =\frac{1}{2}\left(
\left\vert 00\right\rangle +\left\vert 01\right\rangle +\left\vert
10\right\rangle +\left\vert 11\right\rangle \right)  $ results in the
classical mixed strategy game in which Alice's and Bob's directional choices
are given by (\ref{product_state_1},~\ref{product_state_2}). These express
their strategies $p$ and $q$ in the classical mixed strategy game in terms of
the angles $\theta_{\mathrm{A}},\phi_{\mathrm{A}};$ $\theta_{\mathrm{B}}%
,\phi_{\mathrm{B}}$---representing player \textrm{A}'s and player \textrm{B}'s
directional choices. For given values of $p$ and $q$, Eqs.
(\ref{product_state_1},~\ref{product_state_2}) therefore represent the
trajectories on the surface of a unit sphere traced by the tips of the unit
vectors $\mathbf{\hat{a}}$ and $\mathbf{\hat{b}}$, respectively. Playing the
game with the maximally entangled state $\left\vert \psi_{\mathrm{ini}%
}\right\rangle =\frac{1}{\sqrt{2}}\left(  \left\vert 00\right\rangle
+i\left\vert 11\right\rangle \right)  $ results in obtaining the players'
payoff relations (\ref{Payoffs}) that cannot be reduced to the classical
mixed-strategy payoff relations. That is interpreted by stating that there do
not exist such trajectories on the unit sphere such that when these
trajectories are followed by the tips of each player's strategic choices, the
quantum game results in the classical mixed-strategy game.

Playing the game with the state $\left\vert \psi_{\mathrm{ini}}\right\rangle
=\frac{1}{2}\left(  \left\vert 00\right\rangle +\left\vert 01\right\rangle
-\left\vert 10\right\rangle +\left\vert 11\right\rangle \right)  $ results in
a number of Nash equilibria appearing as the edge cases. For the non-edge
cases, we determine that there exist an infinite number of Nash equilibria. At
these Nash equilibria we consider $\phi_{\mathrm{A}}^{\ast},\phi_{\mathrm{B}%
}^{\ast}\in\lbrack0,2\pi)$ as independent variables from which the angles
$\theta_{\mathrm{A}}^{\ast},\theta_{\mathrm{B}}^{\ast}\in\lbrack0,\pi]$ can be
obtained using Eqs.~(\ref{arccots}). Corresponding to these angles, the
players' payoffs at the Nash equilibria i.e. $\Pi_{\mathrm{A},\mathrm{B}}%
(\phi_{\mathrm{A}}^{\ast};\phi_{\mathrm{B}}^{\ast})$ are obtained by
Eq.~(\ref{5_}). The $\phi_{\mathrm{A}}^{\ast},\phi_{\mathrm{B}}^{\ast}$ plane
is found to be divided into rectangular patches with the corresponding
variation of the players' payoffs into two distinct values.

We agree with the perspective that if quantum advantage (or an improved
game-theoretical outcome) does not emerge in a quantum game, it does not
necessarily change a quantum game to a classical game. The games considered in
this paper are truly quantum as they involve quantum superposition and
entanglement. In particular, the players' payoff relations are defined from
underlying quantum mechanical probability distributions and that the
corresponding classical games are recoverable by restricting players'
directional choices along specific trajectories in three dimensions.

Considering Bell-CHSH inequality for the directional choice $(\pi/4,\pi
/4;\pi/2,\pi/4)$ on behalf of two players, we show that Bell's inequalities
are violated. For these directional choices, the players' payoffs in the
quantum game are shown to have no mapping within the classical mixed-strategy
game. An EPR setting provides the route for the players' access to quantum
probability distributions that can violate Bell's inequalities. As the quantum
game involves classical strategy sets, Enk and Pike's argument \cite{EnkPike}
is circumvented.

\section{Conclusion}

Game theory is widely used in a number of disciplines and this paper presents
a scheme for two-player quantum games that establishes a more direct link
between a classical game and its quantum version. Players in the quantum game
have access to classical strategy sets as is the case in the corresponding
classical game, allowing us to circumvent Enk and Pike's argument. As the
contribution of this paper to the theory of quantum games is built on the EPR
paradox, a possible future research direction can be to interpret the EPR
paradox as a strategic quantum game. Also, the proposed scheme motivates
studying refinements of the NE concept using an EPR settinga with players'
moves consisting of directional choices.

\section{Data Availability Statement}

Data sharing not applicable to this article as no datasets were generated or
analysed during the current study.

\section{Appendix A}

The first term in the payoff relations (\ref{Payoffs_gen}) when the game is
played with the state $\left\vert \psi_{\mathrm{ini}}\right\rangle =\frac
{1}{2}\left(  \left\vert 00\right\rangle +\left\vert 01\right\rangle
+\left\vert 10\right\rangle +\left\vert 11\right\rangle \right)  $ is given as%

\begin{gather}
\frac{(\alpha,\alpha)}{16(1+\cos\theta_{A})(1+\cos\theta_{B})}{\Large \{}%
[(1+\cos\theta_{A})(1+\cos\theta_{B})+(1+\cos\theta_{A})\sin\theta_{B}\cos
\phi_{B}+\nonumber\\
(1+\cos\theta_{B})\sin\theta_{A}\cos\phi_{A}+\sin\theta_{A}\sin\theta_{B}%
\cos(\phi_{A}+\phi_{B})]^{2}+\nonumber\\
\lbrack(1+\cos\theta_{A})\sin\theta_{B}\sin\phi_{B}+(1+\cos\theta_{B}%
)\sin\theta_{A}\sin\phi_{A}+\sin\theta_{A}\sin\theta_{B}\sin(\phi_{A}+\phi
_{B})]^{2}{\large \}}\label{1}%
\end{gather}
and consider its part%

\begin{align}
&  [(1+\cos\theta_{A})(1+\cos\theta_{B})+(1+\cos\theta_{A})\sin\theta_{B}%
\cos\phi_{B}+\nonumber\\
&  (1+\cos\theta_{B})\sin\theta_{A}\cos\phi_{A}+\sin\theta_{A}\sin\theta
_{B}\cos(\phi_{A}+\phi_{B})]^{2}\nonumber\\
&  =[1+\cos\theta_{B}+\cos\theta_{A}+\cos\theta_{A}\cos\theta_{B}+\sin
\theta_{B}\cos\phi_{B}+\cos\theta_{A}\sin\theta_{B}\cos\phi_{B}+\sin\theta
_{A}\cos\phi_{A}+\nonumber\\
&  \cos\theta_{B}\sin\theta_{A}\cos\phi_{A}+\sin\theta_{A}\sin\theta_{B}%
(\cos\phi_{A}\cos\phi_{B}-\sin\phi_{A}\sin\phi_{B})]^{2}%
\end{align}

\begin{align}
&  =[1+\cos\theta_{B}+\cos\theta_{A}+\cos\theta_{A}\cos\theta_{B}+\sin
\theta_{B}\cos\phi_{B}+\nonumber\\
&  \cos\theta_{A}\sin\theta_{B}\cos\phi_{B}+\sin\theta_{A}\cos\phi_{A}%
+\sin\theta_{A}\cos\theta_{B}\cos\phi_{A}+\nonumber\\
&  \sin\theta_{A}\sin\theta_{B}\cos\phi_{A}\cos\phi_{B}-\sin\theta_{A}%
\sin\theta_{B}\sin\phi_{A}\sin\phi_{B}]^{2}%
\end{align}

\begin{align}
&  =[(1+\cos\theta_{A})(1+\cos\theta_{B})+(1+\cos\theta_{A})\sin\theta_{B}%
\cos\phi_{B}+(1+\cos\theta_{B})\sin\theta_{A}\cos\phi_{A}+\nonumber\\
&  \sin\theta_{A}\sin\theta_{B}\cos(\phi_{A}+\phi_{B})]\ast\lbrack
(1+\cos\theta_{A})(1+\cos\theta_{B})+(1+\cos\theta_{A})\sin\theta_{B}\cos
\phi_{B}+\nonumber\\
&  (1+\cos\theta_{B})\sin\theta_{A}\cos\phi_{A}+\sin\theta_{A}\sin\theta
_{B}\cos(\phi_{A}+\phi_{B})]
\end{align}

\begin{align}
&  =(1+\cos\theta_{A})^{2}(1+\cos\theta_{B})^{2}+(1+\cos\theta_{A})^{2}%
\sin^{2}\theta_{B}\cos^{2}\phi_{B}+\nonumber\\
&  (1+\cos\theta_{B})^{2}\sin^{2}\theta_{A}\cos^{2}\phi_{A}+\sin^{2}\theta
_{A}\sin^{2}\theta_{B}\cos^{2}(\phi_{A}+\phi_{B})+\nonumber\\
&  2(1+\cos\theta_{A})^{2}(1+\cos\theta_{B})\sin\theta_{B}\cos\phi
_{B}+\nonumber\\
&  2(1+\cos\theta_{A})\sin\theta_{B}\cos\phi_{B}(1+\cos\theta_{B})\sin
\theta_{A}\cos\phi_{A}+\nonumber\\
&  2(1+\cos\theta_{B})\sin\theta_{A}\cos\phi_{A}\sin\theta_{A}\sin\theta
_{B}\cos(\phi_{A}+\phi_{B})+\nonumber\\
&  2\sin\theta_{A}\sin\theta_{B}\cos(\phi_{A}+\phi_{B})(1+\cos\theta
_{A})(1+\cos\theta_{B})+\nonumber\\
&  2(1+\cos\theta_{A})(1+\cos\theta_{B})^{2}\sin\theta_{A}\cos\phi
_{A}+\nonumber\\
&  2(1+\cos\theta_{A})\sin\theta_{B}\cos\phi_{B}\sin\theta_{A}\sin\theta
_{B}\cos(\phi_{A}+\phi_{B})
\end{align}

\begin{align}
&  =(1+\cos\theta_{A})^{2}(1+\cos\theta_{B})^{2}+\sin^{2}\theta_{A}\sin
^{2}\theta_{B}\cos^{2}(\phi_{A}+\phi_{B})+\nonumber\\
&  2(1+\cos\theta_{A})^{2}(1+\cos\theta_{B})\sin\theta_{B}\cos\phi
_{B}+\nonumber\\
&  (1+\cos\theta_{A})^{2}\sin^{2}\theta_{B}\cos^{2}\phi_{B}+(1+\cos\theta
_{B})^{2}\sin^{2}\theta_{A}\cos^{2}\phi_{A}+\nonumber\\
&  2(1+\cos\theta_{A})(1+\cos\theta_{B})\sin\theta_{B}\cos\phi_{B}\sin
\theta_{A}\cos\phi_{A}+\nonumber\\
&  2(1+\cos\theta_{B})\sin^{2}\theta_{A}\cos\phi_{A}\sin\theta_{B}\cos
(\phi_{A}+\phi_{B})+\nonumber\\
&  2(1+\cos\theta_{A})(1+\cos\theta_{B})\sin\theta_{A}\sin\theta_{B}\cos
(\phi_{A}+\phi_{B})+\nonumber\\
&  2(1+\cos\theta_{A})(1+\cos\theta_{B})^{2}\sin\theta_{A}\cos\phi
_{A}+\nonumber\\
&  2(1+\cos\theta_{A})\sin^{2}\theta_{B}\cos\phi_{B}\sin\theta_{A}\cos
(\phi_{A}+\phi_{B})
\end{align}

\begin{align}
&  =(1+\cos\theta_{A})(1+\cos\theta_{B}){\LARGE [}(1+\cos\theta_{A}%
)(1+\cos\theta_{B})+2\sin\theta_{A}\sin\theta_{B}\cos(\phi_{A}+\phi
_{B})+\nonumber\\
&  2(1+\cos\theta_{B})\sin\theta_{A}\cos\phi_{A}+2\sin\theta_{B}\cos\phi
_{B}\sin\theta_{A}\cos\phi_{A}+2(1+\cos\theta_{A})\sin\theta_{B}\cos\phi
_{B}{\LARGE ]}+\nonumber\\
&  \sin^{2}\theta_{A}\sin^{2}\theta_{B}\cos^{2}(\phi_{A}+\phi_{B}%
)+(1+\cos\theta_{A})^{2}\sin^{2}\theta_{B}\cos^{2}\phi_{B}+\nonumber\\
&  (1+\cos\theta_{B})^{2}\sin^{2}\theta_{A}\cos^{2}\phi_{A}+2(1+\cos\theta
_{B})\sin^{2}\theta_{A}\cos\phi_{A}\sin\theta_{B}\cos(\phi_{A}+\phi
_{B})+\nonumber\\
&  2(1+\cos\theta_{A})\sin^{2}\theta_{B}\cos\phi_{B}\sin\theta_{A}\cos
(\phi_{A}+\phi_{B}).\label{A_Eq_1}%
\end{align}

Now consider the 2nd term%

\begin{align}
&  [(1+\cos\theta_{A})\sin\theta_{B}\sin\phi_{B}+(1+\cos\theta_{B})\sin
\theta_{A}\sin\phi_{A}+\nonumber\\
&  \sin\theta_{A}\sin\theta_{B}\sin(\phi_{A}+\phi_{B})]^{2}\nonumber\\
&  =(1+\cos\theta_{A})^{2}\sin^{2}\theta_{B}\sin^{2}\phi_{B}+(1+\cos\theta
_{B})^{2}\sin^{2}\theta_{A}\sin^{2}\phi_{A}+\nonumber\\
&  \sin^{2}\theta_{A}\sin^{2}\theta_{B}\sin^{2}(\phi_{A}+\phi_{B}%
)+2(1+\cos\theta_{A})(1+\cos\theta_{B})\sin\theta_{A}\sin\theta_{B}\sin
\phi_{A}\sin\phi_{B}+\nonumber\\
&  2(1+\cos\theta_{B})\sin^{2}\theta_{A}\sin\theta_{B}\sin(\phi_{A}+\phi
_{B})\sin\phi_{A}+\nonumber\\
&  2(1+\cos\theta_{A})\sin\theta_{A}\sin^{2}\theta_{B}\sin(\phi_{A}+\phi
_{B})\sin\phi_{B}.\label{A_Eq_2}%
\end{align}

Add Eqs.~(\ref{A_Eq_1},~\ref{A_Eq_2}) to obtain%

\begin{align}
&  =(1+\cos\theta_{A})(1+\cos\theta_{B}){\LARGE [}(1+\cos\theta_{A}%
)(1+\cos\theta_{B})+2\sin\theta_{A}\sin\theta_{B}\cos(\phi_{A}+\phi
_{B})+\nonumber\\
&  2(1+\cos\theta_{B})\sin\theta_{A}\cos\phi_{A}+2\sin\theta_{B}\cos\phi
_{B}\sin\theta_{A}\cos\phi_{A}+\nonumber\\
&  2(1+\cos\theta_{A})\sin\theta_{B}\cos\phi_{B}+2\sin\theta_{A}\sin\theta
_{B}\sin\phi_{A}\sin\phi_{B}{\LARGE ]}+\nonumber\\
&  \sin^{2}\theta_{A}\sin^{2}\theta_{B}\cos^{2}(\phi_{A}+\phi_{B})+\sin
^{2}\theta_{A}\sin^{2}\theta_{B}\sin^{2}(\phi_{A}+\phi_{B})+\nonumber\\
&  (1+\cos\theta_{A})^{2}[\sin^{2}\theta_{B}\cos^{2}\phi_{B}+\sin^{2}%
\theta_{B}\sin^{2}\phi_{B}]+\nonumber\\
&  (1+\cos\theta_{B})^{2}[\sin^{2}\theta_{A}\cos^{2}\phi_{A}+\sin^{2}%
\theta_{A}\sin^{2}\phi_{A}]+\nonumber\\
&  2(1+\cos\theta_{B})[\sin^{2}\theta_{A}\cos\phi_{A}\sin\theta_{B}\cos
(\phi_{A}+\phi_{B})+\nonumber\\
&  \sin^{2}\theta_{A}\sin\theta_{B}\sin(\phi_{A}+\phi_{B})\sin\phi
_{A}]+\nonumber\\
&  2(1+\cos\theta_{A})[\sin^{2}\theta_{B}\cos\phi_{B}\sin\theta_{A}\cos
(\phi_{A}+\phi_{B})+\nonumber\\
&  \sin\theta_{A}\sin^{2}\theta_{B}\sin(\phi_{A}+\phi_{B})\sin\phi_{B}]
\end{align}

\begin{align}
&  =(1+\cos\theta_{A})(1+\cos\theta_{B}){\LARGE [}(1+\cos\theta_{A}%
)(1+\cos\theta_{B})+2\sin\theta_{A}\sin\theta_{B}\cos(\phi_{A}+\phi
_{B})+\nonumber\\
&  2(1+\cos\theta_{B})\sin\theta_{A}\cos\phi_{A}+2\sin\theta_{A}\sin\theta
_{B}(\cos\phi_{A}\cos\phi_{B}+\sin\phi_{A}\sin\phi_{B})+\nonumber\\
&  2(1+\cos\theta_{A})\sin\theta_{B}\cos\phi_{B}{\LARGE ]}+\nonumber\\
&  \sin^{2}\theta_{A}\sin^{2}\theta_{B}+(1+\cos\theta_{A})^{2}[\sin^{2}%
\theta_{B}]+(1+\cos\theta_{B})^{2}[\sin^{2}\theta_{A}]+\nonumber\\
&  2(1+\cos\theta_{B})\sin^{2}\theta_{A}\sin\theta_{B}[\cos(\phi_{A}+\phi
_{B})\cos\phi_{A}+\sin(\phi_{A}+\phi_{B})\sin\phi_{A}]+\nonumber\\
&  2(1+\cos\theta_{A})\sin\theta_{A}\sin^{2}\theta_{B}[\cos(\phi_{A}+\phi
_{B})\cos\phi_{B}+\sin(\phi_{A}+\phi_{B})\sin\phi_{B}]
\end{align}

\begin{align}
&  =(1+\cos\theta_{A})(1+\cos\theta_{B}){\LARGE [}(1+\cos\theta_{A}%
)(1+\cos\theta_{B})+2\sin\theta_{A}\sin\theta_{B}\cos(\phi_{A}+\phi
_{B})+\nonumber\\
&  2(1+\cos\theta_{B})\sin\theta_{A}\cos\phi_{A}+2\sin\theta_{A}\sin\theta
_{B}\cos(\phi_{A}-\phi_{B})+2(1+\cos\theta_{A})\sin\theta_{B}\cos\phi
_{B}{\LARGE ]}+\nonumber\\
&  \sin^{2}\theta_{A}\sin^{2}\theta_{B}+(1+\cos\theta_{A})^{2}[\sin^{2}%
\theta_{B}]+(1+\cos\theta_{B})^{2}[\sin^{2}\theta_{A}]+\nonumber\\
&  2(1+\cos\theta_{B})\sin^{2}\theta_{A}\sin\theta_{B}[\cos(\phi_{A}+\phi
_{B})\cos\phi_{A}+\sin(\phi_{A}+\phi_{B})\sin\phi_{A}]+\nonumber\\
&  2(1+\cos\theta_{A})\sin\theta_{A}\sin^{2}\theta_{B}[\cos(\phi_{A}+\phi
_{B})\cos\phi_{B}+\sin(\phi_{A}+\phi_{B})\sin\phi_{B}]\nonumber\\
&  =(1+\cos\theta_{A})(1+\cos\theta_{B}){\LARGE [}(1+\cos\theta_{A}%
)(1+\cos\theta_{B})+2\sin\theta_{A}\sin\theta_{B}\cos(\phi_{A}+\phi
_{B})+\nonumber\\
&  2(1+\cos\theta_{B})\sin\theta_{A}\cos\phi_{A}+2\sin\theta_{A}\sin\theta
_{B}\cos(\phi_{A}-\phi_{B})+2(1+\cos\theta_{A})\sin\theta_{B}\cos\phi
_{B}{\LARGE ]}+\nonumber\\
&  \sin^{2}\theta_{A}\sin^{2}\theta_{B}+(1+\cos\theta_{A})^{2}[\sin^{2}%
\theta_{B}]+(1+\cos\theta_{B})^{2}[\sin^{2}\theta_{A}]+\nonumber\\
&  2(1+\cos\theta_{B})\sin^{2}\theta_{A}\sin\theta_{B}\cos\phi_{B}%
+2(1+\cos\theta_{A})\sin\theta_{A}\sin^{2}\theta_{B}\cos\phi_{A}%
\end{align}

\begin{align}
&  =(1+\cos\theta_{A})^{2}(1+\cos\theta_{B})^{2}+2(1+\cos\theta_{A}%
)(1+\cos\theta_{B})\sin\theta_{A}\sin\theta_{B}\cos(\phi_{A}+\phi
_{B})+\nonumber\\
&  2(1+\cos\theta_{A})(1+\cos\theta_{B})^{2}\sin\theta_{A}\cos\phi
_{A}+\nonumber\\
&  2(1+\cos\theta_{A})(1+\cos\theta_{B})\sin\theta_{A}\sin\theta_{B}\cos
(\phi_{A}-\phi_{B})+\nonumber\\
&  2(1+\cos\theta_{A})^{2}(1+\cos\theta_{B})\sin\theta_{B}\cos\phi_{B}%
+(1+\cos\theta_{A})^{2}\sin^{2}\theta_{B}+(1+\cos\theta_{B})^{2}\sin^{2}%
\theta_{A}+\nonumber\\
&  2(1+\cos\theta_{B})\sin^{2}\theta_{A}\sin\theta_{B}\cos\phi_{B}%
+2(1+\cos\theta_{A})\sin\theta_{A}\sin^{2}\theta_{B}\cos\phi_{A}+\sin
^{2}\theta_{A}\sin^{2}\theta_{B}\nonumber\\
&  =(1+\cos\theta_{A})^{2}(1+\cos\theta_{B})^{2}+2(1+\cos\theta_{A}%
)(1+\cos\theta_{B})\sin\theta_{A}\sin\theta_{B}\ast\nonumber\\
&  \{\cos(\phi_{A}+\phi_{B})+\cos(\phi_{A}-\phi_{B})\}+2(1+\cos\theta
_{A})(1+\cos\theta_{B})^{2}\sin\theta_{A}\cos\phi_{A}+\nonumber\\
&  2(1+\cos\theta_{A})^{2}(1+\cos\theta_{B})\sin\theta_{B}\cos\phi_{B}%
+(1+\cos\theta_{A})^{2}\sin^{2}\theta_{B}+(1+\cos\theta_{B})^{2}\sin^{2}%
\theta_{A}+\nonumber\\
&  2(1+\cos\theta_{B})\sin^{2}\theta_{A}\sin\theta_{B}\cos\phi_{B}+\nonumber\\
&  +2(1+\cos\theta_{A})\sin\theta_{A}\sin^{2}\theta_{B}\cos\phi_{A}+\sin
^{2}\theta_{A}\sin^{2}\theta_{B}%
\end{align}

which is%

\begin{align}
&  =(1+\cos\theta_{A})^{2}(1+\cos\theta_{B})^{2}+\sin^{2}\theta_{A}\sin
^{2}\theta_{B}+\nonumber\\
&  4(1+\cos\theta_{A})(1+\cos\theta_{B})\sin\theta_{A}\sin\theta_{B}\cos
\phi_{A}\cos\phi_{B}+\nonumber\\
&  2(1+\cos\theta_{A})(1+\cos\theta_{B})^{2}\sin\theta_{A}\cos\phi
_{A}+\nonumber\\
&  2(1+\cos\theta_{A})^{2}(1+\cos\theta_{B})\sin\theta_{B}\cos\phi
_{B}+\nonumber\\
&  (1+\cos\theta_{A})^{2}\sin^{2}\theta_{B}+2(1+\cos\theta_{A})\sin\theta
_{A}\sin^{2}\theta_{B}\cos\phi_{A}+\nonumber\\
&  (1+\cos\theta_{B})^{2}\sin^{2}\theta_{A}+2(1+\cos\theta_{B})\sin^{2}%
\theta_{A}\sin\theta_{B}\cos\phi_{B}\nonumber\\
&  =(1+\cos\theta_{A})^{2}(1+\cos\theta_{B})^{2}+4(1+\cos\theta_{A}%
)(1+\cos\theta_{B})\sin\theta_{A}\sin\theta_{B}\cos\phi_{A}\cos\phi
_{B}+\nonumber\\
&  2(1+\cos\theta_{A})(1+\cos\theta_{B})^{2}\sin\theta_{A}\cos\phi
_{A}+2(1+\cos\theta_{A})^{2}(1+\cos\theta_{B})\sin\theta_{B}\cos\phi
_{B}+\nonumber\\
&  (1+\cos\theta_{A})(1-\cos^{2}\theta_{B})[(1+\cos\theta_{A})+2\sin\theta
_{A}\cos\phi_{A}]+\nonumber\\
&  (1+\cos\theta_{B})(1-\cos^{2}\theta_{A})[(1+\cos\theta_{B})+2\sin\theta
_{B}\cos\phi_{B}]+(1-\cos^{2}\theta_{A})(1-\cos^{2}\theta_{B})\nonumber\\
&  =(1+\cos\theta_{A})^{2}(1+\cos\theta_{B})^{2}+4(1+\cos\theta_{A}%
)(1+\cos\theta_{B})\sin\theta_{A}\sin\theta_{B}\cos\phi_{A}\cos\phi
_{B}+\nonumber\\
&  2(1+\cos\theta_{A})(1+\cos\theta_{B})^{2}\sin\theta_{A}\cos\phi
_{A}+2(1+\cos\theta_{A})^{2}(1+\cos\theta_{B})\sin\theta_{B}\cos\phi
_{B}+\nonumber\\
&  (1+\cos\theta_{A})(1+\cos\theta_{B})(1-\cos\theta_{B})[(1+\cos\theta
_{A})+2\sin\theta_{A}\cos\phi_{A}]+\nonumber\\
&  (1+\cos\theta_{B})(1+\cos\theta_{A})(1-\cos\theta_{A})[(1+\cos\theta
_{B})+2\sin\theta_{B}\cos\phi_{B}]+(1-\cos^{2}\theta_{A})(1-\cos^{2}\theta
_{B})\nonumber\\
&
\end{align}

\begin{align}
&  =(1+\cos\theta_{A})(1+\cos\theta_{B})\{(1+\cos\theta_{A})(1+\cos\theta
_{B})+4\sin\theta_{A}\sin\theta_{B}\cos\phi_{A}\cos\phi_{B}\nonumber\\
&  +2(1+\cos\theta_{B})\sin\theta_{A}\cos\phi_{A}+2(1+\cos\theta_{A}%
)\sin\theta_{B}\cos\phi_{B}+\nonumber\\
&  (1-\cos\theta_{B})[(1+\cos\theta_{A})+2\sin\theta_{A}\cos\phi
_{A}]\nonumber\\
&  +(1-\cos\theta_{A})[(1+\cos\theta_{B})+2\sin\theta_{B}\cos\phi_{B}%
]+(1-\cos\theta_{A})(1-\cos\theta_{B})\}\nonumber\\
&  =(1+\cos\theta_{A})(1+\cos\theta_{B})\ast\{(1+\cos\theta_{A})(1+\cos
\theta_{B})+\nonumber\\
&  4\sin\theta_{A}\sin\theta_{B}\cos\phi_{A}\cos\phi_{B}+\nonumber\\
&  2(1+\cos\theta_{B})\sin\theta_{A}\cos\phi_{A}+2(1+\cos\theta_{A})\sin
\theta_{B}\cos\phi_{B}+\nonumber\\
&  (1-\cos\theta_{B})[(1+\cos\theta_{A})+2\sin\theta_{A}\cos\phi_{A}%
]+(1-\cos\theta_{A})[(1+\cos\theta_{B})+\nonumber\\
&  2\sin\theta_{B}\cos\phi_{B}+(1-\cos\theta_{B})]\}\nonumber\\
&  =(1+\cos\theta_{A})(1+\cos\theta_{B})\{(1+\cos\theta_{A})(1+\cos\theta
_{B})+4\sin\theta_{A}\sin\theta_{B}\cos\phi_{A}\cos\phi_{B}+\nonumber\\
&  2(1+\cos\theta_{B})\sin\theta_{A}\cos\phi_{A}+2(1+\cos\theta_{A})\sin
\theta_{B}\cos\phi_{B}+\nonumber\\
&  \lbrack(1-\cos\theta_{B})(1+\cos\theta_{A})+\nonumber\\
&  2(1-\cos\theta_{B})\sin\theta_{A}\cos\phi_{A}]+2(1-\cos\theta_{A}%
)[1+\sin\theta_{B}\cos\phi_{B}]\}
\end{align}

\begin{align}
&  =(1+\cos\theta_{A})(1+\cos\theta_{B})\{(1+\cos\theta_{B}+\cos\theta
_{A}+\cos\theta_{A}\cos\theta_{B})+\nonumber\\
&  4\sin\theta_{A}\sin\theta_{B}\cos\phi_{A}\cos\phi_{B}+2(1+\cos\theta
_{B})\sin\theta_{A}\cos\phi_{A}+2(1+\cos\theta_{A})\sin\theta_{B}\cos\phi
_{B}+\nonumber\\
&  [1+\cos\theta_{A}-\cos\theta_{B}-\cos\theta_{A}\cos\theta_{B}%
+2(1-\cos\theta_{B})\sin\theta_{A}\cos\phi_{A}]+\nonumber\\
&  2(1-\cos\theta_{A})[1+\sin\theta_{B}\cos\phi_{B}]\}\nonumber\\
&  =(1+\cos\theta_{A})(1+\cos\theta_{B})\{4\sin\theta_{A}\sin\theta_{B}%
\cos\phi_{A}\cos\phi_{B}+2(1+\cos\theta_{B})\sin\theta_{A}\cos\phi
_{A}+\nonumber\\
&  2(1+\cos\theta_{A})\sin\theta_{B}\cos\phi_{B}+[1+\cos\theta_{A}-\cos
\theta_{B}-\cos\theta_{A}\cos\theta_{B}+\nonumber\\
&  2(1-\cos\theta_{B})\sin\theta_{A}\cos\phi_{A}+(1+\cos\theta_{B}+\cos
\theta_{A}+\cos\theta_{A}\cos\theta_{B})]+\nonumber\\
&  2(1-\cos\theta_{A})[1+\sin\theta_{B}\cos\phi_{B}]\}\nonumber\\
&  =(1+\cos\theta_{A})(1+\cos\theta_{B})\{4\sin\theta_{A}\sin\theta_{B}%
\cos\phi_{A}\cos\phi_{B}+2(1+\cos\theta_{B})\sin\theta_{A}\cos\phi
_{A}+\nonumber\\
&  2(1+\cos\theta_{A})\sin\theta_{B}\cos\phi_{B}+[2+2\cos\theta_{A}%
+2(1-\cos\theta_{B})\sin\theta_{A}\cos\phi_{A}]+\nonumber\\
&  2(1-\cos\theta_{A})[1+\sin\theta_{B}\cos\phi_{B}]\}\nonumber\\
&  =(1+\cos\theta_{A})(1+\cos\theta_{B})\{4\sin\theta_{A}\sin\theta_{B}%
\cos\phi_{A}\cos\phi_{B}+2(1+\cos\theta_{B})\sin\theta_{A}\cos\phi
_{A}+\nonumber\\
&  2(1+\cos\theta_{A})\sin\theta_{B}\cos\phi_{B}+2[1+\cos\theta_{A}%
+(1-\cos\theta_{B})\sin\theta_{A}\cos\phi_{A}]+\nonumber\\
&  2(1-\cos\theta_{A})[1+\sin\theta_{B}\cos\phi_{B}]\}
\end{align}

which is%

\begin{align}
&  =(1+\cos\theta_{A})(1+\cos\theta_{B})\{4\sin\theta_{A}\sin\theta_{B}%
\cos\phi_{A}\cos\phi_{B}+2(1+\cos\theta_{B})\sin\theta_{A}\cos\phi
_{A}+\nonumber\\
&  2(1+\cos\theta_{A})\sin\theta_{B}\cos\phi_{B}+2(1+\cos\theta_{A}%
)+2(1-\cos\theta_{B})\sin\theta_{A}\cos\phi_{A}+\nonumber\\
&  2(1-\cos\theta_{A})[1+\sin\theta_{B}\cos\phi_{B}]\}\nonumber\\
&  =(1+\cos\theta_{A})(1+\cos\theta_{B})\{4\sin\theta_{A}\sin\theta_{B}%
\cos\phi_{A}\cos\phi_{B}+2(1+\cos\theta_{B})\sin\theta_{A}\cos\phi
_{A}+\nonumber\\
&  2(1+\cos\theta_{A})(1+\sin\theta_{B}\cos\phi_{B})+2(1-\cos\theta_{B}%
)\sin\theta_{A}\cos\phi_{A}+\nonumber\\
&  2(1-\cos\theta_{A})[1+\sin\theta_{B}\cos\phi_{B}]\}\nonumber\\
&  =(1+\cos\theta_{A})(1+\cos\theta_{B})\{4\sin\theta_{A}\sin\theta_{B}%
\cos\phi_{A}\cos\phi_{B}+2(1+\cos\theta_{B})\sin\theta_{A}\cos\phi
_{A}+\nonumber\\
&  2(1+\cos\theta_{A})(1+\sin\theta_{B}\cos\phi_{B})+2(1-\cos\theta
_{A})(1+\sin\theta_{B}\cos\phi_{B})+\nonumber\\
&  2(1-\cos\theta_{B})\sin\theta_{A}\cos\phi_{A}\}\nonumber\\
&  =(1+\cos\theta_{A})(1+\cos\theta_{B})\{4\sin\theta_{A}\sin\theta_{B}%
\cos\phi_{A}\cos\phi_{B}+2(1+\cos\theta_{B})\sin\theta_{A}\cos\phi
_{A}+\nonumber\\
&  2(1+\sin\theta_{B}\cos\phi_{B})[(1+\cos\theta_{A})+(1-\cos\theta
_{A})]+2(1-\cos\theta_{B})\sin\theta_{A}\cos\phi_{A}\}
\end{align}

\begin{align}
&  =(1+\cos\theta_{A})(1+\cos\theta_{B})\{4\sin\theta_{A}\sin\theta_{B}%
\cos\phi_{A}\cos\phi_{B}+2(1+\cos\theta_{B})\sin\theta_{A}\cos\phi
_{A}+\nonumber\\
&  2(1-\cos\theta_{B})\sin\theta_{A}\cos\phi_{A}+4(1+\sin\theta_{B}\cos
\phi_{B})\}\nonumber\\
&  =(1+\cos\theta_{A})(1+\cos\theta_{B})\{4\sin\theta_{A}\sin\theta_{B}%
\cos\phi_{A}\cos\phi_{B}+2\sin\theta_{A}\cos\phi_{A}[(1+\cos\theta
_{B})+\nonumber\\
&  (1-\cos\theta_{B})]+4(1+\sin\theta_{B}\cos\phi_{B})\}\nonumber\\
&  =4(1+\cos\theta_{A})(1+\cos\theta_{B})\{\sin\theta_{A}\sin\theta_{B}%
\cos\phi_{A}\cos\phi_{B}+\sin\theta_{A}\cos\phi_{A}+(1+\sin\theta_{B}\cos
\phi_{B})\}\nonumber\\
&  =4(1+\cos\theta_{A})(1+\cos\theta_{B})\{\sin\theta_{A}\sin\theta_{B}%
\cos\phi_{A}\cos\phi_{B}+\sin\theta_{A}\cos\phi_{A}+(1+\sin\theta_{B}\cos
\phi_{B})\}\nonumber\\
&  =4(1+\cos\theta_{A})(1+\cos\theta_{B})\{\sin\theta_{A}\sin\theta_{B}%
\cos\phi_{A}\cos\phi_{B}+\sin\theta_{A}\cos\phi_{A}+1+\sin\theta_{B}\cos
\phi_{B}\}\nonumber\\
&  =4(1+\cos\theta_{A})(1+\cos\theta_{B})\{(1+\sin\theta_{B}\cos\phi_{B}%
)+\sin\theta_{A}\cos\phi_{A}(1+\sin\theta_{B}\cos\phi_{B})\}\nonumber\\
&  =4(1+\cos\theta_{A})(1+\cos\theta_{B})(1+\sin\theta_{B}\cos\phi_{B}%
)(1+\sin\theta_{A}\cos\phi_{A})
\end{align}
Eq.~(\ref{A_Eq_1}) is therefore reduced to%

\begin{align}
&  =\frac{(\alpha,\alpha)}{16(1+\cos\theta_{A})(1+\cos\theta_{B})}%
[4(1+\cos\theta_{A})(1+\cos\theta_{B})(1+\sin\theta_{A}\cos\phi_{A}%
)(1+\sin\theta_{B}\cos\phi_{B})]\nonumber\\
&  =\frac{(\alpha,\alpha)}{4}(1+\sin\theta_{A}\cos\phi_{A})(1+\sin\theta
_{B}\cos\phi_{B}).
\end{align}

\section{Appendix B}

The second term in the payoff relations (\ref{Payoffs_gen}) when the game is
played with the state $\left\vert \psi_{\mathrm{ini}}\right\rangle =\frac
{1}{2}\left(  \left\vert 00\right\rangle +\left\vert 01\right\rangle
+\left\vert 10\right\rangle +\left\vert 11\right\rangle \right)  $ is given as%

\begin{gather}
\frac{(\beta,\gamma)}{16(1+\cos\theta_{A})(1-\cos\theta_{B})}{\Large \{}%
[(1+\cos\theta_{A})(1-\cos\theta_{B})-(1+\cos\theta_{A})\sin\theta_{B}\cos
\phi_{B}+\nonumber\\
(1-\cos\theta_{B})\sin\theta_{A}\cos\phi_{A}-\sin\theta_{A}\sin\theta_{B}%
\cos(\phi_{A}+\phi_{B})]^{2}+[(1+\cos\theta_{A})\sin\theta_{B}\sin\phi
_{B}-\nonumber\\
(1-\cos\theta_{B})\sin\theta_{A}\sin\phi_{A}+\sin\theta_{A}\sin\theta_{B}%
\sin(\phi_{A}+\phi_{B})]^{2}{\large \}}.\label{2nd_part}%
\end{gather}
Consider its part%

\begin{align}
&  =[(1+\cos\theta_{A})(1-\cos\theta_{B})-(1+\cos\theta_{A})\sin\theta_{B}%
\cos\phi_{B}+(1-\cos\theta_{B})\sin\theta_{A}\cos\phi_{A}-\nonumber\\
&  \sin\theta_{A}\sin\theta_{B}\cos(\phi_{A}+\phi_{B})]\ast\lbrack
(1+\cos\theta_{A})(1-\cos\theta_{B})-(1+\cos\theta_{A})\sin\theta_{B}\cos
\phi_{B}+\nonumber\\
&  (1-\cos\theta_{B})\sin\theta_{A}\cos\phi_{A}-\sin\theta_{A}\sin\theta
_{B}\cos(\phi_{A}+\phi_{B})]\nonumber\\
&  =(1+\cos\theta_{A})(1-\cos\theta_{B}){\LARGE [}(1+\cos\theta_{A}%
)(1-\cos\theta_{B})-(1+\cos\theta_{A})\sin\theta_{B}\cos\phi_{B}+\nonumber\\
&  (1-\cos\theta_{B})\sin\theta_{A}\cos\phi_{A}-\sin\theta_{A}\sin\theta
_{B}\cos(\phi_{A}+\phi_{B}){\LARGE ]}-(1+\cos\theta_{A})\ast\nonumber\\
&  \sin\theta_{B}\cos\phi_{B}{\LARGE [}(1+\cos\theta_{A})(1-\cos\theta
_{B})-(1+\cos\theta_{A})\sin\theta_{B}\cos\phi_{B}+\nonumber\\
&  (1-\cos\theta_{B})\sin\theta_{A}\cos\phi_{A}-\sin\theta_{A}\sin\theta
_{B}\cos(\phi_{A}+\phi_{B}){\LARGE ]}+\nonumber\\
&  (1-\cos\theta_{B})\sin\theta_{A}\cos\phi_{A}{\LARGE [}(1+\cos\theta
_{A})(1-\cos\theta_{B})-(1+\cos\theta_{A})\sin\theta_{B}\cos\phi
_{B}+\nonumber\\
&  (1-\cos\theta_{B})\sin\theta_{A}\cos\phi_{A}-\sin\theta_{A}\sin\theta
_{B}\cos(\phi_{A}+\phi_{B}){\LARGE ]}-\nonumber\\
&  \sin\theta_{A}\sin\theta_{B}\cos(\phi_{A}+\phi_{B}){\LARGE [}(1+\cos
\theta_{A})(1-\cos\theta_{B})-(1+\cos\theta_{A})\sin\theta_{B}\cos\phi
_{B}+\nonumber\\
&  (1-\cos\theta_{B})\sin\theta_{A}\cos\phi_{A}-\sin\theta_{A}\sin\theta
_{B}\cos(\phi_{A}+\phi_{B}){\LARGE ]}%
\end{align}

\begin{align}
&  =(1+\cos\theta_{A}){\LARGE [}(1+\cos\theta_{A})(1-\cos\theta_{B}%
)(1-\cos\theta_{B})-\nonumber\\
&  (1+\cos\theta_{A})(1-\cos\theta_{B})\sin\theta_{B}\cos\phi_{B}+\nonumber\\
&  (1-\cos\theta_{B})(1-\cos\theta_{B})\sin\theta_{A}\cos\phi_{A}%
-(1-\cos\theta_{B})\sin\theta_{A}\sin\theta_{B}\cos(\phi_{A}+\phi
_{B}){\LARGE ]}-\nonumber\\
&  (1+\cos\theta_{A}){\LARGE [}(1+\cos\theta_{A})(1-\cos\theta_{B})\sin
\theta_{B}\cos\phi_{B}-\nonumber\\
&  (1+\cos\theta_{A})\sin\theta_{B}\cos\phi_{B}\sin\theta_{B}\cos\phi
_{B}+\nonumber\\
&  (1-\cos\theta_{B})\sin\theta_{A}\cos\phi_{A}\sin\theta_{B}\cos\phi_{B}%
-\cos(\phi_{A}+\phi_{B})\sin\theta_{A}\sin\theta_{B}\sin\theta_{B}\cos\phi
_{B}{\LARGE ]}+\nonumber\\
&  {\LARGE [}(1+\cos\theta_{A})(1-\cos\theta_{B})(1-\cos\theta_{B})\sin
\theta_{A}\cos\phi_{A}-\nonumber\\
&  (1+\cos\theta_{A})\sin\theta_{B}\cos\phi_{B}(1-\cos\theta_{B})\sin
\theta_{A}\cos\phi_{A}+\nonumber\\
&  (1-\cos\theta_{B})\sin\theta_{A}\cos\phi_{A}(1-\cos\theta_{B})\sin
\theta_{A}\cos\phi_{A}-\nonumber\\
&  \sin\theta_{A}\sin\theta_{B}\cos(\phi_{A}+\phi_{B})(1-\cos\theta_{B}%
)\sin\theta_{A}\cos\phi_{A}{\LARGE ]}-\nonumber\\
&  \sin\theta_{A}\sin\theta_{B}\cos(\phi_{A}+\phi_{B}){\LARGE [}(1+\cos
\theta_{A})(1-\cos\theta_{B})-\nonumber\\
&  (1+\cos\theta_{A})\sin\theta_{B}\cos\phi_{B}+(1-\cos\theta_{B})\sin
\theta_{A}\cos\phi_{A}-\sin\theta_{A}\sin\theta_{B}\cos(\phi_{A}+\phi
_{B}){\LARGE ]}\nonumber\\
&
\end{align}

\begin{align}
&  =(1+\cos\theta_{A}){\LARGE [}(1+\cos\theta_{A})(1-\cos\theta_{B}%
)^{2}-(1+\cos\theta_{A})(1-\cos\theta_{B})\sin\theta_{B}\cos\phi
_{B}+\nonumber\\
&  (1-\cos\theta_{B})^{2}\sin\theta_{A}\cos\phi_{A}-(1-\cos\theta_{B}%
)\sin\theta_{A}\sin\theta_{B}\cos(\phi_{A}+\phi_{B}){\LARGE ]}-\nonumber\\
&  (1+\cos\theta_{A}){\LARGE [}(1+\cos\theta_{A})(1-\cos\theta_{B})\sin
\theta_{B}\cos\phi_{B}-(1+\cos\theta_{A})\sin^{2}\theta_{B}\cos^{2}\phi
_{B}+\nonumber\\
&  (1-\cos\theta_{B})\sin\theta_{A}\sin\theta_{B}\cos\phi_{A}\cos\phi_{B}%
-\cos(\phi_{A}+\phi_{B})\sin\theta_{A}\sin^{2}\theta_{B}\cos\phi_{B}%
{\LARGE ]}+\nonumber\\
&  {\LARGE [}(1+\cos\theta_{A})(1-\cos\theta_{B})^{2}\sin\theta_{A}\cos
\phi_{A}-(1+\cos\theta_{A})(1-\cos\theta_{B})\sin\theta_{A}\sin\theta_{B}%
\cos\phi_{A}\cos\phi_{B}+\nonumber\\
&  (1-\cos\theta_{B})^{2}\sin^{2}\theta_{A}\cos^{2}\phi_{A}-(1-\cos\theta
_{B})\sin^{2}\theta_{A}\sin\theta_{B}\cos\phi_{A}\cos(\phi_{A}+\phi
_{B}){\LARGE ]}-\nonumber\\
&  \sin\theta_{A}\sin\theta_{B}\cos(\phi_{A}+\phi_{B}){\LARGE [}(1+\cos
\theta_{A})(1-\cos\theta_{B})-(1+\cos\theta_{A})\sin\theta_{B}\cos\phi
_{B}+\nonumber\\
&  (1-\cos\theta_{B})\sin\theta_{A}\cos\phi_{A}-\sin\theta_{A}\sin\theta
_{B}\cos(\phi_{A}+\phi_{B}){\LARGE ]}%
\end{align}

\begin{align}
&  =(1+\cos\theta_{A}){\LARGE [}(1+\cos\theta_{A})(1-\cos\theta_{B}%
)^{2}-(1+\cos\theta_{A})(1-\cos\theta_{B})\sin\theta_{B}\cos\phi
_{B}+\nonumber\\
&  (1-\cos\theta_{B})^{2}\sin\theta_{A}\cos\phi_{A}-(1-\cos\theta_{B}%
)\sin\theta_{A}\sin\theta_{B}\cos(\phi_{A}+\phi_{B}){\LARGE ]}+\nonumber\\
&  (1+\cos\theta_{A}){\LARGE [}-(1+\cos\theta_{A})(1-\cos\theta_{B})\sin
\theta_{B}\cos\phi_{B}+(1+\cos\theta_{A})\sin^{2}\theta_{B}\cos^{2}\phi
_{B}-\nonumber\\
&  (1-\cos\theta_{B})\sin\theta_{A}\sin\theta_{B}\cos\phi_{A}\cos\phi_{B}%
+\cos(\phi_{A}+\phi_{B})\sin\theta_{A}\sin^{2}\theta_{B}\cos\phi_{B}%
{\LARGE ]}+\nonumber\\
&  {\LARGE [}(1+\cos\theta_{A})(1-\cos\theta_{B})^{2}\sin\theta_{A}\cos
\phi_{A}-(1+\cos\theta_{A})(1-\cos\theta_{B})\sin\theta_{A}\sin\theta_{B}%
\cos\phi_{A}\cos\phi_{B}+\nonumber\\
&  (1-\cos\theta_{B})^{2}\sin^{2}\theta_{A}\cos^{2}\phi_{A}-(1-\cos\theta
_{B})\sin^{2}\theta_{A}\sin\theta_{B}\cos\phi_{A}\cos(\phi_{A}+\phi
_{B})-\nonumber\\
&  (1+\cos\theta_{A})(1-\cos\theta_{B})\sin\theta_{A}\sin\theta_{B}\cos
(\phi_{A}+\phi_{B})+(1+\cos\theta_{A})\sin^{2}\theta_{B}\sin\theta_{A}\cos
\phi_{B}\cos(\phi_{A}+\phi_{B})-\nonumber\\
&  (1-\cos\theta_{B})\sin^{2}\theta_{A}\sin\theta_{B}\cos\phi_{A}\cos(\phi
_{A}+\phi_{B})+\sin^{2}\theta_{A}\sin^{2}\theta_{B}\cos^{2}(\phi_{A}+\phi
_{B}){\LARGE ]}%
\end{align}

\begin{align}
&  =(1+\cos\theta_{A}){\LARGE [}(1+\cos\theta_{A})(1-\cos\theta_{B}%
)^{2}-(1+\cos\theta_{A})(1-\cos\theta_{B})\sin\theta_{B}\cos\phi
_{B}+\nonumber\\
&  (1-\cos\theta_{B})^{2}\sin\theta_{A}\cos\phi_{A}-(1-\cos\theta_{B}%
)\sin\theta_{A}\sin\theta_{B}\cos(\phi_{A}+\phi_{B}){\LARGE ]}\nonumber\\
&  +(1+\cos\theta_{A}){\LARGE [}-(1+\cos\theta_{A})(1-\cos\theta_{B}%
)\sin\theta_{B}\cos\phi_{B}+(1+\cos\theta_{A})\sin^{2}\theta_{B}\cos^{2}%
\phi_{B}-\nonumber\\
&  (1-\cos\theta_{B})\sin\theta_{A}\sin\theta_{B}\cos\phi_{A}\cos\phi_{B}%
+\cos(\phi_{A}+\phi_{B})\sin\theta_{A}\sin^{2}\theta_{B}\cos\phi_{B}%
{\LARGE ]}\nonumber\\
&  +{\LARGE [}(1+\cos\theta_{A})(1-\cos\theta_{B})^{2}\sin\theta_{A}\cos
\phi_{A}-(1+\cos\theta_{A})(1-\cos\theta_{B})\sin\theta_{A}\sin\theta_{B}%
\cos\phi_{A}\cos\phi_{B}+\nonumber\\
&  (1-\cos\theta_{B})^{2}\sin^{2}\theta_{A}\cos^{2}\phi_{A}-2(1-\cos\theta
_{B})\sin^{2}\theta_{A}\sin\theta_{B}\cos\phi_{A}\cos(\phi_{A}+\phi
_{B})\nonumber\\
&  -(1+\cos\theta_{A})(1-\cos\theta_{B})\sin\theta_{A}\sin\theta_{B}\cos
(\phi_{A}+\phi_{B})+\nonumber\\
&  (1+\cos\theta_{A})\sin^{2}\theta_{B}\sin\theta_{A}\cos\phi_{B}\cos(\phi
_{A}+\phi_{B})+\sin^{2}\theta_{A}\sin^{2}\theta_{B}\cos^{2}(\phi_{A}+\phi
_{B}){\LARGE ]}%
\end{align}

\begin{align}
&  =(1+\cos\theta_{A}){\LARGE [}(1+\cos\theta_{A})(1-\cos\theta_{B}%
)^{2}-2(1+\cos\theta_{A})(1-\cos\theta_{B})\sin\theta_{B}\cos\phi
_{B}+\nonumber\\
&  (1-\cos\theta_{B})^{2}\sin\theta_{A}\cos\phi_{A}-(1-\cos\theta_{B}%
)\sin\theta_{A}\sin\theta_{B}\cos(\phi_{A}+\phi_{B})+\nonumber\\
&  (1+\cos\theta_{A})\sin^{2}\theta_{B}\cos^{2}\phi_{B}-(1-\cos\theta_{B}%
)\sin\theta_{A}\sin\theta_{B}\cos\phi_{A}\cos\phi_{B}+\nonumber\\
&  \cos(\phi_{A}+\phi_{B})\sin\theta_{A}\sin^{2}\theta_{B}\cos\phi_{B}%
+(1-\cos\theta_{B})^{2}\sin\theta_{A}\cos\phi_{A}-\nonumber\\
&  (1-\cos\theta_{B})\sin\theta_{A}\sin\theta_{B}\cos\phi_{A}\cos\phi
_{B}-(1-\cos\theta_{B})\sin\theta_{A}\sin\theta_{B}\cos(\phi_{A}+\phi
_{B})+\nonumber\\
&  \sin^{2}\theta_{B}\sin\theta_{A}\cos\phi_{B}\cos(\phi_{A}+\phi
_{B}){\LARGE ]}+{\LARGE [}+(1-\cos\theta_{B})^{2}\sin^{2}\theta_{A}\cos
^{2}\phi_{A}-\nonumber\\
&  2(1-\cos\theta_{B})\sin^{2}\theta_{A}\sin\theta_{B}\cos\phi_{A}\cos
(\phi_{A}+\phi_{B})+\sin^{2}\theta_{A}\sin^{2}\theta_{B}\cos^{2}(\phi_{A}%
+\phi_{B}){\LARGE ]}%
\end{align}

\begin{align}
&  =(1+\cos\theta_{A}){\LARGE [}(1+\cos\theta_{A})(1-\cos\theta_{B}%
)^{2}-2(1+\cos\theta_{A})(1-\cos\theta_{B})\sin\theta_{B}\cos\phi
_{B}+\nonumber\\
&  2(1-\cos\theta_{B})^{2}\sin\theta_{A}\cos\phi_{A}-2(1-\cos\theta_{B}%
)\sin\theta_{A}\sin\theta_{B}\cos(\phi_{A}+\phi_{B})+\nonumber\\
&  (1+\cos\theta_{A})\sin^{2}\theta_{B}\cos^{2}\phi_{B}-2(1-\cos\theta
_{B})\sin\theta_{A}\sin\theta_{B}\cos\phi_{A}\cos\phi_{B}+\nonumber\\
&  2\sin\theta_{A}\sin^{2}\theta_{B}\cos\phi_{B}\cos(\phi_{A}+\phi
_{B}){\LARGE ]}+\nonumber\\
&  {\LARGE [}+(1-\cos\theta_{B})^{2}\sin^{2}\theta_{A}\cos^{2}\phi
_{A}-2(1-\cos\theta_{B})\sin^{2}\theta_{A}\sin\theta_{B}\cos\phi_{A}\cos
(\phi_{A}+\phi_{B})+\nonumber\\
&  \sin^{2}\theta_{A}\sin^{2}\theta_{B}\cos^{2}(\phi_{A}+\phi_{B}){\LARGE ]}%
\end{align}

\begin{align}
&  =(1+\cos\theta_{A}){\LARGE [}(1+\cos\theta_{A})(1-\cos\theta_{B}%
)^{2}-2(1+\cos\theta_{A})(1-\cos\theta_{B})\sin\theta_{B}\cos\phi
_{B}+\nonumber\\
&  2(1-\cos\theta_{B})^{2}\sin\theta_{A}\cos\phi_{A}-2(1-\cos\theta_{B}%
)\sin\theta_{A}\sin\theta_{B}\cos(\phi_{A}+\phi_{B})+\nonumber\\
&  (1+\cos\theta_{A})\sin^{2}\theta_{B}\cos^{2}\phi_{B}-2(1-\cos\theta
_{B})\sin\theta_{A}\sin\theta_{B}\cos\phi_{A}\cos\phi_{B}+\nonumber\\
&  2\sin\theta_{A}\sin^{2}\theta_{B}\cos\phi_{B}\cos(\phi_{A}+\phi
_{B}){\LARGE ]}+\nonumber\\
&  {\LARGE [}(1-\cos\theta_{B})^{2}(1-\cos^{2}\theta_{A})\cos^{2}\phi
_{A}-2(1-\cos\theta_{B})(1-\cos^{2}\theta_{A})\sin\theta_{B}\cos\phi_{A}%
\cos(\phi_{A}+\phi_{B})+\nonumber\\
&  (1-\cos^{2}\theta_{A})\sin^{2}\theta_{B}\cos^{2}(\phi_{A}+\phi
_{B}){\LARGE ]}%
\end{align}

\begin{align}
&  =(1+\cos\theta_{A}){\LARGE [}(1+\cos\theta_{A})(1-\cos\theta_{B}%
)^{2}-2(1+\cos\theta_{A})(1-\cos\theta_{B})\sin\theta_{B}\cos\phi
_{B}+\nonumber\\
&  2(1-\cos\theta_{B})^{2}\sin\theta_{A}\cos\phi_{A}-2(1-\cos\theta_{B}%
)\sin\theta_{A}\sin\theta_{B}\cos(\phi_{A}+\phi_{B})+\nonumber\\
&  (1+\cos\theta_{A})\sin^{2}\theta_{B}\cos^{2}\phi_{B}-2(1-\cos\theta
_{B})\sin\theta_{A}\sin\theta_{B}\cos\phi_{A}\cos\phi_{B}+\nonumber\\
&  2\sin\theta_{A}\sin^{2}\theta_{B}\cos\phi_{B}\cos(\phi_{A}+\phi
_{B}){\LARGE ]}+(1+\cos\theta_{A}){\LARGE [}(1-\cos\theta_{B})^{2}%
(1-\cos\theta_{A})\cos^{2}\phi_{A}-\nonumber\\
&  2(1-\cos\theta_{B})(1-\cos\theta_{A})\sin\theta_{B}\cos\phi_{A}\cos
(\phi_{A}+\phi_{B})+(1-\cos\theta_{A})\sin^{2}\theta_{B}\cos^{2}(\phi_{A}%
+\phi_{B}){\LARGE ]}\nonumber\\
&
\end{align}

\begin{align}
&  =(1+\cos\theta_{A}){\LARGE [}(1+\cos\theta_{A})(1-\cos\theta_{B}%
)^{2}+(1-\cos\theta_{B})^{2}(1-\cos\theta_{A})\cos^{2}\phi_{A}\nonumber\\
&  -2(\sin\theta_{B}\cos\phi_{B}-\cos\theta_{B}\sin\theta_{B}\cos\phi_{B}%
+\cos\theta_{A}\sin\theta_{B}\cos\phi_{B}-\cos\theta_{A}\cos\theta_{B}%
\sin\theta_{B}\cos\phi_{B})\nonumber\\
&  +2(1-\cos\theta_{B})^{2}\sin\theta_{A}\cos\phi_{A}\nonumber\\
&  -2\sin\theta_{A}\sin\theta_{B}\cos(\phi_{A}+\phi_{B})+2\sin\theta_{A}%
\sin\theta_{B}\cos\theta_{B}\cos(\phi_{A}+\phi_{B})\nonumber\\
&  +\sin^{2}\theta_{B}\cos^{2}\phi_{B}+\cos\theta_{A}\sin^{2}\theta_{B}%
\cos^{2}\phi_{B}-2(1-\cos\theta_{B})\sin\theta_{A}\sin\theta_{B}\cos\phi
_{A}\cos\phi_{B}\nonumber\\
&  +2\sin\theta_{A}\sin^{2}\theta_{B}\cos\phi_{B}\cos(\phi_{A}+\phi
_{B})-2(1-\cos\theta_{A}-\cos\theta_{B}+\nonumber\\
&  \cos\theta_{A}\cos\theta_{B})\sin\theta_{B}\cos\phi_{A}\cos(\phi_{A}%
+\phi_{B})\nonumber\\
&  +\sin^{2}\theta_{B}\cos^{2}(\phi_{A}+\phi_{B})-\sin^{2}\theta_{B}\cos
^{2}(\phi_{A}+\phi_{B})\cos\theta_{A}{\LARGE ]}%
\end{align}

\begin{align}
&  =(1+\cos\theta_{A}){\LARGE [}(1+\cos\theta_{A})(1-\cos\theta_{B}%
)^{2}+(1-\cos\theta_{B})^{2}(1-\cos\theta_{A})\cos^{2}\phi_{A}+\nonumber\\
&  \sin^{2}\theta_{B}\cos^{2}(\phi_{A}+\phi_{B})-\nonumber\\
&  \sin^{2}\theta_{B}\cos^{2}(\phi_{A}+\phi_{B})\cos\theta_{A}+\sin^{2}%
\theta_{B}\cos^{2}\phi_{B}+\cos\theta_{A}\sin^{2}\theta_{B}\cos^{2}\phi
_{B}\nonumber\\
&  -2(\sin\theta_{B}\cos\phi_{B}-\sin\theta_{B}\cos\theta_{B}\cos\phi_{B}%
+\sin\theta_{B}\cos\theta_{A}\cos\phi_{B}-\cos\theta_{A}\cos\theta_{B}%
\sin\theta_{B}\cos\phi_{B})\nonumber\\
&  +2(1-\cos\theta_{B})^{2}\sin\theta_{A}\cos\phi_{A}-2\sin\theta_{A}%
\sin\theta_{B}\cos(\phi_{A}+\phi_{B})+\nonumber\\
&  2\sin\theta_{A}\sin\theta_{B}\cos\theta_{B}\cos(\phi_{A}+\phi
_{B})\nonumber\\
&  -2(\sin\theta_{A}\sin\theta_{B}\cos\phi_{A}\cos\phi_{B}-\sin\theta_{A}%
\cos\theta_{B}\sin\theta_{B}\cos\phi_{A}\cos\phi_{B})\nonumber\\
&  -2{\LARGE \{}\sin\theta_{B}\cos\phi_{A}-\cos\theta_{A}\sin\theta_{B}%
\cos\phi_{A}-\cos\theta_{B}\sin\theta_{B}\cos\phi_{A}+\nonumber\\
&  \cos\theta_{A}\cos\theta_{B}\sin\theta_{B}\cos\phi_{A}-\sin\theta_{A}%
\sin^{2}\theta_{B}\cos\phi_{B}{\LARGE \}}\cos(\phi_{A}+\phi_{B}){\LARGE ]}%
\end{align}

\begin{align}
&  =(1+\cos\theta_{A}){\LARGE [}(1+\cos\theta_{A})(1+\cos^{2}\theta_{B}%
-2\cos\theta_{B})+(1+\cos\theta_{A})\sin^{2}\theta_{B}\cos^{2}\phi
_{B}+\nonumber\\
&  (1-\cos\theta_{A})(1-\cos\theta_{B})^{2}\cos^{2}\phi_{A}+(1-\cos\theta
_{A})\sin^{2}\theta_{B}\cos^{2}(\phi_{A}+\phi_{B})\nonumber\\
&  -2\sin\theta_{B}\cos\phi_{B}(1-\cos\theta_{B})\{(1+\cos\theta_{A}%
)+\sin\theta_{A}\cos\phi_{A}\}+2(1-\cos\theta_{B})^{2}\sin\theta_{A}\cos
\phi_{A}\nonumber\\
&  -2\cos(\phi_{A}+\phi_{B}){\LARGE \{}\sin\theta_{B}\cos\phi_{A}(1-\cos
\theta_{A})-\nonumber\\
&  \sin\theta_{B}\cos\theta_{B}\cos\phi_{A}(1-\cos\theta_{A})-\sin\theta
_{A}\sin^{2}\theta_{B}\cos\phi_{B}+\sin\theta_{A}\sin\theta_{B}(1-\cos
\theta_{B}){\LARGE \}]}%
\end{align}

\begin{align}
&  =(1+\cos\theta_{A}){\LARGE [}(1+\cos\theta_{A})(1-\cos\theta_{B}%
)^{2}+(1+\cos\theta_{A})\sin^{2}\theta_{B}\cos^{2}\phi_{B}+\nonumber\\
&  (1-\cos\theta_{A})(1-\cos\theta_{B})^{2}\cos^{2}\phi_{A}+(1-\cos\theta
_{A})\sin^{2}\theta_{B}\cos^{2}(\phi_{A}+\phi_{B})\nonumber\\
&  -2\sin\theta_{B}\cos\phi_{B}(1-\cos\theta_{B})\{(1+\cos\theta_{A}%
)+\sin\theta_{A}\cos\phi_{A}\}\nonumber\\
&  +2(1-\cos\theta_{B})^{2}\sin\theta_{A}\cos\phi_{A}\nonumber\\
&  -2\cos(\phi_{A}+\phi_{B}){\LARGE \{}\sin\theta_{B}\cos\phi_{A}(1-\cos
\theta_{A})(1-\cos\theta_{B})-\nonumber\\
&  \sin\theta_{A}\sin^{2}\theta_{B}\cos\phi_{B}+\sin\theta_{A}\sin\theta
_{B}(1-\cos\theta_{B}){\LARGE \}]}.\label{part_1}%
\end{align}
In Eq.~(\ref{2nd_part}), now consider the second part%

\begin{align}
&  =[(1+\cos\theta_{A})\sin\theta_{B}\sin\phi_{B}-(1-\cos\theta_{B})\sin
\theta_{A}\sin\phi_{A}+\sin\theta_{A}\sin\theta_{B}\sin(\phi_{A}+\phi
_{B})]^{2}\nonumber\\
&  =(1+\cos\theta_{A})^{2}\sin^{2}\theta_{B}\sin^{2}\phi_{B}+(1-\cos\theta
_{B})^{2}\sin^{2}\theta_{A}\sin^{2}\phi_{A}+\sin^{2}\theta_{A}\sin^{2}%
\theta_{B}\sin^{2}(\phi_{A}+\phi_{B})\nonumber\\
&  -2(1+\cos\theta_{A})\sin\theta_{B}\sin\phi_{B}(1-\cos\theta_{B})\sin
\theta_{A}\sin\phi_{A}\nonumber\\
&  -2(1-\cos\theta_{B})\sin\theta_{A}\sin\phi_{A}\sin\theta_{A}\sin\theta
_{B}\sin(\phi_{A}+\phi_{B})\nonumber\\
&  2\sin\theta_{A}\sin\theta_{B}\sin(\phi_{A}+\phi_{B})(1+\cos\theta_{A}%
)\sin\theta_{B}\sin\phi_{B}%
\end{align}%
\begin{align}
&  =(1+\cos\theta_{A})^{2}\sin^{2}\theta_{B}\sin^{2}\phi_{B}+(1-\cos\theta
_{B})^{2}\sin^{2}\theta_{A}\sin^{2}\phi_{A}+\sin^{2}\theta_{A}\sin^{2}%
\theta_{B}\sin^{2}(\phi_{A}+\phi_{B})\nonumber\\
&  -2(1+\cos\theta_{A})(1-\cos\theta_{B})\sin\theta_{B}\sin\phi_{B}\sin
\theta_{A}\sin\phi_{A}\nonumber\\
&  -2(1-\cos\theta_{B})\sin^{2}\theta_{A}\sin\phi_{A}\sin\theta_{B}\sin
(\phi_{A}+\phi_{B})\nonumber\\
&  2(1+\cos\theta_{A})\sin\theta_{A}\sin^{2}\theta_{B}\sin(\phi_{A}+\phi
_{B})\sin\phi_{B}.
\end{align}
Now add this to (\ref{part_1}) from above%

\begin{align}
&  =(1+\cos\theta_{A}){\LARGE [}(1+\cos\theta_{A})(1-\cos\theta_{B}%
)^{2}+(1+\cos\theta_{A})\sin^{2}\theta_{B}\cos^{2}\phi_{B}+\nonumber\\
&  (1-\cos\theta_{A})(1-\cos\theta_{B})^{2}\cos^{2}\phi_{A}+(1-\cos\theta
_{A})\sin^{2}\theta_{B}\cos^{2}(\phi_{A}+\phi_{B})\nonumber\\
&  -2\sin\theta_{B}\cos\phi_{B}(1-\cos\theta_{B})\{(1+\cos\theta_{A}%
)+\sin\theta_{A}\cos\phi_{A}\}\nonumber\\
&  +2(1-\cos\theta_{B})^{2}\sin\theta_{A}\cos\phi_{A}\nonumber\\
&  -2\cos(\phi_{A}+\phi_{B}){\LARGE \{}\sin\theta_{B}\cos\phi_{A}(1-\cos
\theta_{A})(1-\cos\theta_{B})-\nonumber\\
&  \sin\theta_{A}\sin^{2}\theta_{B}\cos\phi_{B}+\sin\theta_{A}\sin\theta
_{B}(1-\cos\theta_{B}){\LARGE \}]}+\nonumber\\
&  (1+\cos\theta_{A})^{2}\sin^{2}\theta_{B}\sin^{2}\phi_{B}+(1-\cos\theta
_{B})^{2}\sin^{2}\theta_{A}\sin^{2}\phi_{A}+\nonumber\\
&  \sin^{2}\theta_{A}\sin^{2}\theta_{B}\sin^{2}(\phi_{A}+\phi_{B}%
)-2(1+\cos\theta_{A})(1-\cos\theta_{B})\sin\theta_{B}\sin\phi_{B}\sin
\theta_{A}\sin\phi_{A}\nonumber\\
&  -2(1-\cos\theta_{B})\sin^{2}\theta_{A}\sin\phi_{A}\sin\theta_{B}\sin
(\phi_{A}+\phi_{B})\nonumber\\
&  2(1+\cos\theta_{A})\sin\theta_{A}\sin^{2}\theta_{B}\sin(\phi_{A}+\phi
_{B})\sin\phi_{B}%
\end{align}

\begin{align}
&  =(1+\cos\theta_{A}){\LARGE [}(1+\cos\theta_{A})(1-\cos\theta_{B}%
)^{2}+(1+\cos\theta_{A})\sin^{2}\theta_{B}\cos^{2}\phi_{B}+\nonumber\\
&  (1-\cos\theta_{A})(1-\cos\theta_{B})^{2}\cos^{2}\phi_{A}+(1-\cos\theta
_{A})\sin^{2}\theta_{B}\cos^{2}(\phi_{A}+\phi_{B})\nonumber\\
&  -2\sin\theta_{B}\cos\phi_{B}(1-\cos\theta_{B})\{(1+\cos\theta_{A}%
)+\sin\theta_{A}\cos\phi_{A}\}\nonumber\\
&  +2(1-\cos\theta_{B})^{2}\sin\theta_{A}\cos\phi_{A}\nonumber\\
&  -2\cos(\phi_{A}+\phi_{B}){\LARGE \{}\sin\theta_{B}\cos\phi_{A}(1-\cos
\theta_{A})(1-\cos\theta_{B})-\sin\theta_{A}\sin^{2}\theta_{B}\cos\phi
_{B}+\nonumber\\
&  \sin\theta_{A}\sin\theta_{B}(1-\cos\theta_{B}){\LARGE \}}+(1+\cos\theta
_{A})\sin^{2}\theta_{B}\sin^{2}\phi_{B}\nonumber\\
&  -2(1-\cos\theta_{B})\sin\theta_{B}\sin\phi_{B}\sin\theta_{A}\sin\phi
_{A}+2\sin\theta_{A}\sin^{2}\theta_{B}\sin(\phi_{A}+\phi_{B})\sin\phi
_{B}{\LARGE ]}+\nonumber\\
&  (1-\cos\theta_{B})^{2}\sin^{2}\theta_{A}\sin^{2}\phi_{A}+\sin^{2}\theta
_{A}\sin^{2}\theta_{B}\sin^{2}(\phi_{A}+\phi_{B})\nonumber\\
&  -2(1-\cos\theta_{B})\sin^{2}\theta_{A}\sin\phi_{A}\sin\theta_{B}\sin
(\phi_{A}+\phi_{B})
\end{align}

\begin{align}
&  =(1+\cos\theta_{A}){\LARGE [}(1+\cos\theta_{A})(1-\cos\theta_{B}%
)^{2}+(1+\cos\theta_{A})\sin^{2}\theta_{B}+\nonumber\\
&  (1-\cos\theta_{A})(1-\cos\theta_{B})^{2}\cos^{2}\phi_{A}\nonumber\\
&  -2\sin\theta_{B}\cos\phi_{B}(1-\cos\theta_{B})\{(1+\cos\theta_{A}%
)+\sin\theta_{A}\cos\phi_{A}\}\nonumber\\
&  +2(1-\cos\theta_{B})^{2}\sin\theta_{A}\cos\phi_{A}\nonumber\\
&  -2\cos(\phi_{A}+\phi_{B}){\LARGE \{}\sin\theta_{B}\cos\phi_{A}(1-\cos
\theta_{A})(1-\cos\theta_{B})-\sin\theta_{A}\sin^{2}\theta_{B}\cos\phi
_{B}+\nonumber\\
&  \sin\theta_{A}\sin\theta_{B}(1-\cos\theta_{B}){\LARGE \}}-2(1-\cos
\theta_{B})\sin\theta_{A}\sin\theta_{B}\sin\phi_{A}\sin\phi_{B}{\LARGE ]}%
+\nonumber\\
&  +\sin^{2}\theta_{A}\sin^{2}\theta_{B}+(1-\cos\theta_{B})^{2}\sin^{2}%
\theta_{A}\sin^{2}\phi_{A}\nonumber\\
&  +2(1+\cos\theta_{A})\sin\theta_{A}\sin^{2}\theta_{B}\sin\phi_{B}\sin
(\phi_{A}+\phi_{B})-\nonumber\\
&  2(1-\cos\theta_{B})\sin^{2}\theta_{A}\sin\phi_{A}\sin\theta_{B}\sin
(\phi_{A}+\phi_{B})
\end{align}

\begin{align}
&  =(1+\cos\theta_{A}){\LARGE [}(1+\cos\theta_{A})(1-\cos\theta_{B}%
)^{2}+(1+\cos\theta_{A})\sin^{2}\theta_{B}\nonumber\\
&  -2\sin\theta_{B}\cos\phi_{B}(1-\cos\theta_{B})\{(1+\cos\theta_{A}%
)+\sin\theta_{A}\cos\phi_{A}\}\nonumber\\
&  +2(1-\cos\theta_{B})^{2}\sin\theta_{A}\cos\phi_{A}\nonumber\\
&  -2\cos(\phi_{A}+\phi_{B}){\LARGE \{}\sin\theta_{B}\cos\phi_{A}(1-\cos
\theta_{A})(1-\cos\theta_{B})-\nonumber\\
&  \sin\theta_{A}\sin^{2}\theta_{B}\cos\phi_{B}+\sin\theta_{A}\sin\theta
_{B}(1-\cos\theta_{B}){\LARGE \}}\nonumber\\
&  -2(1-\cos\theta_{B})\sin\theta_{A}\sin\theta_{B}\sin\phi_{A}\sin\phi
_{B}{\LARGE ]}+\nonumber\\
&  \sin^{2}\theta_{A}\sin^{2}\theta_{B}+\sin^{2}\theta_{A}(1-\cos\theta
_{B})^{2}\nonumber\\
&  +2\sin\theta_{A}\sin\theta_{B}\sin(\phi_{A}+\phi_{B}){\LARGE \{}%
(1+\cos\theta_{A})\sin\theta_{B}\sin\phi_{B}-(1-\cos\theta_{B})\sin\theta
_{A}\sin\phi_{A}{\LARGE \}}\nonumber\\
&
\end{align}

\begin{align}
&  =(1+\cos\theta_{A}){\LARGE [}(1+\cos\theta_{A})(1-\cos\theta_{B}%
)^{2}+(1+\cos\theta_{A})\sin^{2}\theta_{B}\nonumber\\
&  -2\sin\theta_{B}\cos\phi_{B}(1-\cos\theta_{B})\{(1+\cos\theta_{A}%
)+\sin\theta_{A}\cos\phi_{A}\}\nonumber\\
&  +2(1-\cos\theta_{B})^{2}\sin\theta_{A}\cos\phi_{A}-2\cos(\phi_{A}+\phi
_{B})\sin\theta_{B}\cos\phi_{A}(1-\cos\theta_{A})(1-\cos\theta_{B})\nonumber\\
&  +2\cos(\phi_{A}+\phi_{B})\sin\theta_{A}\sin^{2}\theta_{B}\cos\phi_{B}%
-2\cos(\phi_{A}+\phi_{B})\sin\theta_{A}\sin\theta_{B}(1-\cos\theta
_{B})\nonumber\\
&  -2(1-\cos\theta_{B})\sin\theta_{A}\sin\theta_{B}\sin\phi_{A}\sin\phi
_{B}{\LARGE ]}+2\sin^{2}\theta_{A}(1-\cos\theta_{B})\nonumber\\
&  +2\sin\theta_{A}\sin\theta_{B}\sin(\phi_{A}+\phi_{B}){\LARGE \{}%
(1+\cos\theta_{A})\sin\theta_{B}\sin\phi_{B}-(1-\cos\theta_{B})\sin\theta
_{A}\sin\phi_{A}{\LARGE \}}\nonumber\\
&
\end{align}

\begin{align}
&  =2[(1+\cos\theta_{A})^{2}(1-\cos\theta_{B})-\sin\theta_{B}\cos\phi
_{B}(1-\cos\theta_{B})(1+\cos\theta_{A})^{2}\nonumber\\
&  -\sin\theta_{B}\cos\phi_{B}(1-\cos\theta_{B})(1+\cos\theta_{A})\sin
\theta_{A}\cos\phi_{A}\nonumber\\
&  +(1+\cos\theta_{A})(1-\cos\theta_{B})^{2}\sin\theta_{A}\cos\phi
_{A}-\nonumber\\
&  \cos(\phi_{A}+\phi_{B})\sin\theta_{B}\cos\phi_{A}\sin^{2}\theta_{A}%
(1-\cos\theta_{B})\nonumber\\
&  +(1+\cos\theta_{A})\cos(\phi_{A}+\phi_{B})\sin\theta_{A}\sin^{2}\theta
_{B}\cos\phi_{B}-\nonumber\\
&  (1+\cos\theta_{A})\cos(\phi_{A}+\phi_{B})\sin\theta_{A}\sin\theta
_{B}(1-\cos\theta_{B})\nonumber\\
&  -(1+\cos\theta_{A})(1-\cos\theta_{B})\sin\theta_{A}\sin\theta_{B}\sin
\phi_{A}\sin\phi_{B}+\sin^{2}\theta_{A}(1-\cos\theta_{B})\nonumber\\
&  +\sin\theta_{A}\sin\theta_{B}\sin(\phi_{A}+\phi_{B})(1+\cos\theta_{A}%
)\sin\theta_{B}\sin\phi_{B}-\nonumber\\
&  \sin\theta_{A}\sin\theta_{B}\sin(\phi_{A}+\phi_{B})(1-\cos\theta_{B}%
)\sin\theta_{A}\sin\phi_{A}]
\end{align}

\begin{align}
&  =2[(1+\cos\theta_{A})^{2}(1-\cos\theta_{B})(1-\sin\theta_{B}\cos\phi
_{B})\nonumber\\
&  +(1+\cos\theta_{A})(1-\cos\theta_{B})^{2}\sin\theta_{A}\cos\phi
_{A}\nonumber\\
&  -(1-\cos\theta_{B})\sin^{2}\theta_{A}\sin\theta_{B}\cos(\phi_{A}+\phi
_{B})\cos\phi_{A}-\nonumber\\
&  (1-\cos\theta_{B})\sin^{2}\theta_{A}\sin\theta_{B}\sin(\phi_{A}+\phi
_{B})\sin\phi_{A}\nonumber\\
&  +(1+\cos\theta_{A})\sin\theta_{A}\sin^{2}\theta_{B}\cos(\phi_{A}+\phi
_{B})\cos\phi_{B}+\nonumber\\
&  (1+\cos\theta_{A})\sin\theta_{A}\sin^{2}\theta_{B}\sin(\phi_{A}+\phi
_{B})\sin\phi_{B}\nonumber\\
&  -2(1+\cos\theta_{A})(1-\cos\theta_{B})\sin\theta_{A}\sin\theta_{B}\cos
\phi_{A}\cos\phi_{B}+\nonumber\\
&  (1-\cos\theta_{B})\sin^{2}\theta_{A}]
\end{align}

\begin{align}
&  =2[(1+\cos\theta_{A})^{2}(1-\cos\theta_{B})(1-\sin\theta_{B}\cos\phi
_{B})+(1+\cos\theta_{A})(1-\cos\theta_{B})^{2}\sin\theta_{A}\cos\phi
_{A}\nonumber\\
&  -(1-\cos\theta_{B})\sin^{2}\theta_{A}\sin\theta_{B}{\Large \{}\cos(\phi
_{A}+\phi_{B})\cos\phi_{A}+\sin(\phi_{A}+\phi_{B})\sin\phi_{A}{\Large \}}%
\nonumber\\
&  +(1+\cos\theta_{A})\sin\theta_{A}\sin^{2}\theta_{B}{\Large \{}\cos(\phi
_{A}+\phi_{B})\cos\phi_{B}+\sin(\phi_{A}+\phi_{B})\sin\phi_{B}{\Large \}}%
\nonumber\\
&  -2(1+\cos\theta_{A})(1-\cos\theta_{B})\sin\theta_{A}\sin\theta_{B}\cos
\phi_{A}\cos\phi_{B}+(1-\cos\theta_{B})\sin^{2}\theta_{A}]
\end{align}

\begin{align}
&  =2[(1+\cos\theta_{A})^{2}(1-\cos\theta_{B})(1-\sin\theta_{B}\cos\phi
_{B})+(1+\cos\theta_{A})(1-\cos\theta_{B})^{2}\sin\theta_{A}\cos\phi
_{A}\nonumber\\
&  -(1-\cos\theta_{B})\sin^{2}\theta_{A}\sin\theta_{B}\cos\phi_{B}%
+(1+\cos\theta_{A})\sin\theta_{A}\sin^{2}\theta_{B}\cos\phi_{A}\nonumber\\
&  -2(1+\cos\theta_{A})(1-\cos\theta_{B})\sin\theta_{A}\sin\theta_{B}\cos
\phi_{A}\cos\phi_{B}+(1-\cos\theta_{B})\sin^{2}\theta_{A}]
\end{align}

\begin{align}
&  =2[(1+\cos\theta_{A})^{2}(1-\cos\theta_{B})(1-\sin\theta_{B}\cos\phi
_{B})+(1-\cos\theta_{B})\sin^{2}\theta_{A}(1-\sin\theta_{B}\cos\phi
_{B})\nonumber\\
&  +(1+\cos\theta_{A})(1-\cos\theta_{B})^{2}\sin\theta_{A}\cos\phi_{A}%
+(1+\cos\theta_{A})\sin\theta_{A}\sin^{2}\theta_{B}\cos\phi_{A}\nonumber\\
&  -2(1+\cos\theta_{A})(1-\cos\theta_{B})\sin\theta_{A}\sin\theta_{B}\cos
\phi_{A}\cos\phi_{B}]
\end{align}

\begin{align}
&  =2[(1-\cos\theta_{B})(1-\sin\theta_{B}\cos\phi_{B})\{(1+\cos\theta_{A}%
)^{2}+\sin^{2}\theta_{A}\}\nonumber\\
&  +(1+\cos\theta_{A})(1-\cos\theta_{B})^{2}\sin\theta_{A}\cos\phi_{A}%
+(1+\cos\theta_{A})\sin\theta_{A}\sin^{2}\theta_{B}\cos\phi_{A}\nonumber\\
&  -2(1+\cos\theta_{A})(1-\cos\theta_{B})\sin\theta_{A}\sin\theta_{B}\cos
\phi_{A}\cos\phi_{B}]
\end{align}

\begin{align}
&  =2[2(1+\cos\theta_{A})(1-\cos\theta_{B})(1-\sin\theta_{B}\cos\phi
_{B})\nonumber\\
&  +(1+\cos\theta_{A})(1-\cos\theta_{B})^{2}\sin\theta_{A}\cos\phi_{A}%
+(1+\cos\theta_{A})\sin\theta_{A}\cos\phi_{A}\sin^{2}\theta_{B}\nonumber\\
&  -2(1+\cos\theta_{A})(1-\cos\theta_{B})\sin\theta_{A}\sin\theta_{B}\cos
\phi_{A}\cos\phi_{B}]
\end{align}

\begin{align}
&  =4[(1+\cos\theta_{A})(1-\cos\theta_{B})(1-\sin\theta_{B}\cos\phi
_{B})\nonumber\\
&  +(1+\cos\theta_{A})(1-\cos\theta_{B})\sin\theta_{A}\cos\phi_{A}%
(1-\sin\theta_{B}\cos\phi_{B})]\nonumber\\
&  =4(1+\cos\theta_{A})(1-\cos\theta_{B})(1-\sin\theta_{B}\cos\phi_{B}%
)(1+\sin\theta_{A}\cos\phi_{A}),
\end{align}
and we obtain%

\begin{align}
&  =\frac{(\beta,\gamma)}{16(1+\cos\theta_{A})(1-\cos\theta_{B})}%
\{4(1+\cos\theta_{A})(1-\cos\theta_{B})(1-\sin\theta_{B}\cos\phi_{B}%
)(1+\sin\theta_{A}\cos\phi_{A})\}\nonumber\\
&  =\frac{(\beta,\gamma)}{4}(1+\sin\theta_{A}\cos\phi_{A})(1-\sin\theta
_{B}\cos\phi_{B}).
\end{align}

\section{Appendix C}

The third term in the payoff relations (\ref{Payoff_asym}) when the game is
played with the state $\left\vert \psi_{\mathrm{ini}}\right\rangle =\frac
{1}{2}\left(  \left\vert 00\right\rangle +\left\vert 01\right\rangle
-\left\vert 10\right\rangle +\left\vert 11\right\rangle \right)  $ is given as%

\begin{gather}
\frac{(\gamma,\beta)}{16(1-\cos\theta_{A})(1+\cos\theta_{B})}{\Large \{}%
[(1-\cos\theta_{A})(1+\cos\theta_{B})+(1-\cos\theta_{A})\sin\theta_{B}\cos
\phi_{B}+\nonumber\\
(1+\cos\theta_{B})\sin\theta_{A}\cos\phi_{A}-\sin\theta_{A}\sin\theta_{B}%
\cos(\phi_{A}+\phi_{B})]^{2}+\nonumber\\
\lbrack(1-\cos\theta_{A})\sin\theta_{B}\sin\phi_{B}+(1+\cos\theta_{B}%
)\sin\theta_{A}\sin\phi_{A}-\sin\theta_{A}\sin\theta_{B}\sin(\phi_{A}+\phi
_{B})]^{2}{\large \}}.
\end{gather}

Consider its first part%

\begin{align}
&  [(1-\cos\theta_{A})(1+\cos\theta_{B})+(1-\cos\theta_{A})\sin\theta_{B}%
\cos\phi_{B}+\nonumber\\
&  (1+\cos\theta_{B})\sin\theta_{A}\cos\phi_{A}-\sin\theta_{A}\sin\theta
_{B}\cos(\phi_{A}+\phi_{B})]^{2}\nonumber\\
&  =(1-\cos\theta_{A})^{2}(1+\cos\theta_{B})^{2}+(1-\cos\theta_{A})^{2}%
\sin^{2}\theta_{B}\cos^{2}\phi_{B}\nonumber\\
&  +(1+\cos\theta_{B})^{2}\sin^{2}\theta_{A}\cos^{2}\phi_{A}+\sin^{2}%
\theta_{A}\sin^{2}\theta_{B}\cos^{2}(\phi_{A}+\phi_{B})\nonumber\\
&  +2(1-\cos\theta_{A})(1+\cos\theta_{B})(1-\cos\theta_{A})\sin\theta_{B}%
\cos\phi_{B}\nonumber\\
&  +2(1-\cos\theta_{A})(1+\cos\theta_{B})\sin\theta_{B}\cos\phi_{B}\sin
\theta_{A}\cos\phi_{A}\nonumber\\
&  -2(1+\cos\theta_{B})\sin^{2}\theta_{A}\cos\phi_{A}\sin\theta_{B}\cos
(\phi_{A}+\phi_{B})\nonumber\\
&  -2(1-\cos\theta_{A})(1+\cos\theta_{B})\sin\theta_{A}\sin\theta_{B}\cos
(\phi_{A}+\phi_{B})\nonumber\\
&  +2(1-\cos\theta_{A})(1+\cos\theta_{B})(1+\cos\theta_{B})\sin\theta_{A}%
\cos\phi_{A}\nonumber\\
&  -2(1-\cos\theta_{A})\sin\theta_{A}\sin^{2}\theta_{B}\cos\phi_{B}\cos
(\phi_{A}+\phi_{B})\label{Eq__1}%
\end{align}

Now consider the second part%

\begin{align}
&  [(1-\cos\theta_{A})\sin\theta_{B}\sin\phi_{B}+(1+\cos\theta_{B})\sin
\theta_{A}\sin\phi_{A}-\sin\theta_{A}\sin\theta_{B}\sin(\phi_{A}+\phi
_{B})]^{2}\nonumber\\
&  =(1-\cos\theta_{A})^{2}\sin^{2}\theta_{B}\sin^{2}\phi_{B}+(1+\cos\theta
_{B})^{2}\sin^{2}\theta_{A}\sin^{2}\phi_{A}\nonumber\\
&  +\sin^{2}\theta_{A}\sin^{2}\theta_{B}\sin^{2}(\phi_{A}+\phi_{B}%
)+2(1-\cos\theta_{A})(1+\cos\theta_{B})\sin\theta_{A}\sin\theta_{B}\sin
\phi_{A}\sin\phi_{B}\nonumber\\
&  -2(1+\cos\theta_{B})\sin^{2}\theta_{A}\sin\theta_{B}\sin\phi_{A}\sin
(\phi_{A}+\phi_{B})-2(1-\cos\theta_{A})\sin\theta_{A}\sin^{2}\theta_{B}%
\sin\phi_{B}\sin(\phi_{A}+\phi_{B})\nonumber\\
& \label{Eq__2}%
\end{align}

Adding Eqs.~(\ref{Eq__1},~\ref{Eq__2})%

\begin{align}
&  =(1-\cos\theta_{A})^{2}(1+\cos\theta_{B})^{2}+(1-\cos\theta_{A})^{2}%
\sin^{2}\theta_{B}\cos^{2}\phi_{B}\nonumber\\
&  +(1+\cos\theta_{B})^{2}\sin^{2}\theta_{A}\cos^{2}\phi_{A}+\sin^{2}%
\theta_{A}\sin^{2}\theta_{B}\cos^{2}(\phi_{A}+\phi_{B})\nonumber\\
&  +2(1-\cos\theta_{A})(1+\cos\theta_{B})(1-\cos\theta_{A})\sin\theta_{B}%
\cos\phi_{B}\nonumber\\
&  +2(1-\cos\theta_{A})(1+\cos\theta_{B})\sin\theta_{B}\cos\phi_{B}\sin
\theta_{A}\cos\phi_{A}\nonumber\\
&  -2(1+\cos\theta_{B})\sin^{2}\theta_{A}\cos\phi_{A}\sin\theta_{B}\cos
(\phi_{A}+\phi_{B})\nonumber\\
&  -2(1-\cos\theta_{A})(1+\cos\theta_{B})\sin\theta_{A}\sin\theta_{B}\cos
(\phi_{A}+\phi_{B})\nonumber\\
&  +2(1-\cos\theta_{A})(1+\cos\theta_{B})(1+\cos\theta_{B})\sin\theta_{A}%
\cos\phi_{A}\nonumber\\
&  -2(1-\cos\theta_{A})\sin\theta_{A}\sin^{2}\theta_{B}\cos\phi_{B}\cos
(\phi_{A}+\phi_{B})\nonumber\\
&  +(1-\cos\theta_{A})^{2}\sin^{2}\theta_{B}\sin^{2}\phi_{B}+(1+\cos\theta
_{B})^{2}\sin^{2}\theta_{A}\sin^{2}\phi_{A}\nonumber\\
&  +\sin^{2}\theta_{A}\sin^{2}\theta_{B}\sin^{2}(\phi_{A}+\phi_{B}%
)+2(1-\cos\theta_{A})(1+\cos\theta_{B})\sin\theta_{A}\sin\theta_{B}\sin
\phi_{A}\sin\phi_{B}\nonumber\\
&  -2(1+\cos\theta_{B})\sin^{2}\theta_{A}\sin\theta_{B}\sin\phi_{A}\sin
(\phi_{A}+\phi_{B})-\nonumber\\
&  2(1-\cos\theta_{A})\sin\theta_{A}\sin^{2}\theta_{B}\sin\phi_{B}\sin
(\phi_{A}+\phi_{B})
\end{align}

\begin{align}
&  =(1-\cos\theta_{A})^{2}(1+\cos\theta_{B})^{2}\nonumber\\
&  +(1-\cos\theta_{A})^{2}\sin^{2}\theta_{B}\cos^{2}\phi_{B}+(1-\cos\theta
_{A})^{2}\sin^{2}\theta_{B}\sin^{2}\phi_{B}\nonumber\\
&  +(1+\cos\theta_{B})^{2}\sin^{2}\theta_{A}\cos^{2}\phi_{A}+(1+\cos\theta
_{B})^{2}\sin^{2}\theta_{A}\sin^{2}\phi_{A}\nonumber\\
&  +\sin^{2}\theta_{A}\sin^{2}\theta_{B}\cos^{2}(\phi_{A}+\phi_{B})+\sin
^{2}\theta_{A}\sin^{2}\theta_{B}\sin^{2}(\phi_{A}+\phi_{B})\nonumber\\
&  +2(1-\cos\theta_{A})(1+\cos\theta_{B})(1-\cos\theta_{A})\sin\theta_{B}%
\cos\phi_{B}\nonumber\\
&  +2(1-\cos\theta_{A})(1+\cos\theta_{B})\sin\theta_{A}\sin\theta_{B}\cos
\phi_{A}\cos\phi_{B}+\nonumber\\
&  2(1-\cos\theta_{A})(1+\cos\theta_{B})\sin\theta_{A}\sin\theta_{B}\sin
\phi_{A}\sin\phi_{B}\nonumber\\
&  -2(1+\cos\theta_{B})\sin^{2}\theta_{A}\sin\theta_{B}\cos\phi_{A}\cos
(\phi_{A}+\phi_{B})-\nonumber\\
&  2(1+\cos\theta_{B})\sin^{2}\theta_{A}\sin\theta_{B}\sin\phi_{A}\sin
(\phi_{A}+\phi_{B})\nonumber\\
&  -2(1-\cos\theta_{A})(1+\cos\theta_{B})\sin\theta_{A}\sin\theta_{B}\cos
(\phi_{A}+\phi_{B})+\nonumber\\
&  2(1-\cos\theta_{A})(1+\cos\theta_{B})(1+\cos\theta_{B})\sin\theta_{A}%
\cos\phi_{A}\nonumber\\
&  -2(1-\cos\theta_{A})\sin\theta_{A}\sin^{2}\theta_{B}\cos\phi_{B}\cos
(\phi_{A}+\phi_{B})-\nonumber\\
&  2(1-\cos\theta_{A})\sin\theta_{A}\sin^{2}\theta_{B}\sin\phi_{B}\sin
(\phi_{A}+\phi_{B})
\end{align}

\begin{align}
&  =(1-\cos\theta_{A})^{2}(1+\cos\theta_{B})^{2}+(1-\cos\theta_{A})^{2}%
\sin^{2}\theta_{B}\nonumber\\
&  +(1+\cos\theta_{B})^{2}\sin^{2}\theta_{A}+\sin^{2}\theta_{A}\sin^{2}%
\theta_{B}\nonumber\\
&  +2(1-\cos\theta_{A})(1+\cos\theta_{B})(1-\cos\theta_{A})\sin\theta_{B}%
\cos\phi_{B}\nonumber\\
&  +2(1-\cos\theta_{A})(1+\cos\theta_{B})\sin\theta_{A}\sin\theta_{B}[\cos
\phi_{A}\cos\phi_{B}+\sin\phi_{A}\sin\phi_{B}]\nonumber\\
&  -2(1+\cos\theta_{B})\sin^{2}\theta_{A}\sin\theta_{B}[\cos\phi_{A}\cos
(\phi_{A}+\phi_{B})+\sin\phi_{A}\sin(\phi_{A}+\phi_{B})]\nonumber\\
&  +2(1-\cos\theta_{A})(1+\cos\theta_{B})(1+\cos\theta_{B})\sin\theta_{A}%
\cos\phi_{A}-\nonumber\\
&  2(1-\cos\theta_{A})(1+\cos\theta_{B})\sin\theta_{A}\sin\theta_{B}\cos
(\phi_{A}+\phi_{B})\nonumber\\
&  -2(1-\cos\theta_{A})\sin\theta_{A}\sin^{2}\theta_{B}[\cos\phi_{B}\cos
(\phi_{A}+\phi_{B})+\sin\phi_{B}\sin(\phi_{A}+\phi_{B})]
\end{align}

\begin{align}
&  =(1-\cos\theta_{A})^{2}(1+\cos\theta_{B})^{2}+(1-\cos\theta_{A})^{2}%
(1-\cos^{2}\theta_{B})\nonumber\\
&  +(1+\cos\theta_{B})^{2}(1-\cos^{2}\theta_{A})+(1-\cos^{2}\theta_{A}%
)(1-\cos^{2}\theta_{B})\nonumber\\
&  +2(1-\cos\theta_{A})(1+\cos\theta_{B})(1-\cos\theta_{A})\sin\theta_{B}%
\cos\phi_{B}\nonumber\\
&  +2(1-\cos\theta_{A})(1+\cos\theta_{B})\sin\theta_{A}\sin\theta_{B}\cos
(\phi_{B}-\phi_{A})\nonumber\\
&  -2(1+\cos\theta_{B})\sin^{2}\theta_{A}\sin\theta_{B}\cos\phi_{B}\nonumber\\
&  +2(1-\cos\theta_{A})(1+\cos\theta_{B})(1+\cos\theta_{B})\sin\theta_{A}%
\cos\phi_{A}\nonumber\\
&  -2(1-\cos\theta_{A})(1+\cos\theta_{B})\sin\theta_{A}\sin\theta_{B}\cos
(\phi_{A}+\phi_{B})\nonumber\\
&  -2(1-\cos\theta_{A})\sin\theta_{A}\sin^{2}\theta_{B}\cos\phi_{A}%
\end{align}

\begin{align}
&  =(1-\cos\theta_{A})(1+\cos\theta_{B})[(1-\cos\theta_{A})(1+\cos\theta
_{B})]\nonumber\\
&  +(1-\cos\theta_{A})(1+\cos\theta_{B})[(1-\cos\theta_{A})(1-\cos\theta
_{B})]\nonumber\\
&  +(1-\cos\theta_{A})(1+\cos\theta_{B})[(1+\cos\theta_{A})(1+\cos\theta
_{B})]\nonumber\\
&  +(1-\cos\theta_{A})(1+\cos\theta_{B})[(1+\cos\theta_{A})(1-\cos\theta
_{B})]\nonumber\\
&  +2(1-\cos\theta_{A})(1+\cos\theta_{B})(1-\cos\theta_{A})\sin\theta_{B}%
\cos\phi_{B}\nonumber\\
&  +2(1-\cos\theta_{A})(1+\cos\theta_{B})\sin\theta_{A}\sin\theta_{B}\cos
(\phi_{B}-\phi_{A})\nonumber\\
&  -2(1-\cos\theta_{A})(1+\cos\theta_{B})(1+\cos\theta_{A})\sin\theta_{B}%
\cos\phi_{B}\nonumber\\
&  +2(1-\cos\theta_{A})(1+\cos\theta_{B})(1+\cos\theta_{B})\sin\theta_{A}%
\cos\phi_{A}\nonumber\\
&  -2(1-\cos\theta_{A})(1+\cos\theta_{B})\sin\theta_{A}\sin\theta_{B}\cos
(\phi_{A}+\phi_{B})\nonumber\\
&  -2(1-\cos\theta_{A})(1+\cos\theta_{B})(1-\cos\theta_{B})\sin\theta_{A}%
\cos\phi_{A}%
\end{align}

\begin{align}
&  =(1-\cos\theta_{A})(1+\cos\theta_{B})[(1-\cos\theta_{A})(1+\cos\theta
_{B})]\nonumber\\
&  +(1-\cos\theta_{A})(1+\cos\theta_{B})[(1-\cos\theta_{A})(1-\cos\theta
_{B})]\nonumber\\
&  +(1-\cos\theta_{A})(1+\cos\theta_{B})[(1+\cos\theta_{A})(1+\cos\theta
_{B})]\nonumber\\
&  +(1-\cos\theta_{A})(1+\cos\theta_{B})[(1+\cos\theta_{A})(1-\cos\theta
_{B})]\nonumber\\
&  +2(1-\cos\theta_{A})(1+\cos\theta_{B})(1-\cos\theta_{A})\sin\theta_{B}%
\cos\phi_{B}\nonumber\\
&  +2(1-\cos\theta_{A})(1+\cos\theta_{B})\sin\theta_{A}\sin\theta_{B}\cos
(\phi_{B}-\phi_{A})\nonumber\\
&  -2(1-\cos\theta_{A})(1+\cos\theta_{B})(1+\cos\theta_{A})\sin\theta_{B}%
\cos\phi_{B}\nonumber\\
&  +2(1-\cos\theta_{A})(1+\cos\theta_{B})(1+\cos\theta_{B})\sin\theta_{A}%
\cos\phi_{A}\nonumber\\
&  -2(1-\cos\theta_{A})(1+\cos\theta_{B})\sin\theta_{A}\sin\theta_{B}\cos
(\phi_{A}+\phi_{B})\nonumber\\
&  -2(1-\cos\theta_{A})(1+\cos\theta_{B})(1-\cos\theta_{B})\sin\theta_{A}%
\cos\phi_{A}.
\end{align}
Dividing by $(1-\cos\theta_{A})(1+\cos\theta_{B})$ gives%

\begin{align}
&  =(1-\cos\theta_{A})(1+\cos\theta_{B})+(1-\cos\theta_{A})(1-\cos\theta
_{B})\nonumber\\
&  +(1+\cos\theta_{A})(1+\cos\theta_{B})+(1+\cos\theta_{A})(1-\cos\theta
_{B})\nonumber\\
&  +2(1-\cos\theta_{A})\sin\theta_{B}\cos\phi_{B}+2\sin\theta_{A}\sin
\theta_{B}\cos(\phi_{B}-\phi_{A})\nonumber\\
&  -2(1+\cos\theta_{A})\sin\theta_{B}\cos\phi_{B}+2(1+\cos\theta_{B}%
)\sin\theta_{A}\cos\phi_{A}\nonumber\\
&  -2\sin\theta_{A}\sin\theta_{B}\cos(\phi_{A}+\phi_{B})-2(1-\cos\theta
_{B})\sin\theta_{A}\cos\phi_{A}%
\end{align}

\begin{align}
&  =2(1-\cos\theta_{A})+2(1+\cos\theta_{A})\nonumber\\
&  +2\sin\theta_{B}\cos\phi_{B}[(1-\cos\theta_{A})-(1+\cos\theta
_{A})]\nonumber\\
&  +2\sin\theta_{A}\sin\theta_{B}\cos(\phi_{B}-\phi_{A})+2\sin\theta_{A}%
\cos\phi_{A}[(1+\cos\theta_{B})-(1-\cos\theta_{B})]\nonumber\\
&  -2\sin\theta_{A}\sin\theta_{B}\cos(\phi_{A}+\phi_{B})
\end{align}

\begin{align}
&  =2(1-\cos\theta_{A})+2(1+\cos\theta_{A})-4\sin\theta_{B}\cos\phi_{B}%
\cos\theta_{A}\nonumber\\
&  +2\sin\theta_{A}\sin\theta_{B}\cos(\phi_{B}-\phi_{A})+4\sin\theta_{A}%
\cos\phi_{A}\cos\theta_{B}\nonumber\\
&  -2\sin\theta_{A}\sin\theta_{B}\cos(\phi_{A}+\phi_{B})
\end{align}

\begin{align}
&  =4-4\sin\theta_{B}\cos\phi_{B}\cos\theta_{A}+2\sin\theta_{A}\sin\theta
_{B}\cos(\phi_{B}-\phi_{A})\nonumber\\
&  -2\sin\theta_{A}\sin\theta_{B}\cos(\phi_{A}+\phi_{B})+4\sin\theta_{A}%
\cos\phi_{A}\cos\theta_{B}%
\end{align}

\begin{align}
&  =4-4\sin\theta_{B}\cos\phi_{B}\cos\theta_{A}+4\sin\theta_{A}\sin\theta
_{B}\sin\phi_{A}\sin\phi_{B}+4\sin\theta_{A}\cos\phi_{A}\cos\theta
_{B}\nonumber\\
&  =4(1-\sin\theta_{B}\cos\phi_{B}\cos\theta_{A}+\sin\theta_{A}\sin\theta
_{B}\sin\phi_{A}\sin\phi_{B}+\sin\theta_{A}\cos\phi_{A}\cos\theta_{B}).
\end{align}
The third term in the payoffs becomes%

\begin{equation}
\frac{(\gamma,\beta)}{4}(1-\sin\theta_{B}\cos\phi_{B}\cos\theta_{A}+\sin
\theta_{A}\sin\theta_{B}\sin\phi_{A}\sin\phi_{B}+\sin\theta_{A}\cos\phi
_{A}\cos\theta_{B}).
\end{equation}

\end{document}